\documentclass[a4paper,12pt]{article}
\usepackage{graphicx}
\usepackage{scrextend}
\addtokomafont{labelinglabel}{\sffamily}
\usepackage{lnfprep}
\usepackage{enumitem}
\usepackage[top=2.5cm, bottom=3cm, left=3cm, right=3cm]{geometry}
\usepackage[normalem]{ulem} 
\usepackage {ulem} 
\usepackage{wrapfig}
\usepackage[greek,english]{babel}
\usepackage{teubner}
 \usepackage[unicode]{hyperref}
 \usepackage{verbatim} 
\usepackage{cleveref}
\usepackage{natbib}
\bibpunct[;]{(}{)}{,}{a}{}{;}
 \usepackage{type1cm}   
 \usepackage{array} 
\hypersetup{ colorlinks=true,  linkcolor=blue,  filecolor=magenta,       urlcolor=blue, citecolor=black}

\usepackage{mathptmx}
\usepackage[intlimits]{amsmath}
\usepackage{helvet}
\usepackage{courier}
\usepackage{type1cm}  
\usepackage[font={small}]{caption}
\usepackage{makeidx}         
\usepackage{multicol}        
\usepackage[bottom]{footmisc}
\def\BT{Bruno Touschek}
\def\RW{Rolf Wider\o e}

\def\LAL{Laboratoire de l'Acc\'el\'erateur Lin\'eaire}
\def\FJ{Fr\'ed\'eric Joliot}

\def\CdF{Coll\`ege de France}

\def\WW2{World War II}

\def\EYT{Elspeth Yonge Touschek}
\def\RLM{Reichsluftfahrtministerium}
\def\Archiv{Archiv f\"ur Elektrotechnik}
\def\Gott{G\"ottingen}
\def\tr{\textcolor{red}}
\def\tm{\textcolor{magenta}}
\def\tb{\textcolor{blue}}
\def\bef{\begin{figure}}
\def\enf{\end{figure}}
\def\befoot{\begin{footnotesize}}
\def\enfoot{\end{footnotesize}}
\def\UGA{University of Glasgow Archives \& Special Collections, University collection, GB 248 DC 157/18/56}
\def\CHA{Churchill Archives Center of   Churchill College,  Cambridge University}
\def\RSUA{``Edoardo Amaldi Archives'', Physics Department, Sapienza University of Rome} 
\def\LNF{INFN Frascati National Laboratories}
\setcitestyle{notesep={, }}
\begin{document}
\begin{titlepage}
\title
 {
   \begin{flushright}
  {\underline{\bf INFN-2020-03/LNF}}\\ 
  \vskip 0,2 cm
 {\underline{\bf MIT-CTP/5201}}    
\end{flushright}
	\vskip 3,5 cm
 {\Large {\bf \bf Bruno Touschek in Glasgow.} \\ The making of a theoretical physicist}
}

\author{ Giulia Pancheri $^1$,
   Luisa Bonolis$^2$\\
{\it ${}^{1)}$INFN, Laboratori Nazionali di Frascati, P.O. Box 13,
I-00044 Frascati, Italy}\\
{\it ${}^{2)}$
Max Planck Institute for the History of Science, Boltzmannstra\ss e 22, 14195 Berlin, Germany}} 

\date{February 28, 2020}
\maketitle

\begin{abstract}
In the history of the  discovery tools of last century  particle physics, central stage is taken by elementary particle accelerators and in particular by colliders.
In their  start and early development, a major role was played by the Austrian born Bruno Touschek, who proposed and built the first electron positron collider, AdA, in Italy, in 1960. In this note, we   present a period of Touschek's life barely explored in the literature, namely the five  years he spent at University of Glasgow, first to obtain his doctorate in 1949 and then as a lecturer.    We shall highlight his formation as a theoretical physicist, his contacts and correspondence with Werner Heisenberg in \Gott \ and  Max Born in Edinburgh, as well as his close involvement with colleagues intent on building modern  particle accelerators in Glasgow, Malvern,  Manchester and Birmingham.    We shall discuss how the Fuchs affair, which unraveled in early 1950,  may have influenced his decision to leave the UK, and how  contacts with the Italian physicist  Bruno Ferretti  led  Touschek to  join the Guglielmo Marconi Physics Institute  of University of Rome   in January  1953.
\end{abstract}
\vskip 0.5cm
\begin{center}
{\it Ich will ein Physiker werden}  
\\{\it I want  to become a physicist}, Bruno Touschek, 1946
\end{center}
\vskip 2cm
\rule{15.0cm}{0.09mm}\\
\begin{footnotesize}
e-mail:lbonolis@mpiwg-berlin.mpg.de,pancheri@lnf.infn.it. Authors' ordering in this and related works alternates to reflect that this work is part of a joint collaboration project with no principal author.
\end{footnotesize}
\end{titlepage}
\pagestyle{plain}

\newpage
\tableofcontents

\section*{Premise}
This note\footnote{A version with higher resolution images and extra figures  can be found at \url{http://www.lnf.infn.it/sis/preprint/search.php}.} is part of a project whose aim is  to contribute to the knowledge of how present day high energy particle accelerators were conceived  and constructed in the second half of the {20th century} in Europe, the United States, Japan, the USSR. One of the early protagonists of this story is the Austrian born  theoretical physicist Bruno Touschek, \begin{wrapfigure}{r}{0.5\textwidth}
\centering
  \includegraphics[scale=0.9]{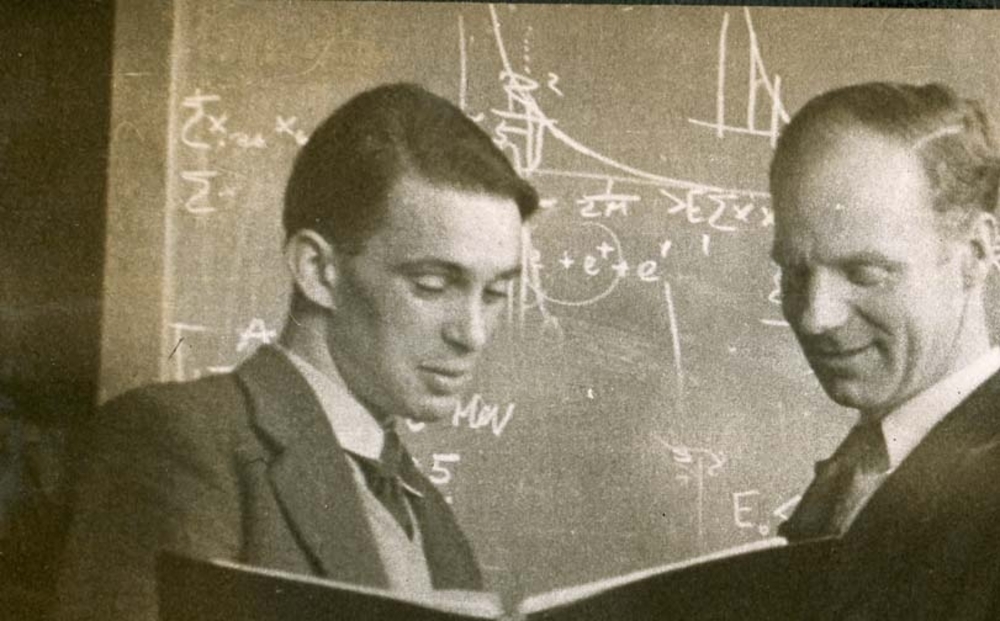}
\caption{Bruno Touschek with Samuel Curran, at University of Glasgow, around 1948. Photograph courtesy of Mrs. \EYT.}
\label{fig:BT-Curran}
 \end{wrapfigure}
 who built the first electron positron collider, in Frascati, Italy, in 1960.

 As we get close to the hundredth anniversary of his birth   on February 3rd,  1921, we are preparing and posting a set of notes, which highlight some of the most important periods in his life. 
In particular, this note will highlight the period he spent at University of Glasgow, from 1947 to December  1952.  We show him in a contemporary photograph with his colleague 
\href{https://www.worldchanging.glasgow.ac.uk/article/?id=39}{Samuel Curran},  in Fig.~\ref{fig:BT-Curran}. 

A  full length biography,  was published three years after Touschek  passed away \citep{Amaldi:1981}. During the years of his maturity,  Touschek was very famous in the particle accelerator community. His work led  to the construction of  powerful machines, which then opened the way to the formulation of the Standard Model (SM)  of elementary particle physics. While Amaldi's biography is  unparalleled and constitutes the  basis for any more recent work,  forty years have  passed since its  publication. Using {newly found material and archival sources}, we  have started examining Touschek's life  in the context of the great scientific developments in the field of particle accelerator physics which took place during last century \citep{Pancheri:2004,Bonolis:2005,Bonolis:2011wa,Bonolis:2018gpn,Pancheri:2018xdl}. This  development has a number of major milestones in particle colliders, marked by the discovery of new particles:  the charm quark in 1974, at Brookhaven with  a traditional accelerator, but also at the  electron-positron collider SPEAR  at Stanford and confirmed at the electron-positron collider ADONE in Frascati  a few days later,  the   bottom quark  in 1977 at the ${\bar p} p$ Tevatron collider  in FermiLab,  the $W$ and $Z$ bosons in 1983 at  the CERN $S{\bar p}pS$,  the top quark  in 1989  in FermiLab,  the precision measurements of the Z-boson at the  CERN Large Electron Positron  (LEP) collider through the 1990's, the gluon at the $e-p$ collider HERA in DESY, and finally the Higgs boson in 2012 at the CERN Large Hadron Collider (LHC).  

Unfortunately, Bruno  died young. His early death, at only 57 years of age, was  a deep loss for all who knew him. In 1987,  one of his friends said that he had contributed to the {\it Quattrocento} of particle physics, but could not see the {\it Rinascimento}, the Renaissance, which was born of it \citep{Salvini:2004aa}. 

Bruno Touschek's work and life cross  Europe in space and time: from  Austria to  Germany, United Kingdom, Italy, and France, through World War II  and the  beginning of the great particle discoveries of the 1970s.

He  was born in Vienna, with a mixed heritage: his mother, Camilla Weltmann,  who died when Bruno was a young boy, came from a Jewish family immersed in the artistic background of the Vienna Secession movement,  his father, Franz Xaver Touschek,   was a Staff  Officer in the Austrian Army.  His  studies were disrupted by the annexation of Austria to Germany, the Anschluss,  which took place in 1938, and  were definitely interrupted in June 1940, when, because of his Jewish origin on the maternal side,  he was expelled from  the University of Vienna, where he had started  physics studies  in September 1939. 
Upon Arnold  Sommerfeld's suggestion, his desire to continue studying physics   led him to move to Germany, where his paternal name could hide his Jewish origin, and allow  him to do some odd jobs, while attending physics classes, held by Sommerfeld's  earlier pupils,  in Berlin and Hamburg. During the war, he came in contact with \RW, the Norwegian scientist who had built the first linear collider and invented the betatron in 1928.  In the last days of the war, he  suffered imprisonment and narrowly escaped death.

 In our most recent note, we have described Touschek's  life after  liberation  by the Allied Forces in April 1945, and the period of his life until the end of 1946 \citep{Bonolis:2019qqh}. In the present note, we shall pick up our story from there. 
 \newpage
\section{Introduction}
In the fall of the academic year 1946-47, 
the Principal of  Glasgow University welcomed a 
\begin{wrapfigure}{r}{0.37\textwidth}
\centering
    \includegraphics[width=0.37\textwidth] {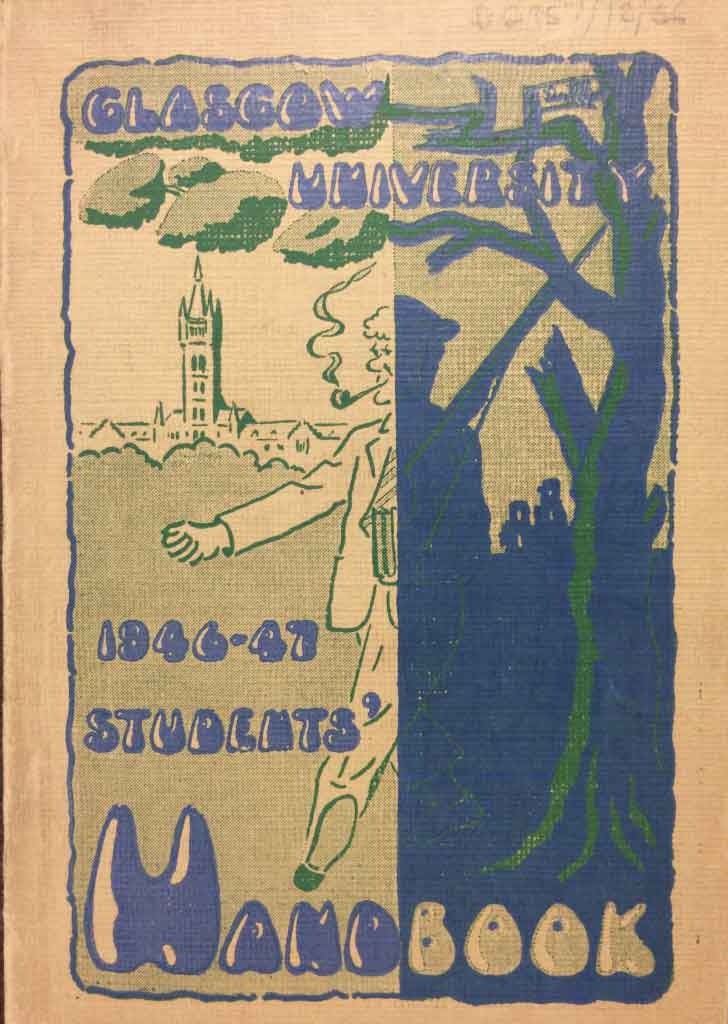}
    \caption{Front page of University of Glasgow 1946-47 Student Handbook. Courtesy of \UGA.}
  \label{fig:studenthandbookfrontpage}
    \end{wrapfigure}
very special  student cohort, the one which  was returning from the war. As he wrote in the preface to the 1946-47 Student Handbook, 
he welcomed not only the new first year students, but equally warmly those who would be  resuming the course of studies interrupted by the war.  ``These are sombre and anxious days'' he added,  ``as they must be after the years of loss and destruction'', but the process of recovery had started and, although possibly  long and hard and disappointing,  he hoped that the new students would set a standard for the future. The Principal's words are poignantly reflected in the Student Hand book cover  in Fig.~\ref{fig:studenthandbookfrontpage}, which  shows  the  transition from the  war scenario on the right side, to the purposeful stride of a confident young man at left,  under the benign eye of the University tower. 

Among this new class, joining in the Spring as a graduate student, there would be Bruno Touschek \citep{Amaldi:1981,Bonolis:2011wa}.   He  had gone through the   disruption of his life, suffered loss and fear, and now
he was moving ahead to a new country, a country which  had won the war. Unlike Germany, still under the weight of restrictions and lacking resources, the UK was  ready to pour energy and money into   new  
roads, which  scientists, coming back to basic research after their  war engagement, were eager to follow.

When the preface was written on July 19th, 1946,  Bruno was still in \Gott, had just  submitted his Diploma thesis and  was worrying about the  future directions his studies would take. 
In the meantime, he was working as assistant to Werner Heisenberg, one of the founders of Quantum Mechanics and one of the most illustrious German scientists,  a key theoretical figure in the German nuclear project,  and, since 1942, official director of the Kaiser Wilhelm Institute for Physics in Berlin-Dahlem \citep{Cassidy:1993aa,Cassidy:2017}.\footnote{ After being taken into custody by the Allied Scientific Intelligence Mission, code-named Alsos, on May 4th, Heisenberg had been kept  secluded  at the country estate Farm Hall in UK with other key members of the ``Uranium Club'' and brought back to G\"ottingen in early spring 1946. In G\"ottingen, in the years to follow, Heisenberg would devote himself  to two large tasks: the reconstruction of the Kaiser Wilhelm Institut f\"ur Physik as a center for experimental and theoretical research in physics and the renewal of scientific research in Germany, where, during these early post-war years,  research was  limited by the directives of the Allied Control Commission.}

The  year 1946   had been very hard for Bruno, who was torn between a desire to remain to study in \Gott, under very dire monetary circumstances,  and the worry about his parents' precarious situation in the city of Vienna, under Soviet occupation. His  unique expertise in accelerator science, gained working with Rolf Wider\o e \citep{Wideroe:1994,Sorheim:2020} during WWII, decided for him.  In early January 1946, in fact, the British accelerator program was being defined,  and a project for constructing an electron synchrotron  at Glasgow University took definite form.  Touschek was one of few experts in the field present in Germany at the time, and his  knowledge was prized by the Allied T-force, who had advised  for him to be moved to the UK \citep{Bonolis:2019qqh}. After interrogations in Wimbledon, Touschek  had then been moved from Kellinghusen to \Gott, where Heisenberg was starting the reconstruction of scientific life in Germany and,  on April 1st 1946,  had been  in Glasgow, where he expected to stay as a research assistant for about a year. However, problems   arising  from his Austrian citizenship,  blocked this appointment,  as discussed in \citep{Bonolis:2019qqh}.
While the paper handling in the UK was being sorted out, he had  to  go back to \Gott\, where   he  obtained his Physics Diploma in June. He    spent the rest of the year as Heisenberg's assistant \citep{Amaldi:1981,Bonolis:2019qqh}, worried about  his immediate future.

 The uncertainty about a future,  which was not  in his power to direct, and the  desire to continue his physics studies made him oscillate between hopes  of   remaining in \Gott \ and  obtain  a doctorate under Heisenberg or  moving to Glasgow, where he would face unknown surroundings and, in some ways, a less prestigious prospect.  At the same time,  Glasgow would provide a clean break from the past, and be financially more rewarding. It is very likely that, once the decision had been taken (by the T-force),   Bruno  was happy to leave Germany, which was  still under the deprivations and gloom of the war. He  had finally made true  his old dream of studying in the UK, would have a reasonable salary, and thus could help his family, as he had wanted all along. He  would  join  a great and ancient  University, and be  part of   the British post-war  effort 
focused on bringing  to peacetime use  the scientific  successes  and technological innovations  which  had won the war to the Western Alliance.

In early 1947 Touschek was  aware  of  the approaching engagement with  Glasgow University, but was also  hoping  to continue in \Gott \ towards  a doctorate under Werner Heisenberg's guidance.
This would not happen. Firstly, the financial situation of Heisenberg's Institute was still a difficult one, most probably
not allowing the possibility of having PhD students, but mostly,
events, which  had started  early in 1946 and are discussed  in Sec.~ \ref{sec:UKbackstage},  began to  unravel.  In January 1947,  the University of Glasgow Senate  submitted to the University Court a proposal for  the construction   of a synchrotron in the University, whose development, supply and erection would be  provided by the  Department of Scientific and Industrial Research (DSIR), together with a    capital grant of 3,000 pounds towards the provision of a high voltage generator. 
The proposal was approved and the decision was taken to build a synchrotron as proposed by the DSIR, and appoint the famous Scottish architect Mr.~Basil Spence to design the expansion to house the synchrotron.
This was part of the project to build an extension to the original  Natural Philosophy Building having the purpose to establish the Department of Natural Philosophy ``as one of the pre-eminent physics departments in the UK.''\footnote{From Minutes of University of Glasgow Court 1946-47,  January 23rd 1947. Courtesy of \UGA Archives. News  from the  Glasgow University Archive Services with a later photo of the Building can be found at \url{https://www.theglasgowstory.com/image/?inum=TGSD00214}. The first phase was completed around 1954 and consisted of a new lecture room and the purpose built accommodation for the synchrotron. A 2007  journalistic assessment  of Spence's work can be found at \url{https://www.theguardian.com/society/2007/oct/16/communities}.}

  Following this decision, Philip I. Dee, the chair  of Department of Natural Philosophy of the University of Glasgow was finally able to carry through the appointment of Bruno, a researcher of exceptional promise, as Dee himself noted in the submission forms a few months later.

 Bruno Touschek spent almost 6 years in Glasgow, from April 1947 until December 1952. Little has been  known until now about this period of Touschek's life and the crucial influence it had on Touschek's later emerging in Rome as full fledged theoretical physicist, whose exceptional intelligence and  sarcastic wit  would soon conquer his Roman colleagues, and make him  one of the major stars of an Institute bent on taking a central role in the physics scenario of the new Europe. In Edoardo Amaldi 's work \citep[9-12]{Amaldi:1981},  only   a few pages describe  the  period Bruno Touschek spent in Glasgow.  Amaldi cites the copious scientific output, more than 15  published articles   with many different co-workers, and, as a personal note, reproduces a long letter by Philip Dee, which describes Bruno as brilliant but temperamental, able to do clean and tidy scientific work, but occasional unrestrained by normal rules of  behaviour.      This period of Bruno's life seemed almost wasted, or at least irrelevant to his future as the undisputed genius of later years. His scientific work in the Glasgow period  appears  even  forgettable, with the  sole exception of the article  with Walter Thirring on the Bloch and Nordsieck theorem \citep{Thirring:1951cz}.  And yet,  these are the years when  Touschek became a theoretical physicist and met the Italian theorist Bruno Ferretti and, through him, found his way to Rome.
 
 Indeed, as we continued in our search  for clues to Touschek's genius and tragic life, and
 to  the genesis of the development of particle  beam colliders, we discovered the importance of  this period  in Touschek's life. In particular, inspection of available archival sources 
highlighted   the relationship Touschek had with Werner Heisenberg and Max Born during his Glasgow years,   and revealed the influence  these two giants of 20th century physics  had on Touschek's  formation as a theorist. 
  
Building on this, and placing  Touschek in the context of historical events around him, such as the Fuchs affair, and the effect  it had on UK physics, we  now present  a
  new insight  on  Touschek's life in Glasgow,  which  shows  how his destiny became entangled and  influenced by his larger surroundings. Far from the image of a studious but nervous and eccentric  young man, as  described by Philip Dee  in the letter  to  Edoardo Amaldi,  we now see him as the young protagonist of a future great adventure, well aware of  the mainstream of world events and  physics developments, bent on his studies and anxious  to recover the 
  years    he had lost when he was expelled in June 1940 from the University of Vienna, because of the Jewish origin from his  maternal side.  These years, lost through the disruptions  arising from  anti-semitism, had already  started in Vienna  with the Anschluss in 1938,  had then culminated in Germany  with his  imprisonment in the Fuhlsb\"uttel prison in March 1945, and  the improbable escape from death on the way to the Kiel concentration camp \citep{Amaldi:1981,Bonolis:2011wa}.

In what follows, our aim will be to unravel what made Bruno into the scientist who in 1960 merged   theory and experimental insight to invent AdA, the first electron-positron collider, have it built  and led  to success in 1964.  We will outline how the theoretical foundations of his  later work  in Italy   were laid during the Glasgow years, and highlight  his life-long interest in  the radiation electrons emit, when accelerated to very high energies. We shall see the many close exchanges with the UK accelerator scientists during this period, which furthered  his   deep   understanding of   how  electrons in a  beam  behave as they move under acceleration,  and his personal exchanges  with two great theoretical physicists such as Werner Heisenberg and Max Born.
All this   would come back to him in the 1960s, when AdA was conceived and built, a proof-of-concept  leading to the construction of new   powerful   electron positron colliders  in Italy, France, the United States, USSR,  and, later,  everywhere else in the world.

In order to adequately  address Touschek's five years in Glasgow, we shall start with an overview of   the UK scientific environment after the war, and  contemporary developments  in particle physics. 
It was a time of great discoveries and great advancements in the field of what was then called nuclear or meson physics, and a vast literature about it exists. We have tried to highlight the parts most relevant to the story we want to present. This said,  Sec.~\ref{sec:UKbackstage}, which follows, implies some pre-existing knowledge  of particle physics  and its language, and  interested readers from other fields  may   prefer to skip it, except to find in the quoted literature longer and more exhaustive descriptions of the physics of this period.

\section{Backstage in UK and crucial particle physics developments elsewhere} 
\label{sec:UKbackstage}
Touschek's move to Glasgow can be seen in the context of the   extensive program of particle accelerator construction in which the UK had embarked  soon after the end of the war. %
Initially motivated by the possibility of having more powerful machines for investigations at a nuclear level, preliminary plans were further boosted by the discovery of new elementary particles in cosmic ray studies which definitely set the stage for subnuclear physics.

Up to 1945, particle accelerators in use were the van de Graaff, the Cockcroft-Walton, the cyclotron and the betatron. A new era in the field was inaugurated with the principle of phase stability, proposed in 1945 by Edwin W. McMillan in the US and, independently, by Vladimir Veksler in the USSR \citep{McMillan:1945aa,Veksler:1946aa}. This idea would allow to accelerate particles into the GeV region, which was out of the reach of cyclotrons. In these machines, which up to that time had been able to produce the highest energies, the relativistic mass increase with high velocities resulted in an energy limit of about 25 MeV for protons. This obstacle was overcome by  the new principle, which opened the way to a completely new type of accelerator, the synchrotron, and also allowed to convert cyclotrons into synchro-cyclotrons, operating them at much higher energies. A remarkable advantage of such device was the possibility of accelerating both protons and electrons, and in any case, all the basic knowledge already acquired through betatrons was instrumental for the functioning of electron synchrotrons. All this, combined with the outstanding role of nuclear physics as a research field, achieved during the war, became a powerful trigger for the construction of new accelerators. As we will see, the new type of machine was immediately included in the British nuclear physics program.

By 1945, eight particles were known: the electron, the positron, the proton, the neutron, the photon, the neutrino, and the so called positive and negative mesotron -- as it was termed the very penetrating component of local cosmic rays because of its mass of 200 electron masses, intermediate between that of electron and proton -- which was wrongly thought at the time to be Hideki Yukawa's meson, the predicted field quantum mediating the strong interaction. As such mesons were believed to be associated with the extraordinary attractive forces binding together the neutrons and protons of the nucleus, the new challenge was to make mesons in the laboratory, and there to study them in the number and the detail which was impossible while their source was still only the cosmic rays. The needed energy was at least 200 MeV, because of their rest energy, known to be 200 times the electron rest energy. But more was of course required to make them in some quantity, and so a popular target for the builders of machines was 300 MeV. As we shall see next, this minimum threshold  clarifies  the energy choices for the Glasgow and  Liverpool synchrotrons which the post-war UK scientists  planned to build.

\subsection{The UK accelerator program in 1945-46}
The UK accelerator program  was part of a wider postwar atomic energy program. It was  started with the decision in October 1945 to establish the Atomic Energy Research Establishment (AERE), to be built in a site near Harwell, which would become the main center for atomic energy research and development in the UK At the time, UK was the leading nuclear power in western Europe, having  the financial, industrial and technical resources to launch such a large-scale project, which would include also a number of accelerators. Applied nuclear physics was the basic task of the new center at Harwell, directed by Sir John Cockcroft.\footnote{In 1929, John Cockcroft and Ernest Walton, working at Ernest Rutherford's Cavendish Laboratory at Cambridge University, started to build the first `Cockcroft-Walton accelerator', as it was named since then, a system of capacitors and thermionic rectifiers capable of 600 KV when it was used in 1932 to bombard lithium and beryllium targets with high-energy protons achieving the first artificial disintegration of an atomic nucleus and the first artificial  transmutation on one element (lithium) into another (helium) \citep{CockcroftWalton:1932aa}. This machine was still used much later to supply voltage in large particle accelerators.}  Cockcroft had come back from war work at Chalk River, in Canada, where a heavy-water nuclear reactor to manufacture plutonium and enriched uranium had been built. At AERE, where the first nuclear reactor in western Europe started up in August 1947, Cockcroft supervised the construction of various reactors.
But in parallel with nuclear reactors, a variety of accelerators had also been planned to be built there, to be used for more applied tasks of pure nuclear nature, like the production of neutrons by photo-disintegration or the determination of cross-section of neutron induced reactions. They were of interest also for radiotherapy and for the study of high energy X-rays.

Between October 1945 and March 1946, the decision was  taken to build accelerators in as many  UK universities which were chosen as adequately equipped for the task.
 In early 1946, the Nuclear Physics Committee of the Ministry  of Supply sent circulars to thirty universities and other institutions asking about their research programs in nuclear physics \citep{Mersits:1987aa,Krige:1989aa}.
Only five of them replied with requests for large scale equipment,   subsequently receiving  grants from the DSIR for the construction and maintenance of the following accelerators: a
 1.3 GeV proton synchrotron at Birmingham, designed by Mark Oliphant, a  300-400 MeV electron linear accelerator at Cambridge (later abandoned), a 300-MeV electron synchrotron at Glasgow, a 400 MeV  proton synchro-cyclotron at Liverpool, a 150 MeV electron synchrotron at Oxford.
Actually, all these machines would go into operation between 1953-1954, when the Bevatron and the Cosmotron, accelerators of much higher energies, were already available in the 
US. 

By the end of 1945 the advantage of a synchrotron for the photo-production of mesons had been recognised, and a research program in synchrotron development had been  prepared. Philip Dee's idea of building a 200 MeV betatron at Glasgow had been abandoned in favour of the mentioned 300 MeV synchrotron. A working  group, under the leadership of Donal Fry,  was created in the government center at Malvern, where a top secret radar group had been hosted during the war. 
Discussions  with  industrial  groups, such as Metropolitan-Vickers, English Electric and British Thomson-Houston, all of which had participated to the war effort,  had started \citep{Lawson:340513}.\footnote{See  also {\it The CERN Synchrotrons} by  Giorgio Brianti  in  \href{https://inis.iaea.org/collection/NCLCollectionStore/_Public/30/002/30002733.pdf}{50 Years of Synchrotrons} where Lawson's recollections appear.}.

The decision was taken to build a 30-MeV electron synchrotron, which should serve as a prototype for the  300 MeV machine planned by Dee in Glasgow, and at the same time could be used to study nuclear photo-disintegration and gamma-neutron reactions. Following Frank Goward's idea, the first step in this direction was done converting a betatron at the Woolwich Arsenal Research Laboratory (the first in UK during the war) into the world's first electron synchrotron. This prototype, first operated in October 1946 by Goward and D. E. Barnes  \citep{Goward:1946}, established the practicability of the synchrotron acceleration, together  with the machine operated by General Electric in US \citep{ElderEtAl:1947aa}. The Woolwich synchrotron was then moved to Malvern where it was modified to be used for more general experiments.  In providing the premise for the 30-MeV machine, which would go into operation in October 1947 \citep{Fry:1948aa}, the Malvern synchrotron in turn became the hotbed for analyzing the problems of the 300 MeV Glasgow machine. The magnet for the Glasgow electron synchrotron would be built by Metropolitan-Vickers Electrical Co., who were to be responsible for overall design and construction together with the Malvern group and Glasgow University. The Malvern--Harwell group also  gave advice to the building of  the Oxford electron synchrotron.\footnote{See Metropolitan-Vickers Electrical Co 1899-1949 by John Dummelow: 1939-1949, \url{https://www.gracesguide.co.uk/Metropolitan-Vickers_Electrical_Co_1899-1949_by_John_Dummelow:_1939-1949}.} 

This new-generation accelerators were required  to provide higher beam currents to make interactions between particles more likely, and provide higher  energies  to investigate in the laboratory new forms of ``nuclear" interactions, previously observed only in cosmic rays experiments. At that time, the  adjective  ``nuclear"   referred to the whole research related to the basic structure of matter and the laws behind it, i.e. nuclear forces, mesons, field theory, etc., and its meaning 
was much broader and quite different from   present-day.

\subsection{1946: Reaching out to the reconstruction of European science}
 
 Alongside the planning for particle accelerators and their transformation  as a new tool  for   exploring  particle physics  in an unprecedented energy range, the postwar period  also brought together many of   the prewar   European scientists  bent on reconstructing their University laboratories and eager to compare   each other's ideas about past and present new directions.
 
The UK was the natural place for such reconstruction to start. English Universities   were now seeing many of their scientists coming back from the United States where they had participated to the Manhattan project and the associated Anglo-Canadian Project in Montreal and Chalk River. Among the scientist returning to the UK from the US 
   or from Canada, and of direct interest for Touschek's story, we note Rudolph  
 Peierls, and Hans von Halban.  Peierls, who would be Touschek's external PhD examiner in Glasgow, came back from Los Alamos, where he had played a major role in the development of the atomic bomb,  and   joined the University of Birmingham \citep{Lee:2007aa}.\footnote{An extensive description of Peierls role in the development of the atomic bomb can be found in \citep{Close:2019}.} Von Halban, returning from Canada,  was to become the first director of the  \LAL \ in Orsay  (LAL) and oversaw   the construction of the linear accelerator which  contributed to AdA's success story \citep{Marin:2009,Pancheri:2018xdl}.\footnote{In 1939, at \CdF\ in Paris,  Hans  von Halban had collaborated with  \FJ \  and Lew Kowarski in the discovery that several neutrons were emitted in the fission of uranium-235  \citep{halban_liberation_1939,halban_number_1939} -- a discovery leading to the possibility of a self-sustaining chain reaction. In 1941  von Halban  fled to England  to subtract  the laboratory heavy water equipment  from the approaching Germans. He  settied in Oxford and  then moved to Canada to join the Manhattan project related  effort. Back to Oxford,   he  returned to France  in 1954 to  become the first director of  LAL.}

 There were also visitors from continental Europe, in particular from Italy, where the process of reconstruction of scientific institutions was taking place. In Italy, because  of  Mussolini's 1938 anti-semitic   laws, many of the best scientists had moved to the United States, as it had been the case of Enrico Fermi \citep{BernardiniBonolis2004,Pontecorvo:1993aa}, Bruno Rossi \citep{Bonolis:2011aa}, Emilio Segr\`e \citep{Segre:1993qy}, Bruno Pontecorvo \citep{Close:2015}, and many others. But some of Fermi's students or young collaborators  had remained, notably Edoardo Amaldi \citep{Rubbia:1991}.\footnote{In 1938, Fermi was awarded the Nobel Prize in Physics ``for his demonstrations of the existence of new radioactive elements produced by neutron irradiation, and for his related discovery of nuclear reactions brought about by slow neutrons."  Fermi's group in Rome   included Franco Rasetti, Edoardo Amaldi,  Ettore Majorana, Bruno Pontecorvo,  and Emilio Segr\`e, who  was awarded the 1959 Nobel prize in Physics for the discovery of the anti-proton.} Soon after the war,   Italian physicists, in particular  from Milan and  Rome,  
in contact with Fermi and Rossi, both in the US, and following their advice, were ready to take up new roads in  particle physics.
They were  restarting from cosmic rays,  which in those days,    did  not need large or costly equipments and could thus be undertaken by many countries, such as  Italy, Germany -- where other forms of ``nuclear'' research were initially forbidden -- or India, under the leadership of Homi J. Bhabha, who had founded in 1945 the Tata Institute of Fundamental Research, which became a major center of cosmic ray studies \citep{Sreekantan:1998aa}.

In Italy,  the great tradition of exploring matter through the particles coming from the sky, the cosmic rays, 
     had continued to be fostered even under  the bombing of Rome and despite of the political divisions between South and North, which took place  after the 1943 armistice between the Italian Government and the Allies. Now, after the war, an alternative way   for exploring nature's atomic scale was  available through particle accelerators. As envisioned already before the war in the major European physics laboratories, in particular by  both \FJ\ and Fermi, this new road, which required much larger equipments and investments,   had been shown to be  scientifically viable and could  be favoured by postwar reconstruction. 

In 1946, a pivotal moment in the scientific reconstruction of Europe had been the international conference on {\it Fundamental Particles and Low Temperatures}, held from 22 to 27 July 1946 at the Cavendish Laboratory in Cambridge  \citep{Proceedings:1947pca}.  As recalled by Amaldi, \citep[62]{Amaldi:1979aa},   ``Contacts between physicists in different parts of the world had been impossible for years and this conference provided a welcome opportunity to renew old friendships, and to hear what others had been doing.''  Many European states, but not Germany for obvious reasons, as well as other countries like the US, USSR, China and India were represented at the Conference. All of them presented their latest work on cosmic ray physics. Large attention was also given to  the theoretical physics side, in particular  the last session was entirely devoted to the S-matrix theory proposed by
Werner Heisenberg during the war \citep{Heisenberg:1943aa,Heisenberg:1943ab,Heisenberg:1944}. Focusing on observables such as cross-sections or energy levels, Heisenberg's   S-matrix theory aimed on building a theory  to calculate  only observables, such as  scattering cross-sections and energy levels. Heisenberg,  although released from enforced stay at Farm Hall \citep{Cassidy:2017}, was not present as  his travels were still restricted, but talks on the subject were presented by  Walter Heitler, Christian M\o eller, and Carl G. Stueckelberg  \citep{Mersits:1987aa}.

 Besides the opportunity of scientific exchanges after the forced isolation that many scientists had experienced during the war, the Cambridge conference  was also an important occasion to renew the strong pre-war relationships,  in particular between physicists from University of Rome, such as  Edoardo Amaldi and Gilberto Bernardini,\footnote{An affectionate and humorous \href{https://cds.cern.ch/record/1734431}{reminiscence of Gilberto Bernardini} by Leon Lederman can be found in  \citep{Ledermann:2009}. See also   \href{https://www.europhysicsnews.org/articles/epn/pdf/1995/05/epn19952605p117.pdf}{Remembering Gilberto Bernardini} in  \citep{Ricci:1995}.}
 and  Patrick Blackett, then  at  the University of Manchester, promoting with his group a broadly based cosmic-ray  program \citep{Butler:1999aa,lovell_patrick_1975}.\footnote{\href{https://www.britannica.com/biography/Patrick-Blackett-Baron-Blackett}{Patrick Blackett} was awarded the Nobel Prize in Physics in 1948 for his discoveries in cosmic rays research.}   
 
  Among the speakers at the Cambridge conference of special interest to Touschek's  story, there  was Bruno Ferretti, a young theoretician in Amaldi's group, who had given a talk on the absorption of slow mesons by atomic nuclei. He was  analyzing the problem of nuclear capture \citep{Ferretti:1947aa}, at stake in cosmic-ray experiments being performed in Rome by Marcello Conversi, Ettore Pancini and Oreste Piccioni, which would soon attract  attention on both sides of the Atlantic Ocean. 
  
In Rome, cosmic ray research had been initiated in the late 1930's under the leadership of Gilberto Bernardini and  had been kept alive  through the war  with excellent results, continuing   a research tradition which  would flourish again in the wider framework of reconstruction. In  1937, Bruno Ferretti, born and educated in Bologna, had joined Fermi's group in Rome. Following Fermi's departure for the US in  December 1938, he had been appointed to  teach  his    course  in  Theoretical Physics for the rest of the academic year.   From 1940, Gian Carlo Wick, who had been closely associated with Fermi, moved on his chair from Padua   \citep{Jacob:1999aa}. After the war, 
  when  Wick as well   emigrated  to the US  \citep{Jacob:1999aa}, Ferretti remained for a while the only senior theoretician, following in detail nuclear physics and cosmic ray developments   and was himself appointed to the chair of theoretical physics in 1947.
  
During the Cambridge conference, Amaldi saw the opportunity to renew the pre-war
 exchanges 
 between Italy and the UK, and was able to secure a British Council fellowship for  Ferretti  to spend ten months at Manchester in 1946, where 
 Ferretti's  theoretical experience in the design and analysis of cosmic ray experiments made him more than welcome. Ferretti was thus in  Manchester  at the time of major cosmic ray discoveries from UK laboratories,   and was invited to teach  a course on cosmic radiation at University of Birmingham   \citep[441]{Amaldi:1979ab}. There,  he established close   contacts with Rudolph Peierls,  future    external examiner  of Touschek's PhD. In 1947,  Ferretti and Peierls  wrote together  an article on the quantum theory of radiation damping applied to the problem of propagation of light \citep{FerrettiPeierls:1947aa},  related to well known divergences in quantum electrodynamics, a highly debated question at the time  and a life-long interest of Touschek's.  
  This common interest and shared acquaintances would be  instrumental   in Touschek's move to Rome in December 1952, as we shall discuss in  Sec.~\ref{ssec:planningbetrayal}.

   \subsection{Post war revolutions in particle physics}
 The Cambridge conference was a major event which re-established collaborations and communication between the UK and the rest of Europe, but in a few months, the new year 1947 would open up completely new perspectives for elementary particle physics. 

\begin{figure}
\centering
\includegraphics[scale=3.16]{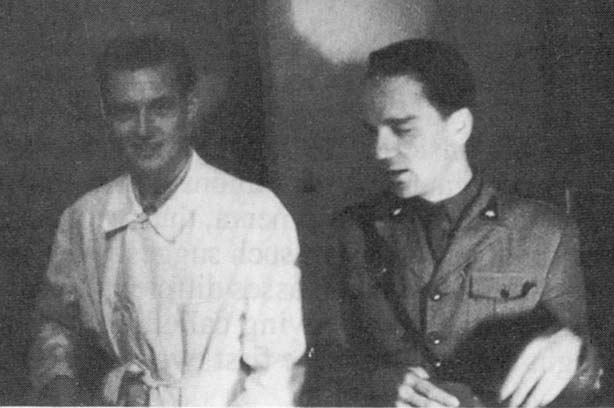}
\includegraphics[scale=0.0745]{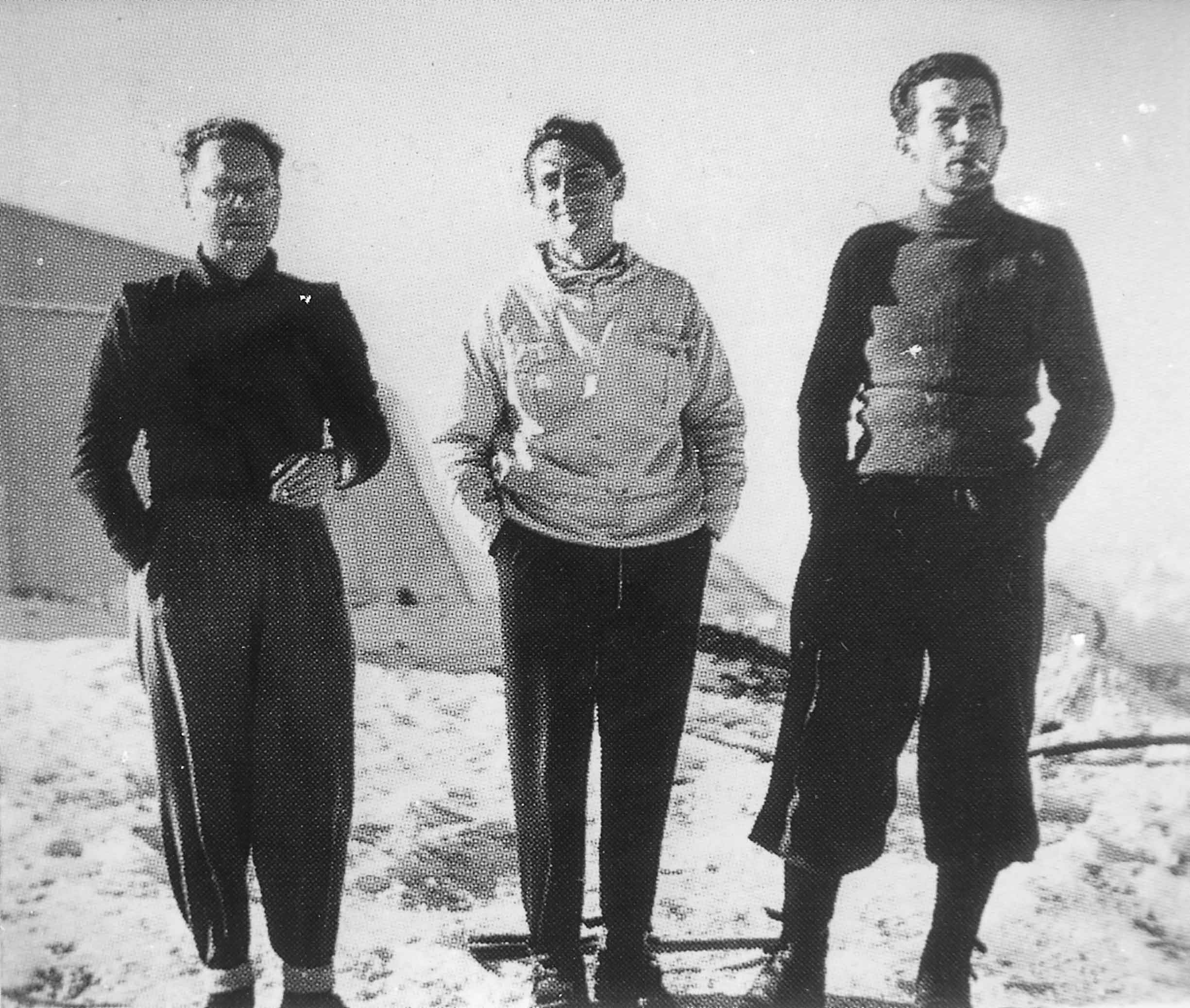}
\caption{From left: Marcello Conversi  and Oreste Piccioni in 1943, Edoardo Amaldi, Gilberto Bernardini and Ettore Pancini in 1947 at the  Testa Grigia cosmic ray Observatory, in the Italian Alps, at the southeast limit of the Plateau Ros\`a glacier. Photographs  from ``Edoardo Amaldi Archives'', Physics Department, Sapienza University of Rome.} 
\label{fig:ConversiPiccioni}
\end{figure}

Based on Yukawa's meson theory, and related theoretical suggestions, slow positive Yukawa mesons, i.e. positive mesotrons of cosmic rays, traversing matter should strongly prefer to decay rather than be absorbed by a nucleus, because of Coulomb repulsion by protons. Negative Yukawa mesons, on the other hand, should strongly prefer absorption to decay. These predictions were blown to bits by a  crucial experiment carried out in Rome during the war by Marcello Conversi, Ettore Pancini and Oreste Piccioni.  In four consecutive steps they had directly investigated  ``the fate of mesons coming to rest in matter'' \citep[p. 11]{Conversi:1988fk}, disclosing that positive cosmic ray mesons behaved as predicted, but a substantial fraction of negative sea-level  cosmic ray mesotrons decayed in a carbon plate \citep{Conversi:1944fk,Conversi:1944ab,Conversi:1945aa,Conversi:1946uq,Conversi:1947aa}. 
{In Fig.~\ref{fig:ConversiPiccioni} we show a snapshot of Marcello Conversi and Oreste Piccioni,  in Rome around the time of the first experiment, and a later (1947) photograph of Ettore Pancini, with Amaldi and G. Bernardini, in the cosmic ray high altitude Testa Grigia Laboratory, near Monte Rosa,  in 1947}. 
See \citep{Piccioni:1988aa} and \citep{Conversi:1988fk} for a detailed overview of the main steps related to the accomplishment of the four consecutive experiments.\footnote{As recalled by Piccioni and Conversi, the first experiment was prepared at the Physics Institute in the Rome University campus, the {\it Citt\`a Universitaria}, close to Termini  Rail road station.  However,  on July 19, 1943, Rome was bombed for the first time by the American Air Force and nearly 80 bombs fell within the perimeter of the University campus, which was located near  to the main station, an area especially subject to raids.
{It was thus}
 decided to transport the experiment to a semi-underground class room of the high school Liceo Virgilio, transforming the class into a laboratory. 
 {The liceo was located in Via Giulia, near the river Tiber, not far from the Vatican City, and hopefully less in danger of being targeted  by American bombs.} 
 Work was however interrupted by the Italian armistice (September 8, 1943) and, following it, by the occupation of Rome by the German troops. Resumed under very difficult and dramatic conditions, only late in 1943 Conversi and Piccioni  finally started to run the first experiment. While they were planning the third experiment, Rome was liberated by the Allied troops (June 5, 1944) and they moved everything back to the Physics Institute at the University. After the liberation of Northern Italy (April 25, 1945), Conversi and Piccioni were joined by Ettore Pancini, a leader in the Partisan movement of the Italian Resistance against German occupation, with whom they performed the third, and then the fourth, decisive, experiment. In this last experiment they replaced the iron with carbon and were able to provide convincing proof that negative mesotrons stopped in carbon did undergo spontaneous decay whereas they did not when stopped in iron. 
 {For a reconstruction of these events see also} 
 \href{https://books.google.it/books?id=FSlhDwAAQBAJ\&pg=PT7\&lpg=PT7\&dq=ereditato+e+liceo+virgilio+1943\&source=bl\&ots=sP87i4M0pk\&sig=ACfU3U2EPPuLIcfxt9kEOo-T5tN2BRjAMQ\&hl=en\&sa=X\&ved=2ahUKEwiTlpXfy7roAhVGxIsKHYSfDasQ6AEwAHoECAoQAQ\#v=onepage\&q=ereditato\%20e\%20liceo\%20virgilio\%201943\&f=false}{Le particelle elementari} 
 { by A. Ereditato, one of Pancini's students.}}

As stressed by Luis W. Alvarez in his Nobel lecture: ``[\dots] modern particle physics started in the last days of World War II, when a group of young Italians, Conversi, Pancini and Piccioni, who were hiding from the German occupying forces, initiated a remarkable experiment\dots''.\footnote{L. W. Alvarez, Recent developments in particle physics, Nobel Lecture, December 11, 1968 \url{http://nobelprize.org/nobel\_prizes/physics/laureates/1968/alvarez-lecture.html}.}
The  surprising results of the ``Rome experiment'' showed that negative cosmic-ray mesotrons  were  almost completely unreactive in a nuclear sense, and  provided  the first demonstration that  this particle  was not  behaving as it should, if it were the meson predicted by Yukawa as the mediator of nuclear forces \citep{Conversi:1947aa}. The final result of the experiment appeared on February 1st, 1947, but Fermi, who had been already informed by Amaldi in winter 1946, immediately reacted writing an article with Edward Teller and Victor Weisskopf, in which a startling conclusion was reached \citep[p. 315]{Fermi:1947aa}: ``If the experimental results are correct \textit{they would necessitate a very drastic change in the form of mesotron interactions} [emphasis added].''\footnote{See also Ferretti's publications analyzing results of the Conversi, Pancini and Piccioni  experiment  \citep{Ferretti:1947aa,Ferretti:1948aa}.} 

{The question was definitively resolved through a  cosmic ray experiment performed by Cecil F. Powell's group in  Bristol  \citep{Lattes:1947kx}.  Improved nuclear emulsions, developed following an idea of the Italian physicist Beppo Occhialini, had enabled Powell's group to establish the existence of a new elementary particle in cosmic rays, the $\pi$-meson, a
strongly  interacting particle of short lifetime, which was identified as the Yukawa meson, the accepted quantum of nuclear forces. The mesotron of cosmic rays, now termed the \textit{$\mu$-meson},  was   recognized to be the product of $\pi$-meson  decay and   was clearly a weakly interacting particle,  according to the Rome experiment confirmed by Fermi's interpretation.  } 

Mesotron  interactions  were still the  focus of 
 discussion at the first Shelter Island conference on  the  Foundations of Quantum Mechanics,  held from June 2--4, 1947  in Shelter Island, near New York. It was the first peace time opportunity  for the leaders of the American physics community  to meet, exchange new ideas and  assess the state of the field, after the Manhattan project \citep{Kaiser:2005}.  
  At this  time,  the May 24 issue of \textit{Nature}  with the  article by Lattes, Occhialini and Powell 
had not yet reached the United States, but, not long  after, in December of 1947, evidence for the existence of new unstable elementary particles detected in cosmic-ray showers -- the so called V-particles because of the characteristic tracks they left in the cloud chamber -- was announced  by Butler and Rochester, members of Blackett's group in Manchester.  \citep{ButlerRochester:1947aa}. Although it  took the particle physics community several years to appreciate the importance of the V-particles,  the modern day K-mesons,  
this discovery,  together with that   of the $\pi$-meson, was  signalling the existence of a large number of hither though unsuspected  sub-nuclear particles, which would open the new field of  particle physics. In parallel with the results of the Conversi, Pancini and Piccioni experiment, these achievements made 1947 a high point in the history of elementary particles, recognized by the Nobel Prize to Blackett, Yukawa and Powell.\footnote{Blackett was awarded the 1948 Nobel prize in physics for his research  on ``cosmic rays, using his invention of the counter-controlled cloud chamber." He had actually built this device  in the early 1930s in collaboration with Giuseppe Occhialini, who had brought from Florence Bruno Rossi's method of electronic coincidences with which they triggered the expansion of the cloud chamber. Together they were able not only to optimize observation of cosmic rays but were the first to expound the pair formation mechanism, putting the positron -- Dirac's hole-theory particle -- into theoretical perspective. Since the early 1930s Blackett had very strong ties with the Italian physicists community and in 1938 he had offered Bruno Rossi a position in Manchester, when the latter was forced to leave his chair in Padua because of the 1938 anti-semitic fascist  laws \citep{Bonolis:2011aa}. In 1939 Rossi settled in the 
US helped by Arthur Compton and Hans Bethe. For the discovery of the $\pi$-meson and its subsequent decay into a muon and a neutrino, Powell  was awarded the 1950 Nobel Prize in Physics ``for his development of the photographic method of studying nuclear processes and his discoveries regarding mesons made with this method," while Yukawa had been awarded the 1949 Prize for his theoretical prediction. Occhialini, who had been closely associated with Blackett and Powell in such important discoveries, was awarded jointly with George Uhlenbeck with the prestigious Wolf Prize only in 1982.}

With the discovery of the $\pi$-meson, the  energy threshold of meson production would become  the frontier of particle physics. ``Meson physics'' would rapidly develop, while the body of experimental knowledge in the field grew even more rapidly as new meson-producing accelerators began to swing into action. Soon after, in 1948, Eugene Gardner and Cesare Lattes,  would be able to artificially produce pions for the first time by bombarding carbon atoms with a 400-MeV proton beam from the 184-in synchrocyclotron at the University of California Radiation Laboratory. For the first time ``nuclear physics'' was challenging cosmic-ray physics. In the meantime, major proton synchrotron projects were being set up at Brookhaven National Laboratory (BNL) in Long Island, under M. Stanley Livingston, where a 3.3 GeV machine, the Cosmotron, was planned, and with the Bevatron, at Lawrence Berkeley National Laboratory in Berkeley, specifically designed to be energetic enough to create antiprotons. In the early 1950s, these accelerators would definitely mark  a new era for   particle physics. From then on,  cosmic rays would no more be   the only tool available  for   discovering    new particles and  probing  their high energy  regime. Throughout the years to follow, accelerators and the observation of the sky  would move together to probe nature at its fundamental level. 
{In the years to  follow, a growing interaction between cosmic ray physics, astronomy and astrophysics, cosmology and particle physics would merge the many diverse ways to tackle fundamental questions of the universe under the general field named {\it astro-particle} physics}.

This was the scenario which Bruno Touschek stepped into, as he arrived in Glasgow in April 1947. He had lived in Germany the terrible `Hungerwinter' 1946/1947, noted as the coldest of the 20th century in Europe. The struggle for survival had been extremely painful, and many thousands of  people had died because of cold and famine. The winter had severe effects also in the United Kingdom, with fuel and food shortage, and floods caused in March by the thawing snow and heavy rain. Agriculture and breeding were dramatically affected, too, and thousands of British people emigrated, especially to Australia. 

At the same time, the initiative of the Marshall Plan, the European Recovery Program, would soon be established by the United States to restart industrial and agricultural production, set up financial stability and expand trade in war-torn Europe.

\section{1947-49: Getting a doctorate in Glasgow}
\label{sec:glasgow4749}
 
\begin{figure}
\vspace{-0.5cm}
\centering
\includegraphics[scale=0.4]{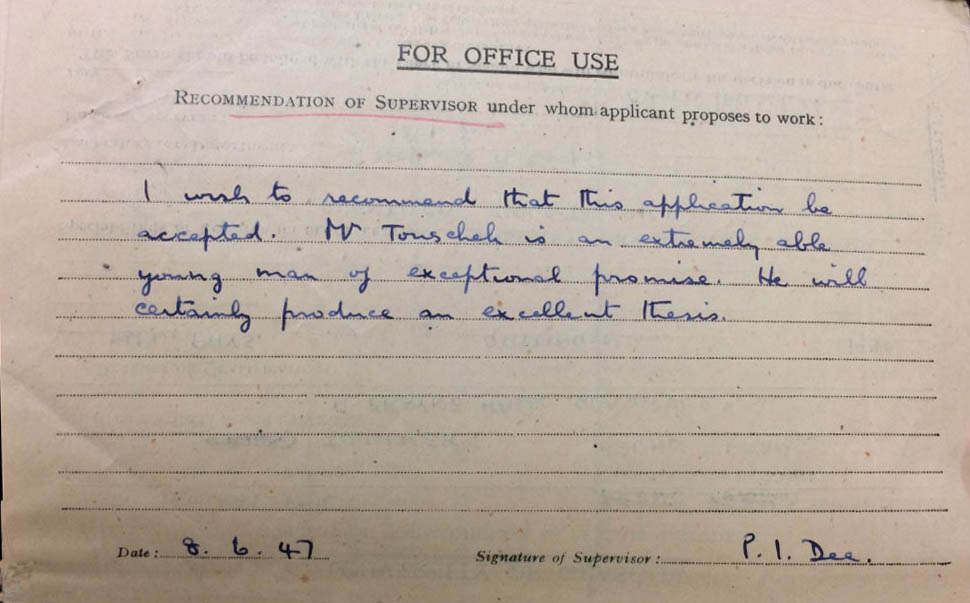}
\caption{Philip Dee's application to the Senate in favour of  a research scholarship for Bruno Touschek. Courtesy of  UG's Archives. }
\label{fig:Dee-exceptional-1947}
\end{figure}

Touschek's time in Glasgow is first presented  by Philip Dee in a 1979 letter to Edoardo Amaldi  \citep[9]{Amaldi:1981}.\footnote{One of Glasgow University colleagues, S. Curran, wrote an insightful biography of Philip I. Dee \citep{Curran:1984aa}.
Of present note in this biographical sketch are two comments. One, at page 7, about Dee's publication record, is the fact that machine builders do not have adequate recognition from publication, since during the construction of the machine, their research output in terms  of papers is often very poor. This is interesting because the same can be said of Touschek, whose publication output dramatically drops  after  he started working on AdA.  Another interesting comment, in page 5,  is that Dee was very kind, and ``often he would make ample allowance for illness and the like". This attitude is confirmed   in his  communications  to Amaldi, after Touschek's death, as Dee acknowledges  Touschek's original  but  sometimes unnerving behaviour. }
 After Touschek's early death in 1978,  Amaldi had written to all the people who had been friends with Touschek asking for their recollections and memories.\footnote{ The answers he received are kept in  Amaldi papers, ``Edoardo Amaldi Archives'', Sapienza University of Rome (from now on EAA), Box 205.} The letter describes  Bruno's  
sparkling intelligence, but also his restlessness and,  occasionally, a behaviour  outside the accepted norm. On 11th April, 1979, Dee wrote to Amaldi:    
\begin{quote}
Dear Professor Amaldi, 

I did not know about Touschek and I am very sad indeed to have your news.

I will write something for you about his life in Glasgow. For quite a while he lived in our house in the University and became in many ways a member of the family [\dots]

I was quickly impressed by Touschek's obvious ability, his extensive knowledge of physics, and his enthusiasm and I arranged [\dots] for him to have  a research appointment in the department, which, at that time, had only one staff member on the theoretical side [\dots] 

He was very clever and original, he was also untiringly energetic and extrovert. Bruno led his life to the full extent in all situations and at all times. His enthusiasms were many and, although often brief, were exploited in a manner which most people would have found exhausting.
\end{quote}
Dee's high opinion of Bruno is clearly expressed in the  application he submitted for a research scholarship for Bruno in June 1947, which is shown in Fig.~\ref{fig:Dee-exceptional-1947}. This probably refers to the renewal of the  existing scholarships, since at this time Touschek had already been in Glasgow for two months.

\subsection{Arriving in Glasgow accompanied by a guard}

When Touschek arrived in Glasgow in April   1947, he was escorted by a guard, as  Philip Dee  would  remember   in the  letter to Edoardo Amaldi \citep{Amaldi:1981}:
\begin{quote}
 My association with Bruno Touschek began in April 1947 when he was brought to my office under guard (!) for an interview. This had been arranged by Dr. Ronald Fraser (a friend of mine) who had met Bruno when serving on a post-war Allied Commission which was visiting laboratories in Germany and elsewhere. Touschek  had expressed the wish to work  in a British laboratory and Fraser knew that I had recently come to Glasgow  to construct a nuclear physics center in the university here.\footnote{Notice that, as we have seen from  our previous note, entitled {\it Bruno Touschek in Germany after the war} and  based on Touschek's two letters from Glasgow in April 1946, Dee must have met Touschek already in April 1946 \citep{Bonolis:2019qqh}. } 
 \end{quote}

Touschek's duties included teaching, research in theoretical physics, and support to the synchrotron program. He had never lectured in English, and this was  a challenge. He started  preparing  a regular course on selected nuclear physics topics, to be held together with Ian Sneddon, the only other theorist in the department at the time.\footnote{I. N. Sneddon (1919-2000) was born in Glasgow and became  a  noted mathematician. After studies in Glasgow and Cambridge, after the war he held a research position in Bristol University, where he worked   with N. F. Mott, and authored with him his first book, {\it  Wave Mechanics and its Applications } - which was published in 1948.  He  returned to Glasgow as a lecturer in physics 1946, or rather natural philosophy as the subject was called in the ancient Scottish Universities at that time. During this period his  next major text was the book entitled {\it Fourier Transforms},  which appeared in 1951, and became such a classic text to be later reissued in 1995. Before the book appeared in print, Sneddon  left Glasgow to take up the chair of mathematics at the University College of North Staffordshire (which later became Keele University), and then returned to Glasgow, in 1956 to take up the Simson Chair of Mathematics. Sneddon travelled widely, particularly in North America where he held a number of visiting professorships. From \url{http://www-history.mcs.st-andrews.ac.uk/Biographies/Sneddon.html}. See also \citep{Chadwick:2002aa}.} They oulined  the topics, then Sneddon would write down the lectures in ``proper english'' as Touschek says, and then lectures would be  delivered, alternating between the two of them. But his first lecture in English  came up as a surprise, as, just on the day before starting his regular course  with Sneddon,  he was suddenly called to substitute for  one of the Professors, who had originally wanted to talk about the synchrotron but had to go to the University Court. Dee asked him to  please  jump in and 
Bruno had to rapidly put together a lecture on angular valencies.
\footnote{Letter to parents, April 23rd, 1947.}
This was a subject he had studied in depth when  in \Gott, and  there would be no problem in preparing the  lecture. The well received delivery in English was a confirmation that he could now confidently start teaching his regular course.  Years later, in Rome, Touschek became famous for his impeccably  clear lectures in Italian, his ability to derive the results as if he had obtained them in that instant, conquering 
heart and minds of   the best students of an already exceptional graduating class.
\footnote{See the  2013 interview given by the theoretical physicist Luciano Maiani, 1999-2003  CERN Director General, in the movie \href{http://www.lnf.infn.it/edu/materiale/video/AdA_in_Orsay.mp4}{Touschek with AdA in Orsay}. See also \citep{Maiani:2017hol}. }

Shortly after giving his first lectures, Bruno  was also  immersed in the  vibrant scientific atmosphere which characterised the early  British post-war period, with frequent visits from scientists from abroad and a general atmosphere of novel expectations and discoveries, both from an experimental and a theoretical point of view.  
Bruno and Sneddon, who was only two years older than Bruno and had joined the department  just a few months before in 1946,  soon became  friends  and collaborators. 
 They started writing a paper together on meson theory, and, in early May, Sneddon   took  Bruno to   Edinburgh to meet Max Born,  the great German born theoretical scientist, one of the fathers of Quantum Mechanics, who had left Germany after Hitler came to power.\footnote{Max Born (1882-1970) had left Germany in 1933 and in  1936 settled in Edinburgh as Tait professor of Applied Mathematics. In 1954 was awarded  the Nobel Prize in Physics ``for his fundamental research in quantum mechanics, especially for his statistical interpretation of the wavefunction", sharing it with   Walther Bothe ``for the coincidence method and his discoveries made therewith". See also \citep{Kemmer:1971} in \href{https://royalsocietypublishing.org/doi/10.1098/rsbm.1971.0002}{Bibliographical Memoirs of the Fellows of the Royal Society} and a translation of the main part of a Memorial address given by Werner Heisenberg on January 12th, 1971 in \Gott \ \citep{Heisenberg:1970aa}.}

Leaving in the early morning from Glasgow, Touschek and Sneddon reached Edinburgh after  1 and 1/2 hour train ride. The train crosses the country side from West to East, until the
approaching  North Sea signals its presence with  a change in the clouds and  increasing light in  the sky.  They  had  time to be tourists before joining the Colloquium. 
In Edinburgh Bruno discovered a beautiful city, in stark contrast with  Glasgow which  in the late 40's was  a rather gloomy place.\footnote{About living conditions in Glasgow, see a recent  article in {\it The Guardian} about the so called  \href{https://www.theguardian.com/cities/2019/oct/16/urban-living-makes-us-miserable-this-city-is-trying-to-change-that}{`Glasgow effect'.}} He thought Edinburgh
  most interesting   and  was enchanted with its hilly setting, the   medieval streets and old state prisons, or Princes Street,  said to be the most beautiful main street in Europe. In the middle of the city, Bruno and Sneddon  visited 
 the ancient castle, from which one  can see the Firth of Forth and on clear days the sea, and   the city Museum, with paintings  by Rembrandt and Brueghel. Bruno's artistic disposition  had been nourished through pre-war visits to the great European Museums in Rome and Vienna  and fostered by a family immersed in the tradition of the Vienna secession movement. After the ravages of the war in Germany, where museums had been bombed, masterpieces hidden or expropriated, walking into a museum was like being reborn.  His life was picking up again. 

\begin{figure}
  \includegraphics[scale=0.372]{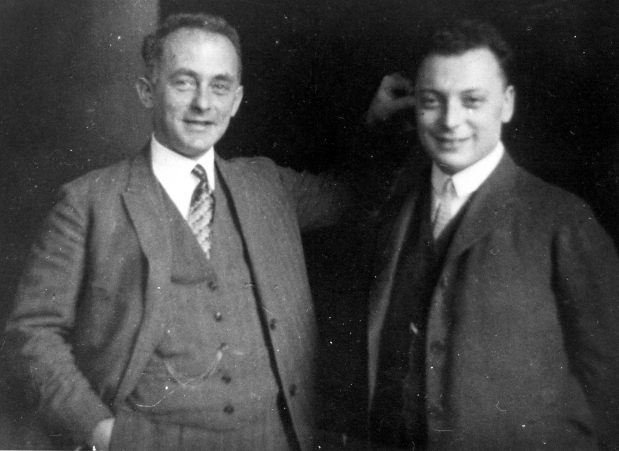}
  \includegraphics[scale=0.63]{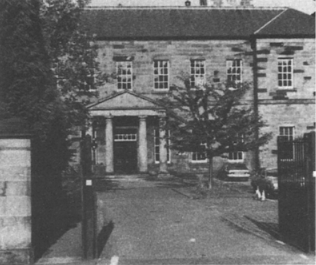}
\caption{Left panel:  Max Born  and Wolfgang Pauli (at right), in Hamburg, circa 1925,  from CERN archives at \url{http://cds.cern.ch/record/42702}. This  same photograph  is  also found at \url{https://calisphere.org/item/ark:/28722/bk0016t4k4m/}, contributed by UC Berkeley, Bancroft Library, has the two scientists mirror inverted. Studying the two photographs, it is likely that the original be the one from UC Berkeley. Right panel: a photograph of the building housing the 
{Applied Mathematics Department} in Drummond Street, in Edinburgh, from \citep{Wolf:1995}.}
\label{fig:BornPauli-drummond}
  \end{figure}

In the afternoon they joined the Colloquium. In Edinburgh Born held a  weekly seminar, which he had  introduced following the German tradition.  In the 1920's in the Institute for theoretical physics at the University of \Gott, one of prominent weekly events had been  the Proseminar, conducted by Max Born and James Franck,\footnote{{The Nobel Prize in Physics 1925 was jointly awarded  to James Franck and Gustav Ludwig Hertz ``for their discovery of the laws governing the impact of an electron upon an atom".}} in which  ideas were debated and young graduate students would present their work. It is remembered that in 1926,  at one of these seminars,  Werner Heisenberg presented a lecture on the development of Quantum Mechanics, and was enthusiastically applauded by the audience \citep[14]{Amaldi:2012}. In Fig.~\ref{fig:BornPauli-drummond}, we show a 1925 photograph of Max Born, together with Wolfgang Pauli, who, in later years, became  good friend of Touschek. We also show the building housing the 
Applied Mathematics Department in Drummond Street, in Edinburgh, where Born held the weekly seminar.
  
 Touschek  and  Born shared many common acquaintances, in particular they both knew  Fritz Houtermans, who had graduated  in \Gott\ with James Frank in the 1920's, when Born was holding the chair of theoretical physics  before leaving  Germany in 1933. Since these  \Gott\ times, Houtermans had been imprisoned by the Soviets, divorced and remarried \citep{Amaldi:2012}. 
Fritz Houtermans' witty and adventurous personality
was an obvious common argument of conversation.  Touschek had been close to Houtermans when in \Gott \ in the previous year, in 1946, in particular sharing evening gatherings  and occasional visits to the Observatory to watch the stars and   had been talking  with Houtermans  on the last day of his stay in \Gott, while waiting for the car to take him away, just a few weeks before. This last meeting with Houtermans in \Gott\  belonged  to  another world now.   
Sharing memories and anecdotes about  their common friend  soon put  Touschek and Born on  close grounds.  Born, appreciative  of  Touschek's brilliant intelligence and  profound interest in theoretical physics,  invited him to the  seminars, which were   held every week, on Monday and Thursday. Edinburgh, being some 1 and 1/2 hour away by train, biweekly attendance of the Seminar could be difficult, but Bruno decided he would do his best to  participate at least once a week.\footnote{Letter to parents, May 3rd,  1947.}
   Thus they started an acquaintance which later became a true collaboration, with Touschek giving a contribution to one of the Appendices of a new edition of Born's classic {\it Atomic Physics} \citep{Born:1951}, whose first English edition had already appeared in 1935.\footnote{{The eighth edition of Born's book was published  in 1969, one year before he  passed away in \Gott, on January 5th, 1970. The eighth edition, including revisions by R. J. Blin-Stoyle \& J. M. Radcliffe,  is available from Dover Publications in paper cover, ISBN 0-486-65984-4.}}

In early May, Spring  was  on its way, and   the trees were just starting
to sprout,  but icy winds would  whistle through the fireplaces when  not  burning. It  was difficult to believe   that summer was just a few weeks away.  
But it arrived and, early in June,  the annual department excursion to the shore took place.\footnote{Letter to parents, June 7th, 1947.} Two buses were rented to take  the 50 students, the research workers and the lecturers, to the Cobbler, a mountain of 884 metres (2,900 ft) height located near the head of Loch Long, a narrow  salt water fjord. The trip went past Loch Lomond, where Bruno had already been during a previous outing. They all climbed to the mountain top, including all the girls and Mrs. Dee. Bruno, although he had never climbed in Austria, was confident in his Alpine genes and reached it first.  It took more than four hours to be back down. Afterwards,  they all lay on the beach along Loch Long, playing ball and throwing stones in the water. Bruno,  hot from the climbing and the descent which followed, jumped in the Loch, unknowing of its true nature,  getting  a good dose  of salty water, to everybody's merriment. He could not remember of having had such  light hearted fun  in a long time. The landscape was astonishingly beautiful, the mood like that of   children on some illegal outing, since the trip  was not a holiday, but a department yearly date. Bruno could finally appreciate the  famed carefree English  humour. 

In the meantime, Touschek was working hard on topics which would later become the subject of his PhD dissertation. {He had also written  an article related to the problem of infinities in quantum field theories  in which he was  commenting on  a work by Walter Heitler's  assistant, Huanwu Peng  \citep{Touschek:1948ac}.\footnote{Peng had worked since 1938 with Max Born at University of Edinburgh, then recommended by the latter went to Dublin Institute for Advanced Studies as a post doc from 1941 to 1943, and later an assistant professor until 1947, when he returned to China.} During the war,  a  radiation damping theory had been proposed  by Heitler and Peng, 
which gave a procedure 
 to calculate scattering amplitudes and extract empirical predictions. The theory   
    could also be used to calculate the nucleon-meson scattering cross sections, or purely electrodynamic scattering processes. Heitler had presented his damping theory at the 1946 Cambridge conference on fundamental particles
   \citep{Heitler:1947aa}.\footnote{Heitler, who had left G\"ottingen for the University of Bristol in 1933, had remained there until 1941, when he became a professor at the Dublin Insitute for Advanced Studies, established by Erwin Schr\"odinger, Director of the School for Theoretical Physics.  In 1946 Heitler  became Director of the School for Theoretical Physics after Schr\"odinger resigned. } However, Heitler's program was now being abandoned after renormalized QED was developed 
in fundamental papers by Tomonaga, Schwinger, Feynman and Dyson \citep{Schweber:1994aa}.\footnote{The Nobel Prize in Physics 1965 was awarded jointly to Sin-Itiro Tomonaga, Julian Schwinger and Richard P. Feynman ``for their fundamental work in quantum electrodynamics, with deep-ploughing consequences for the physics of elementary particles." from \url{https://www.nobelprize.org/prizes/physics/1965/summary/}.} Touschek  began also to analyze the use of electrons as particles to produce interactions at a nuclear and subnuclear level \citep{Touschek:1947aa} \citep{TouschekSneddon:1948aa}  and, at the same time, worked on theoretical issues related to the nuclear structure  \citep{TouschekSneddon:1948ab} \citep{TouschekSneddon:1948aaa}, on the eve of the formulation of the  nuclear shell model theory.\footnote{The nuclear shell model was formulated  by  Maria Goeppert Mayer and, independently, by Hans Jensen, Otto Haxel and Hans Suess, Touschek's good friends since Hamburg and G\"ottingen times.  Goeppert's first article summarizing the evidence for a shell model of the nucleus appeared in August 1948 \citep{Goeppert:1948}. Her second decisive paper \citep{Goeppert:1949aa} was published together with the one by Jensen and colleagues, in June 1949 \citep{Haxel:1949aa}. Goeppert Mayer and Jensen shared the 1963 Nobel Prize in Physics, jointly with Eugene Wigner for unrelated work.}

In August, he developed an idea for the new synchrotron building and started discussing it with colleagues in the synchrotron group.  
 There were  trips  to Manchester and Edinburgh, and he was  busy moving  from the University Halls to private lodgings.\footnote{In September his address was c/o Fisher, 16 South Park Ave, Glasgow W2. Letter to parents September 2nd, 1947.} Not all was work, however. He  was  also frequently bathing in the Loch Lomond and other nearby places, such as Gerloch, the Clyde and even in the Atlantic, getting tanned as he had never been before.\footnote{Letter to parents, September 2nd, 1947.}  
  In Fig.~\ref{fig:touschek-UK} we indicate some of  the location   Touschek visited  in 1947.
  \begin{figure}
  \centering
  \includegraphics[scale=0.5]{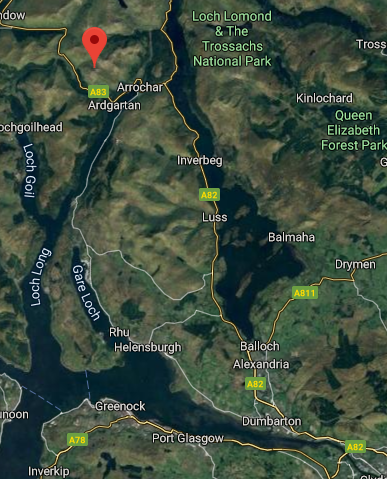}
  \includegraphics[scale=0.495]{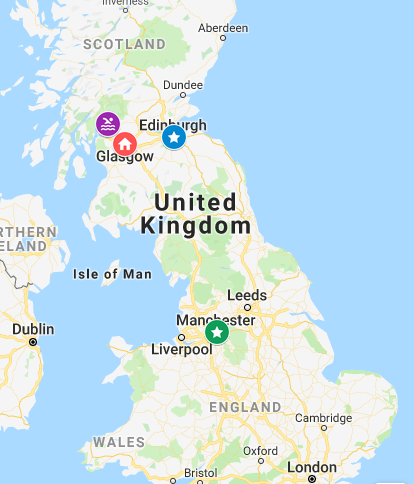}
  \caption{A view of places where Touschek  moved in 1947: at left for the excursion to the Cobbler, with Loch Lomond in the center, Glasgow -not shown- just outside the lower right hand corner. At right a larger view from Scotland to Manchester, down in the middle of England.  }
  \label{fig:touschek-UK}
  \end{figure}
Born in a country with no access to the sea, swimming became  a long life passion for Touschek. When, in later years, he would be working in Frascati on the electron positron colliding beam accelerators he proposed and had built, his frequent trips to nearby  Lake of Castel Gandolfo were memorable. His maternal aunt Ada had a villa 
nearby,  and bathing and fishing in the lake  would be   a favourite pastime.  He would just leave the laboratories in the early afternoon and  escape to Albano.  He also loved the sea. After Aunt Ada passed away in 1960, he started spending some time on the  south of Rome, in Sperlonga \citep{Amaldi:1981}, or  on the Amalfi coast \citep{Pancheri:2004}, swimming in the crystal clear waters of  Positano, which  was chosen for family holidays in September, 
 when  the vacation crowd has left, the Mediterranean sea is calm, and the water still warm.

The first summer in Glasgow being over, he dutifully kept up the correspondence with his family in Vienna, where life was still difficult, worrying about their well being,  proud to help his parents with his earnings,  sending both  money as well as packages. Then a crack appeared in this otherwise almost idyllic picture. 
   Because of administrative requests, some problems had arisen with the DSIR about his contract  after the  first six months in Glasgow.\footnote{A copy of his birth certificate, attesting to his Austrian citizenship, was required to renew his contract. The problem was soon solved, though, and the birth certificate received by the DSIR on September 1st, 1947. }  As the renewal of his fellowship ran into some delay, he started  wishing he could leave the UK.\footnote{Letter  to  parents, October 11th, 1947. The letter ends with two verses, paraphrasing a popular song, Sowieso: ``Egal was kommt, es wird gut, sowieso/Immer geht 'ne neue T\"ur auf, irgendwo/Auch wenn's grad nicht so l\"auft, wie gewohnt/Egal, es wird gut, sowieso [No matter what, it's going to be fine, anyway, a new door always opens, even if it's not going as usual, it's going to be fine, anyway.]'' \url{https://lyricstranslate.com/en/mark-forster-sowieso-lyrics.html}. The letter  implies that  happiness may be  elsewhere.}  
  He was very annoyed, as he would  often  be throughout his life when similar  impediments  forced him to interrupt his work and studies. His expectations had been  be forever forged  by  the war   experience, first on the betatron project under 
  German military control,   and then in \Gott, under the Allied Forces administration, when civil regulations would be easily overruled by the military. 
   Later in his life, he 
     never understood, nor accepted, the slow   methodical process  which regulates ordinary dealings of University life.\footnote{See the difficulties he had with applying for professorship in Rome in 1974 as described in \citep{Amaldi:1981}.}  

All through this first year, Bruno   worked hard with the newly established synchrotron group in Glasgow, which included   Samuel Curran, Walter Mc Farlane, A. C. Robb and Philip Dee. As one can see from \citep{Curran:1948aa}, where Touschek's collaboration is acknowledged,
he worked with Samuel Curran on a problem related to  measurements and results obtained with the proportional counter, a new
device Curran was developing with his student John Angus \citep[113-115]{Close:2015}.\footnote{Before the war, Samuel Curran (1912-1998) had been  at the Cavendish Laboratories with Rutherford in Cambridge.  In late 1940, he joined Philip Dee, Bernard Lovell, Alan Hodgkin and others at Worth Matravers, {in Dorset. A recollection of this period by  Bernard Lovell can be found in the 28th October 1982 issue of the {\it The New Scientist}, pag. 246.} After the invention of the cavity magnetron, Curran's development of spark gap modulators was critical to the success of the magnetron transmitters. In 1944 he went to the United States with a number of other UK  other scientists to work on the Manhattan Project, and  in 1946 he went to Glasgow University to join  Philip Dee.  He assisted  Dee and   McFarlane in the installation of the 300 MeV Glasgow Synchrotron, which became operational in 1954. Later, as Principal and vice chancellor of the University of Strathclyde, he took the lead in developing Britain's first technological university, as from  \url{http://www.purbeckradar.org.uk/biography/curran_sam.htm} and \url{https://www.worldchanging.glasgow.ac.uk/article/?id=39}. See also \citep{Fletcher:1999}.} This work with Curran  is also described in the December 1947 research report Touschek submitted to the University.\footnote{BTA Research Reports (Glasgow), Dic. 1947-Apr. 1948, Box 3, Folder 1.}
.\footnote{Before the war, Samuel Curran (1912-1998) had been  at the Cavendish Laboratories with Rutherford in Cambridge.  In late 1940, he joined Philip Dee, Bernard Lovell, Alan Hodgkin and others at Worth Matravers, {in Dorset. A recollection of this period by  Bernard Lovell can be found in the 28th October 1982 issue of the {\it The New Scientist}, pag. 246.}After the invention of the cavity magnetron, Curran's development of spark gap modulators was critical to the success of the magnetron transmitters. In 1944 he went to the United States with a number of other UK  other scientists to work on the Manhattan Project, and  in 1946 he went to Glasgow University to join  Philip Dee.  He assisted  Dee and   McFarlane in the installation of the 300 MeV Glasgow Synchrotron, which became operational in 1954. Later, as Principal and vice chancellor of the University of Strathclyde, he took the lead in developing Britain's first technological university, as from  \url{http://www.purbeckradar.org.uk/biography/curran_sam.htm} and \url{https://www.worldchanging.glasgow.ac.uk/article/?id=39}. See also \citep{Fletcher:1999}.  }
 
 The Glasgow  group started preparing  plans for  the synchrotron and  meetings were held with  other UK groups working on synchrotrons.\footnote{He writes about his work on the 300 MeV synchrotron in the October 11th, 1947  letter to his parents.} 
  The list of attendees of the 8th Meeting of the Glasgow synchrotron group held on November 4th, 1947   in Manchester, at Trafford Park, on the grounds of the Metropolitan-Vickers company, shows that all  the major actors in electron synchrotron preparations in the UK were present, among them  Frank Goward, who  had succeeded in building the world first electron synchrotron, just over one year before.\footnote{The list of attendees and agenda of the 8th meeting of the Glasgow Synchrotron, held in Manchester, at the Metropolitan-Vickers factory in Trafford Park is kept in Bruno Touschek papers, ``Edoardo Amaldi Archives'', Physics Department, Sapienza University of Rome, Box 3, Folder 5). Agenda for a previous meeting held on September 7th is also available at same location.}

In December Rudolf Kollath and Gerhard Schumann, with whom he had collaborated at the 15 MeV-betatron designed by Wider\o e and built in Hamburg, published a review article on the work done during the war and in the early post-war period at Kellinghusen. It included all the important information and many details, making Bruno Touschek's contribution to the project officially known \citep{KollathSchumann:1947}.\footnote{Touschek sent his parents  a copy of the issue of the Journal of Applied Physics containing the article (letter to parents, October 11th, 1947).}

As an expert in the field, Touschek continued to be involved with betatrons, as a consultant for the 20 MeV machine operating at the High Voltage Laboratory of the Metropolitan-Vickers at Manchester.\footnote{See letter by John D. Craggs of January 27th, 1948: ``You may remember that we had a short discussion on the peculiar spectrum we had found for our 20 MeV betatron [\dots] you may have been able to do some thinking about the problem. We should be most grateful for any light you can throw on the problem.'' Bruno Touschek papers, ``Edoardo Amaldi Archives'', Physics Department, Sapienza University of Rome (from now on BTA), Box 1, Folder 1. After graduating from King's College London in 1938, Craggs had joined the staff in the High Voltage Research Laboratories of Metropolitan Vickers Ltd., where -- except for a period in Berkeley, California, in 1944-45 -- he stayed until moving to Liverpool University in 1946 as a lecturer in Electrical Engineering. His work in Manchester on the neutron generator resulted in the building of the first full-size Van de Graaff in UK (News from the Archive of the University of Liverpool, \url{http://sca-arch.liv.ac.uk/ead/search?operation=full&recid=gb141unistaffc-d-d835}).} However, his knowledge on accelerators was quickly evolving. From his research report on work carried out during the months December 1947--January 1948,\footnote{BTA, Research Reports (Glasgow), Dic. 1947-Apr. 1948, Box 3, Folder 1.} we learn that he was preparing a lecture ``on the family tree of accelerators (Cyclotron, Betatron, Synchrotron, Synchro-Cyclotron)'' as well as a related paper on the synchrotron for the newly founded {\it Acta Physica Austriaca}, published by the Austrian Academy of Science, of which Hans Thirring was co-director at the time \citep{Touschek:1949aa}.\footnote{See also manuscript ``Zur Theorie des Synchrotrons'', BTA, Box 3, Folder 6.} 

Back in Europe, among his mentors and friends, there was appreciation for his achievements, as it appears from a postcard, sent by Arnold Sommerfeld to Paul Urban, on November 2nd: ``Touschek has had great success, in \Gott\ he passed all his exams one after the other, is now sought after by Hamburg and Hannover and is presently in England [sic!], well paid by the British".\footnote{EAA, Box 205.}

 \subsection{Bruno Touschek and Werner Heisenberg}
All along during the doctorate years, Touschek's  published theoretical physics  output  is remarkable.  {In what follows, we shall highlight contacts and, so far unpublished, correspondence  with Werner Heisenberg,  which shed light  on  his formation as a theoretical physicist.} Such scientific correspondence was a natural continuation of the relationship  established before Touschek's  arrival in Glasgow, during  his stay in G\"ottingen, while getting his Diploma    and afterwards, when he was,  for some time, {one of }Heisenberg's assistants at the Kaiser Wilhelm Institute (KWI) for Physics.

 Touschek had seen Heisenberg  for the first time in 1939, giving a public lecture in Vienna,  when Heisenberg was already one of the most prominent and influential German physicists. During the war Heisenberg had become director of the Kaiser Wilhelm Institute for Physics in Berlin-Dahlem, leading the German nuclear project and it is in Berlin that Touschek met him for the second time: ``[\dots] hatless hurrying to the KWI \& I asked him the way because I wanted to visit him \& had not recognised him. He brought me to his office [\dots]."  After the war, in G\"ottingen, Touschek attended his lectures on  quantum field theory, and later commented on them, in  the same undated manuscript: ``It was not a good course of lectures, but there was one among them, which for me was a complete eye-opener: the harmonic oscillator \& its quantization. I had learned Q.T. [Quantum Mechanics]  from Sommerfeld's ``wellenmechanisches Erg\"anzungsband'' \& I had tried Dirac's famous book both of which lean heavily on wave mechanics. H.'s lecture opened my understanding to the mechanical approach.''\footnote{Quotations are from manuscript document, courtesy of Mrs. Elspeth Yonge Touschek.} 
   
Among Touschek's {unrecorded}  work there are his studies of the analyticity properties of Heisenberg's S-matrix.\footnote{Correspondence with Heisenberg cited here is  partly kept in BTA, partly  in Touschek's personal paper, courtesy Mrs. \EYT.}
Touschek  was bent on understanding  Heisenberg's proposed theoretical approach to  particle dynamics \citep{Heisenberg:1943ab}. Heisenberg's seminal work,  {\it Die `beobachtbaren Gr\"o\ss en' in der Theorie der Elementarteilchenphysik}, namely {\it The 'observable quantities' in the theory of elementary particle physics}, was focused on dealing with observable quantities, rather than order by order perturbation theory calculations, plagued by divergences at  a given   fixed  order.\footnote{See  \citep{Rechenberg:1989} for a historical view of the S-matrix development from 1942 to 1952.} This work had a profound influence on Touschek and is echoed in his later work about infra-red Radiative corrections to electron-positron experiments, where Touschek reflected about the incompatibility between   ``the picture of an experiment [as] drawn by theory and reality'' \citep{Etim:1967}.

  The analyticity properties of the S-matrix were a strongly debated  topic in the theoretical physics community and  Touschek had discussed the argument with Heisenberg when he was still  in  \Gott \ and then again after joining Glasgow.  In particular, since fall of 1947, Touschek exchanged letters with Werner Heisenberg focused on the S-matrix and other theoretical issues which he was studying at the moment.\footnote{See letters dated October 6th (Touschek to Heisenberg), October 10th, 1947 (Touschek to Heisenberg), BTA, Box 1, Folder 1.} A contemporary photograph of Werner Heisenberg is shown in Fig.~\ref{fig:Heisenberg}.  

In January 1948, Heisenberg involved Touschek in a discussion about a work on the S-matrix, which he   had just received 
  from Ning Hu, 
  one of Walter Heitler's collaborators  in Dublin, but at the moment in Niels Bohr's Institute in Copenhagen.\footnote{After graduation from  Tsinghua University in Beijing, Ning Hu had moved to the United States. He had obtained his PhD from Caltech and during 1944-1945 had studied quantum field theory under Wolfgang Pauli at the Institute for Advanced Study in Princeton. During his stay in Europe from 1946 to 1949 he visited Walter Heitler in Dublin and Niels Bohr in Copenhagen. In 1980 he   went back to  China, and was a long-time professor in the department of physics at Peking University.}
\begin{figure}
\centering
\includegraphics[scale=0.8]{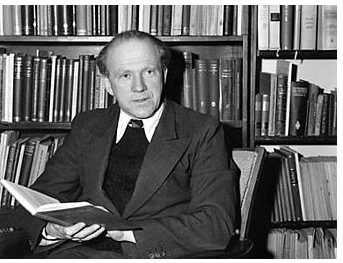}
\caption{A 1947 photograph of Werner Heisenberg from AP Archives,  in  Florian Hildebrand's article in \url{https://www.deutschlandfunkkultur.de/heisenbergs-traum.984.de.html?dram:article_id=153450}.}
\label{fig:Heisenberg}
\end{figure}
Heisenberg wanted Touschek to look into Ning Hu's work because he thought it to be in some contradiction with Touschek's unpublished notes.\footnote{The January 12th, 1948  Heisenberg's letter  to Touschek, also announcing his forthcoming visit to the UK, is in Touschek's  personal papers, courtesy of  Mrs. \EYT.} In asking him to check Hu's calculations, Heisenberg suggested him to directly contact Hu.
In the same letter, he also announced 
  that he would be in Cambridge for six weeks, starting on January 26th.\footnote{{During Heisenberg's visit to Cambridge, Peierls  extended an invitation to Heisenberg to come to Birmingham, letter [445] in  \citep{Peierls:2009}, and discuss the matter of German scientist responsibility and different attitudes towards the Nazi regime.  To this Heisenberg replied in letter [447] in \citep{Peierls:2009}: ``Dear Peierls!
Thank you very much for your letter. I like to come to Birmingham
and I can talk  at  your seminar
about the little thing I think I know about the theory of elementary particles.
I also thank you very much for being open to your opinion  about a difficult political problem. It is
as you suspect: I do not agree with you. But the
 fact that you wrote to me so openly gives me hope
that in a conversation, if not an approximation of the
viewpoints so as to understand the other viewpoint,
can come.
As to the timing: From 10th to 12th I am at Blackett's [in Manchester] in March;
I could come to B[irmingham] on the 12th and stay until the evening of
13.  Possibly also March 8 and 9, if from the planned visit to  Oxford would  come to  nothing. Would that be ok for you?
Goodbye".} }  

The correspondence continued through January, with Heisenberg writing from Cambridge on January 28th, 1948. 
 Other letters followed.\footnote{On January 26 and 27 (Touschek to Heisenberg), on January 28 (Heisenberg to Touschek, from Cambridge, UK), on January 31 (Touschek to Heisenberg) on February 29 (Heisenberg to Touschek), on February 23 (Heisenberg to Touschek from Cambridge, Cavendish Lab), on February 29 (Touschek to Heisenberg), on April 20 (Heisenberg to Touschek), on May 2 (Touschek to Heisenberg), BTA, Box 1, Folder 1.} 
 
 The exchanges on the S-matrix properties  between Touschek and Heisenberg
 appear in the research report Touschek submitted to the University of Glasgow, for the periods December 1947\& January 1948.
  In the report, one of the mentioned items is  a `triangular discussion' between Heisenberg, Hu and Touschek, reflecting a correspondence with Hu, mostly  lost. A  second report was  submitted on May 2nd for the period February 1st to April 30th 1948, and 
   we  learn that Touschek met Heisenberg in Manchester about  the matter.\footnote{Touschek's research reports submitted to University of Glasgow in 1948  are available in  BTA, Box 3, Folder 1.} In the meantime, on April 15, Hu's article ``On the Application of Heisenberg's Theory of S-Matrix to the Problems of Resonance Scattering and Reactions in Nuclear Physics'' was submitted to the {\it Physical Review} and appeared in the July 15 issue of the journal \citep{Hu:1948aa}. On May 25 Touschek thanked Hu for his manuscript, that he was carefully reading, also mentioning previous letters they had exchanged (but which are not present in Touschek's papers in Rome). Finally, on July 27, Touschek wrote to  Heisenberg that he had a rather lengthy correspondence with Hu and at the moment he did not think that Hu's work was correct.\footnote{BTA, Box 1, Folder 1. Touschek's work remained  unpublished, but it is a very interesting  result  quoted by the noted theoretical physicist R. J. Eden in  a   Royal Society communication   presented by Dirac \citep{Eden:1949}.} 
From the available documents, we see that the exchanges on this subject between Touschek and Heisenberg continued through part of the summer 1948. His closeness to Heisenberg is also seen in  a  sequence of letters between Touschek  and a colleague,  B.J.Warren, about a position in Vancouver offered  to  Dr. H. Koppe, a member of Heisenberg's group in
 in \Gott. However, after July 27th, 1948, 
 no letters are present in Bruno Touschek Archives in Rome up to 1958, when correspondence with Heisenberg started again.

The correspondence is proof of how  the young theoretician was trusted by Heisenberg, how much Bruno  would  be  learning  from this exchange of ideas, and how much he was deeply involved in strongly debated  theoretical questions of his time. The available documents indicate that the exchanges with Heisenberg  during the Glasgow period lasted about 2 years. There is no doubt they  had a profound influence on Touschek. The direct confrontation with one of the proponents of Quantum Mechanics, a scientist of great  intellectual and scientific stature,  did influence his formation. He could not avoid feeling proud of Heisenberg's clearly good opinion, and his own self-esteem could now be enforced. He would still make mistakes, as we shall see from his correspondence with Born or with Neville Mott \citep{Pippard:1998}, but,  encouraged by his exchanges with Heisenberg, he had no reasons  to doubt anymore his capacity to do physics. 

We shall now step back, to see how Bruno's other physics interests and personal life unravelled during 1948, the first complete year he spent in Glasgow.




\subsection{1948: Settling in Glasgow}

At end of 1947,  Bruno was finally able to travel to Vienna and see his parents, probably for the first time after the end of the war. He left London on December 16th, spent Christmas and New Year with them, and was back to Glasgow on January 6th 1948.
 In between, combining work and family, on his way back from Vienna in early January, Touschek passed through Malvern,  to see the synchrotron with Donald Fry.\footnote{An expense note to the DSIR on January 7th, 1948 mentions travelling through  to London   onto Malvern.}
The trip to London is also mentioned  in  letter to parents on January 28th, 1948.

The long sought  reunion with his parents  fortified his spirits and carried him through the harshness of the Glasgow winter and his frequent travels to England or to Edinburgh.
His research report for December 1947 and January 1948
describes many different physics projects,  about accelerator physics and possible  experiments,  
and, of course, theoretical physics. 
 In addition to giving lectures and working with the Glasgow group on designing and planning the 300 MeV synchrotron, 
 he was   involved  with the operation of  the 30 MeV  electron synchrotron in Manchester, and giving support to the group taking measurements with the 20 MeV betatron, as also seen by    the  intense exchange of letters between Touschek and the colleagues in Manchester, which   took place through the first  months of the year.\footnote{Letter to F.K. Goward January 8th, 1948, and exchanges with  Bosley in February, BTA, Box 1, Folder 1.} Both Bosley and Cragg from Metropolitan-Vickers  in Manchester came up to Glasgow to discuss with him about their results on X-radiation from the 20 MeV betatron.\footnote{In Manchester,  the Metropolitan-Vickers Electrical Company  had participated intensively to the war effort, including having some of its scientists released to work in the United States for the atomic bomb effort. The company was also active  in the field of nuclear physics, and a High Voltage (HV) laboratory had been opened in 1930  by  Ernest Rutherford,   director of Cavendish laboratory at the time, who  had been professor in Victoria University in Manchester from 1907 until 1917.  At the HV laboratory, cyclotrons had been constructed in 1938, one with  Cockcroft for the  Cavendish Laboratory, one to be installed at Liverpool for James Chadwick, the discoverer of the neutron.  After the war, a research group, finalised to work on accelerators, was started   in 1946, and a 20 MeV betatron, the first in the country, was designed and built. A collaboration, also involving the Telecommunication Research Establishment (TRE) in Malvern, was  started  with Philip  Dee in the design of the 300 MeV synchrotron in Glasgow, as described by John Dummelow in  the section about {\it Nuclear Physics}  in  \href{https://www.gracesguide.co.uk/Metropolitan-Vickers_Electrical_Co_1899-1949_by_John_Dummelow:_1939-1949}{Metropolitan-Vickers Electrical Co 1899-1949}. The British industrial enterprise asa M-V E. also  provided diffusion pumps for the Malvern synchrotron.}
 Touschek's  contribution was later acknowledged by  the authors  as their being ``greatly indebted to Mr. B. Touschek" (and F. K. Goward) for  ``instructive discussions on the work"  \citep{Bosley:1948}. He also was in correspondence with  Goward in Malvern.\footnote{{The extraordinary history of the contribution to science and technology from Malvern is kept alive by the  Malvern Radar and Technology History Society (MRATHS),  a registered charity No 1183001.}} 

But  he was keen to  continue on with  theoretical physics. In January he submitted a paper on the double $\beta$-decay to the {\it Zeitschrift f\"ur Physik} for  a special issue  prepared to celebrate Lenz' 60th birthday \citep{Touschek:1948aa}.\footnote{Letter to parents, January 28th, 1948. Already in November 1946, Touschek had announced to his parents, that he had come back from one of his preliminary travels to UK with an article on such topic ``in his pocket''. However he also told having found an error and that it took him a couple of months to set the problem and that it would appear in the {\it Zeitschrift f\"ur Physik}, but further problems must have hindered publication, probably also a normal delay as it happened for all scientific journals at the end of the war. The submission date is marked December 1946, but the article appeared only in early 1948 and was included in the issue dedicated to Lenz. In a later manuscript note, Touschek wrote: ``The problem which bothered H. [Heisenberg] \& which he asked me to unravel was `double $\beta$-decay''. Haxel felt he could just do it (experimentally) \& I ran into the difficulty of distinguishing what was arbitrary \& what was sound in Fermi's theory of `weak interactions.' I saw that clearly in 1947, but what I wrote then was riddled by stupid mistakes, which H. Did not -- or did not want to -- see'' (manuscript, courtesy of Mrs. \EYT). It even appears that this research argument had been already suggested by Paul Urban when Touschek moved from Vienna to Hamburg during the war, as claimed by the former in a letter written much later to Amaldi,  who was writing Touschek's biography (Urban to Amaldi, June 3rd, 1980, EAA, Box 205, Folder 4).} 
Touschek had been very close to Lenz during the war period,  attending his lectures in  Hamburg as an unregistered student, and also living with him for a period of time \citep[4]{Amaldi:1981}.\footnote{{Wilhelm Lenz (1988-1957) was Professor of Theoretical Physics at University of Hamburg, since 1921. His major scientific accomplishment is the formulation of the Ising model \citep{1967RvMP...39..883B}. He was a student of Arnold Sommerfeld, and protected Touschek during the war years.   \RW \ mentions  to Amaldi after  Touschek's death: ``Touschek lived in the flat of Professor Lenz in Hamburg [\dots] and he had considerable difficulty bringing the old and often sick man to the cellar when the bombers came". We notice that Lenz was not that old at the time, being only 55, but may very well have been weak or sick.}} 
His    contribution  had been asked by Hans Thirring, the former professor of Theoretical physics at
 University of Vienna until 1938, later reinstated to his position after the war.\footnote{{After the end of the war, Hans Thirring (1988-1976)  became dean of the Philosophical Faculty at University of Vienna in 1946-47, and  a member of the Socialist Party of Austria.}}  Touschek had   been  in touch with Hans Thirring  during the war years, whenever he could go to Vienna. In particular, we note that sometime, late 1942 or early 1943,  Bruno had been part  of  a discussion held with Hans Thirring and Hans Suess about relativistic effects when electrons reach high energies. This discussion is mentioned in a letter to his parents, and was    the   problem which Touschek  had noted  in   Wider\o e's article  (about   a  15 MeV betatron), which he   was reviewing for the \Archiv\   \citep[26]{Bonolis:2011wa}.\footnote{Letter to parents, February 15th,1943.} {This paper, accepted but never published, started the secret betatron project financed by the \RLM, the Aviation Ministery of the Reich  \citep{Bernardini:2015wja}.}


Notwithstanding  occasional bouts of annoyance, such as  we mentioned at the end of 1947,  towards a world which, to him, appeared to be moving too slowly,   he was not only working  on his doctorate, but also enjoying  the new environment. 

His involvement with the social life of the department is aptly described by Philip Dee in a letter he sent to Touschek's step mother, Rosa Touschek. Busy between lecturing and doing synchrotron work in Glasgow, traveling to Manchester, Malvern and Birmingham, attending Born's Seminar in Edinburgh, meeting and writing to Heisenberg, it is no wonder that     Bruno  may have neglected to write to his parents as frequently as had been his custom since his young age, when away from home.  Thus, one day,  on April 7th, 1948, his  step mother took the extraordinary step to write to Dee and inquire about Bruno's well being. Dee's answer is reproduced in Fig~\ref{fig:DeetorosaTouschek-missmeric}.
\begin{figure}
\centering
\includegraphics[scale=0.11]{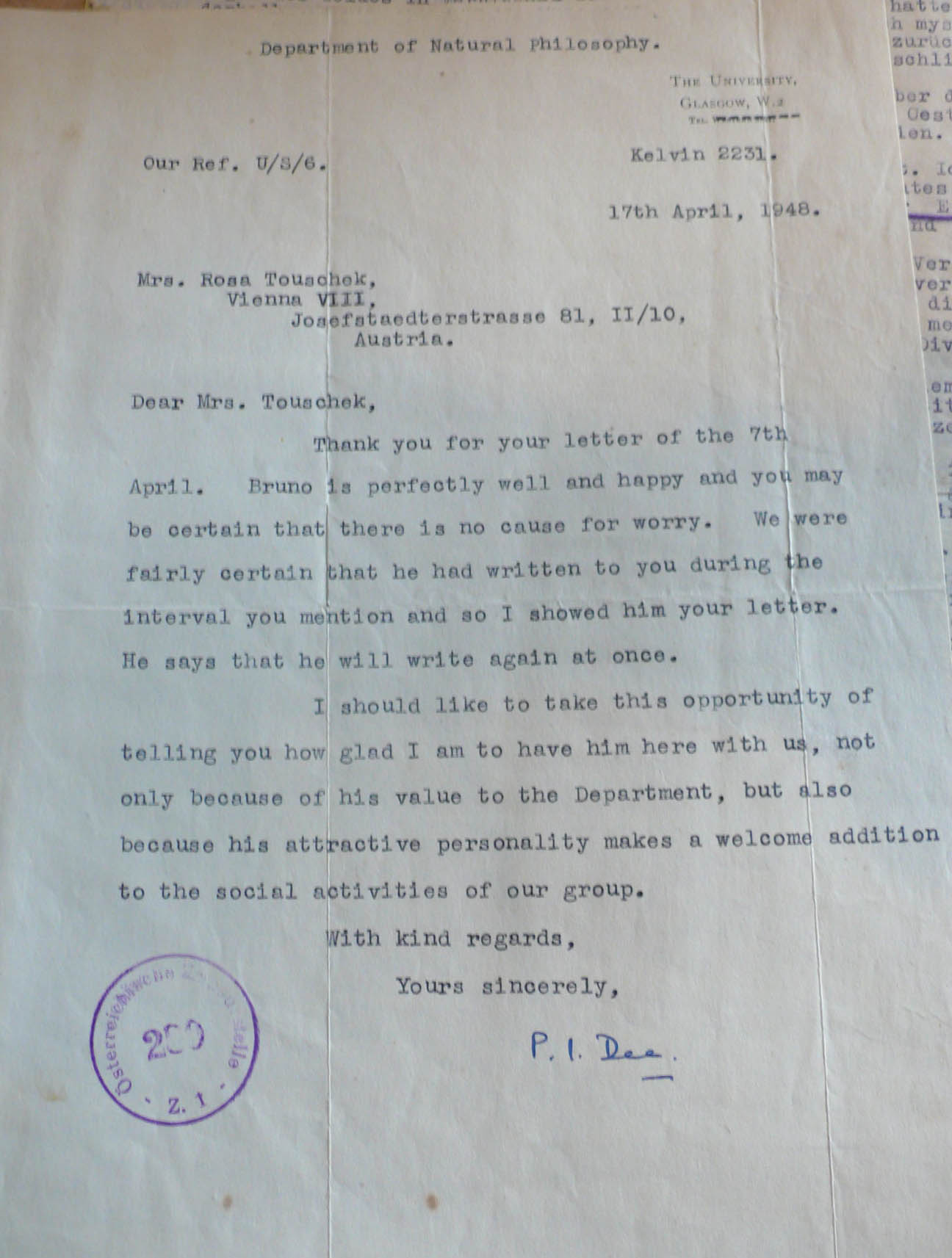}
\includegraphics[scale=0.41]{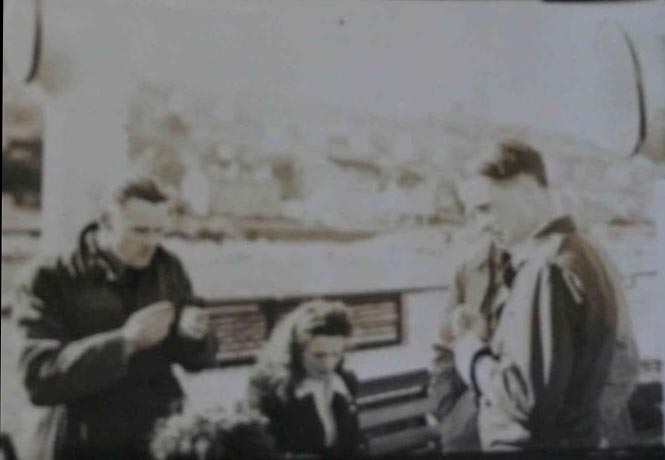}
\caption{The letter which Philip Dee wrote to Rosa Touschek  reassuring her that Bruno was fine and striving. At right, a snapshot  
taken on board returning from  an excursion to Rothsay, with Walter MacFarlane at   left and Touschek at right. Samuel Curran was also part of the same excursion, but is hidden behind Touschek, and  Miss Meri\c c is at the center. Courtesy of the late  Mrs. \EYT.}
\label{fig:DeetorosaTouschek-missmeric}
\end{figure} 
In this letter, Bruno is described by Dee as ``perfectly well and happy", adding how  glad he is to have him there, ``not only because of his value to the department, but   also because his attractive personality makes him  a welcome addition to the social activities of our group." This comment by Dee is confirmed by a snapshot taken during a later excursion which we show in the right hand panel of Fig.~\ref{fig:DeetorosaTouschek-missmeric}.  The group is identified   by Touschek in  the back of the photo,  the young woman in the center of the photograph  being Miss Merri\c c,   a graduate student from University of Instanbul, who received her PhD  in physics in November 1949, in the   same session as did Touschek.\footnote{Courtesy of \UGA.}

As the work  on the Glasgow synchrotron progressed,  rumors spread through the town and the countryside about  University professors planning to build some `atomic' project. To reassure the public of  lack of any danger, Philip Dee had to give an interview to the Glasgow Herald, as we can see from a contemporary  newspaper cutting 
 shown in 
Fig.~\ref{fig:cutting2}. The date of this article is not known, but it is likely to have been published  in mid  1948, when Touschek was {still} strongly involved   in the synchrotron project.
\footnote{Letter to parents, July  3rd, 1948.}
\begin{figure}[htb]
\centering
\includegraphics[scale=0.08]{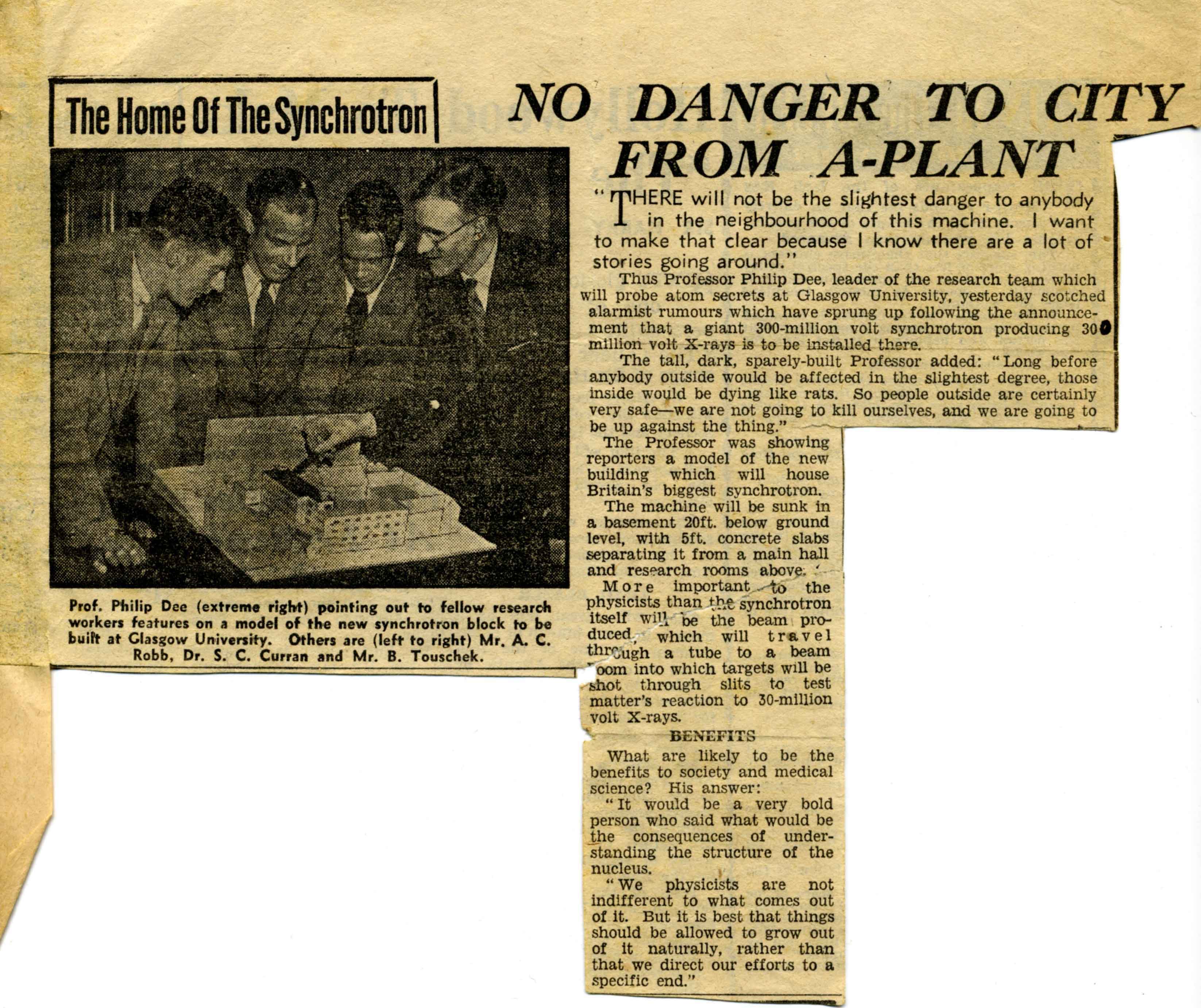}
\includegraphics[scale=0.08]{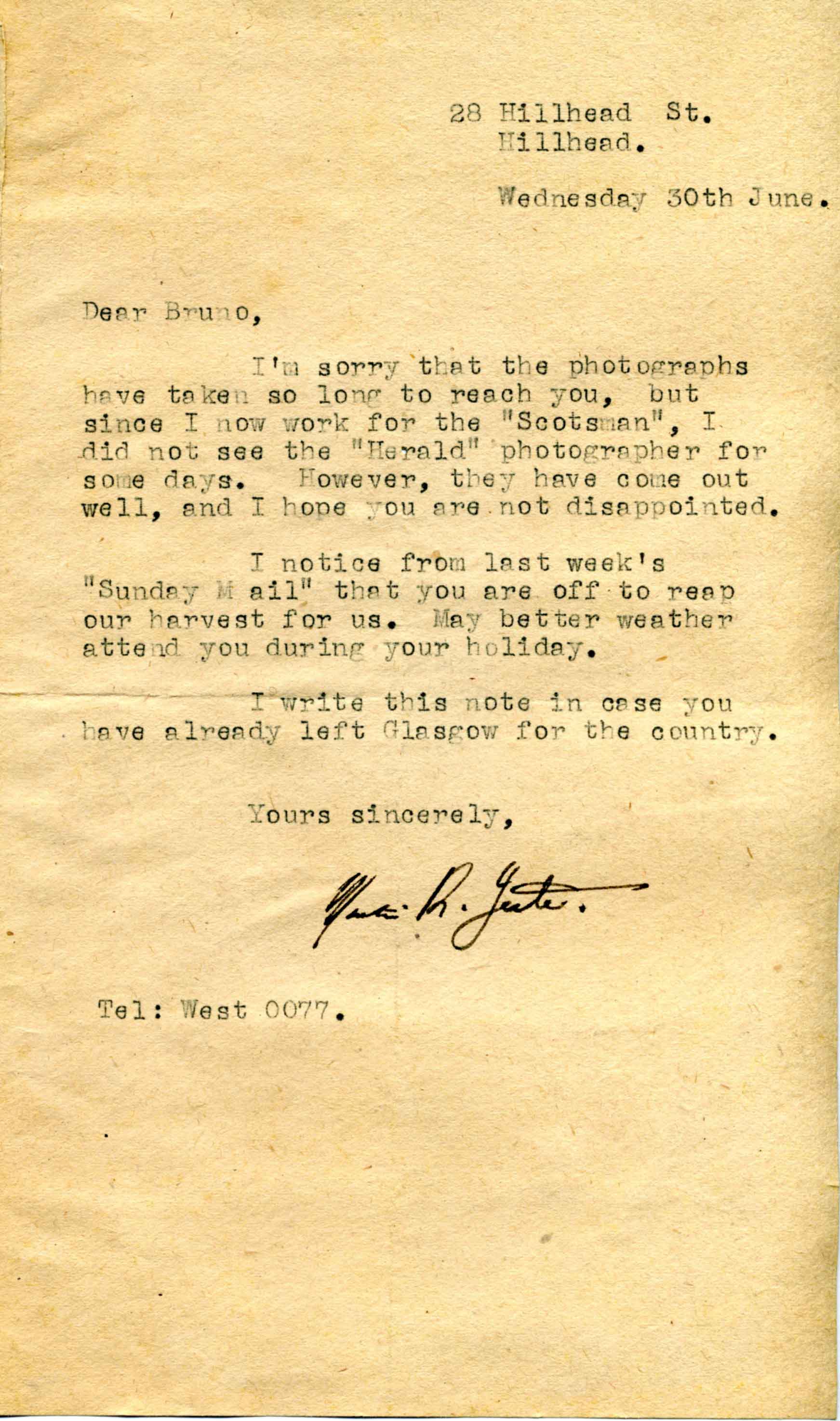}
\caption{ At left we show  
a cutting 
of an article in the local Glasgow daily. The photograph  is likely to have been  among those included by a journalist from the 'Scotsman', in a  June 30th, 1948 letter,  shown at right, where Touschek's harvesting in Northern Scotland is also mentioned. 
 Courtesy of the late Mrs. \EYT.}
\label{fig:cutting2}
\end{figure}
 {Part of his activity includes exchanges with  Emlyn Rhoderick, who was working on the Cavendish cyclotron in Cambridge at the time,   
 and would  join Glasgow University, shortly later.\footnote{ Emlyn H. Rodherick (1920 Ð  2007) worked at the Royal Signals and Radar Establishment during the Second World War on coastal defence radar, and studied physics at Trinity College, Cambridge. He then taught at Glasgow University, and  went on to become professor of solid-state electronics at Manchester. } Bruno was concerned about the treatment of Coulomb  interaction in meson scattering, and the related divergences. This problem was strongly debated, and of interest also to Rudolph Peierls, who visited the Cambridge group in August, as we learn from Rhoderick's August letter to Touschek.\footnote{BTA, Box 1, Folder 1.} 
Physics was not all encompassing, however. Always a lover of nature, Bruno  joined his colleagues in  excursions to the islands, to Rothsay as we have seen, and, in summer 1948,  went harvesting  to Wick,  a town in Caithness county,  very far up North. In Fig.~\ref{fig:DSIRlettera-trafilettoRaccolto}
  we show  the letter from the D.S.I.R. allowing him to take leave to participate to the harvest in Northern Scotland (BTA, Box 1, Folder 1), and, to the right, a  notice from the Glasgow Herald, which mentions these activities. 
  \begin{figure}
\includegraphics[scale=0.089]{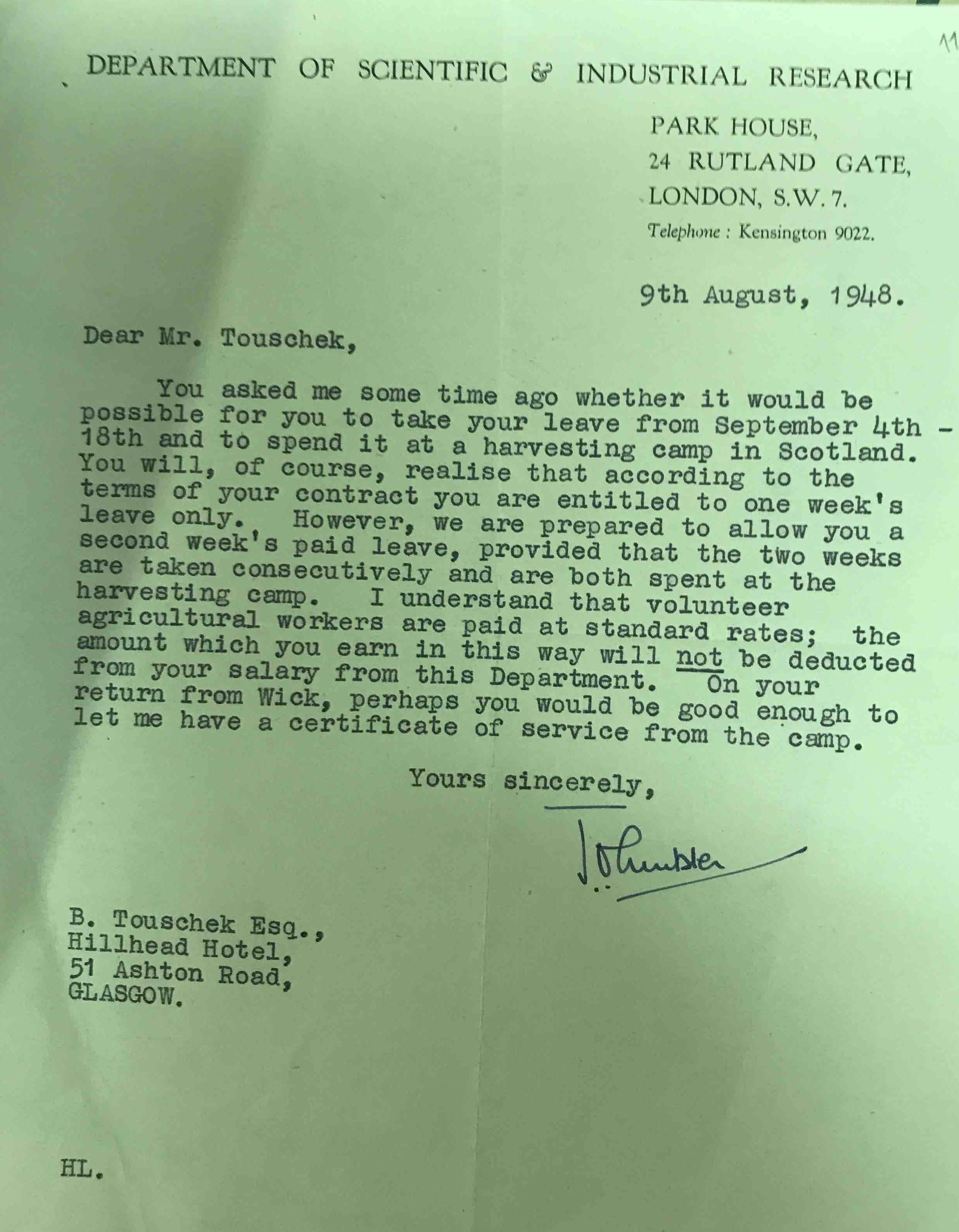}
\hspace{1cm}
 \includegraphics[scale=0.166]{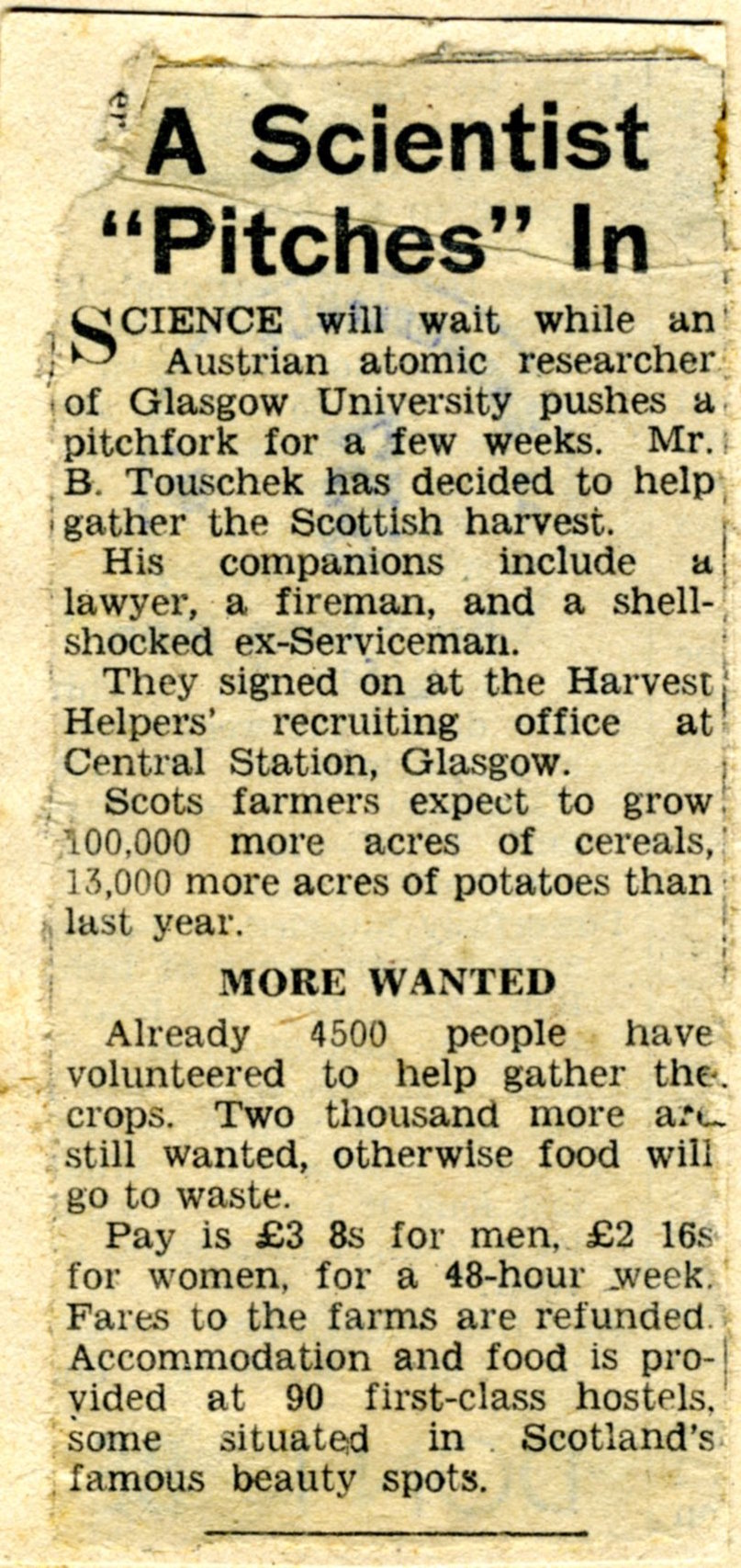}
 \caption{At left, copy of the DSIR letter of August 9th, 1948, granting Touschek leave to participate to the harvest. At right a  
 {newspaper}  article about Touschek's participation to the summer harvest in Northern Scotland. 
 Courtesy of  Mrs. \EYT.}
\label{fig:DSIRlettera-trafilettoRaccolto}
 \end{figure}
 
 Bruno's spirits in this period were high, and he started including little drawings in  the letters to his parents, as he had done since he was a little boy, and had continued doing through the war years until  September 1944. Since then, however, no drawings are  present  in the home letters until summer 1948, when he  related to his parents some  adventures of the summer harvest, humorously  drawing   his engagement in potato picking}.\footnote{One can find some anguished scribbles  or doodles   on the back of 1947 or 1948 letters he kept in his office  at University of Rome and presently in BTA, folder 1,  Box 1, {\it Corrispondenza varia, anni 1947-1949.}}
 We reproduce  them   in Fig.~\ref{fig:kilt}.\footnote{Letter to parents, September 3rd 1948 [our dating], with written date  3/8/48, probably but  a  typo, in place of  3/9/48.}
 \begin{figure}
\vspace{-1cm}
\includegraphics[scale=0.4]{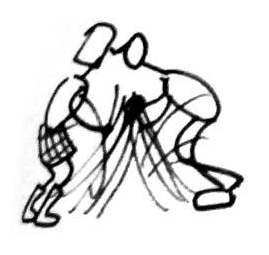}
  \includegraphics[scale=0.35]{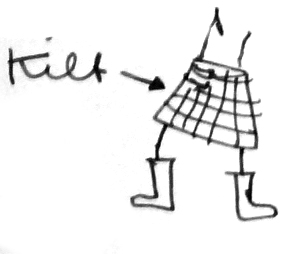}
  \includegraphics[scale=0.4]{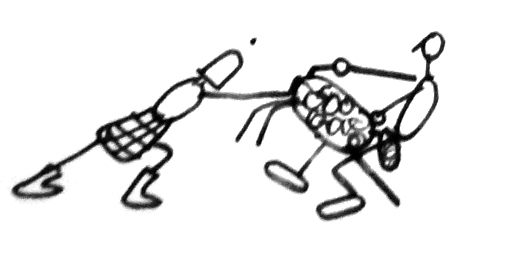}
  \centering
\includegraphics[scale=0.4]{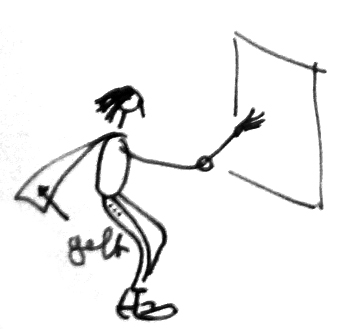} 
 \caption{Drawings included by Touschek in one of his letters home, describing his harvesting time in  Northern Scotland in summer 1948, graphics by A. Ianiro. Courtesy of Mrs. \EYT.}
 \label{fig:kilt}
  \end{figure}
{The reappearance of his  drawings and the humorous nature of their content signal a renewed confidence  in  his  abilities.  After the traumas of imprisonment, the tragedy of immediate post-war months in Hamburg, the  displacement to \Gott, and the move to Glasgow, he was relaxing, engaging with fellow harvesters, and enjoying, it seems, the company of a painter, something to which he had  been very much exposed in Vienna.\footnote{Touschek's mother  was rather good at drawing and his maternal uncle Oskar Weltmann was a painter of some renown in Vienna \citep[81]{Amaldi:1981}, and  one of Touschek's maternal aunts had married an architect, Emmanuel Joseph  Margold, assistant to Joseph Hoffmann, co-founder of the Vienna Workshop.  The family  artistic bend is  witnessed by one of Touschek's  war time letters to parents, where he asks they send him a picture by Egon Schiele, which used to hang in his own room in Vienna \citep{Bernardini:2015wja}.} }

Returning from the harvest, he found an unpleasant surprise. In his recollections to Amaldi,  Dee related the incident, writing \citep[9]{Amaldi:1981}: 
\begin{quote}
Naturally, after this  early period he gave me many problems! The first was his housing. A small lodging house seemed satisfactory for a while, but after a short 'holiday', which he spent potato picking in the north of Scotland, under spartan conditions, but fortified by the prospect of an early return to his comfortable room in Glasgow, this arrangement came to an abrupt end. On his return he found that the landlady had changed his curtains without prior consultation and, enraged by this destruction of  his anticipated homecoming, he immediately returned the curtains to the manageress with a demand for instant restoration of the original ones.  
 \end{quote}
 This request was obviously refused, and  one more  attempt  to find a suitable lodging took Bruno from 51 Ashton Road to   Kew terrace.\footnote{ Adresses given in May 23rd, 1948 letter to parents and   c/o Mrs. Boyle in October 29th, 1948  letter.} 

Shortly after the harvesting holiday,  physics was once more at the center of Touschek's life. Between 14th and 18th September,  a conference was organized in Birmingham by Rudolph Peierls.\footnote{See Peierls'  letter to Hans Bethe  [448], in \citep{Lee:2007aa}.} In Fig.~\ref{fig:1948Peierls-kerst}  see  Rudolf E. Peierls with Paul A.M. Dirac, at left, and Wolfgang Pauli, at the 1948 International Conference on Nuclear Physics, in Birmingham.
\begin{figure}
\includegraphics[scale=0.18]{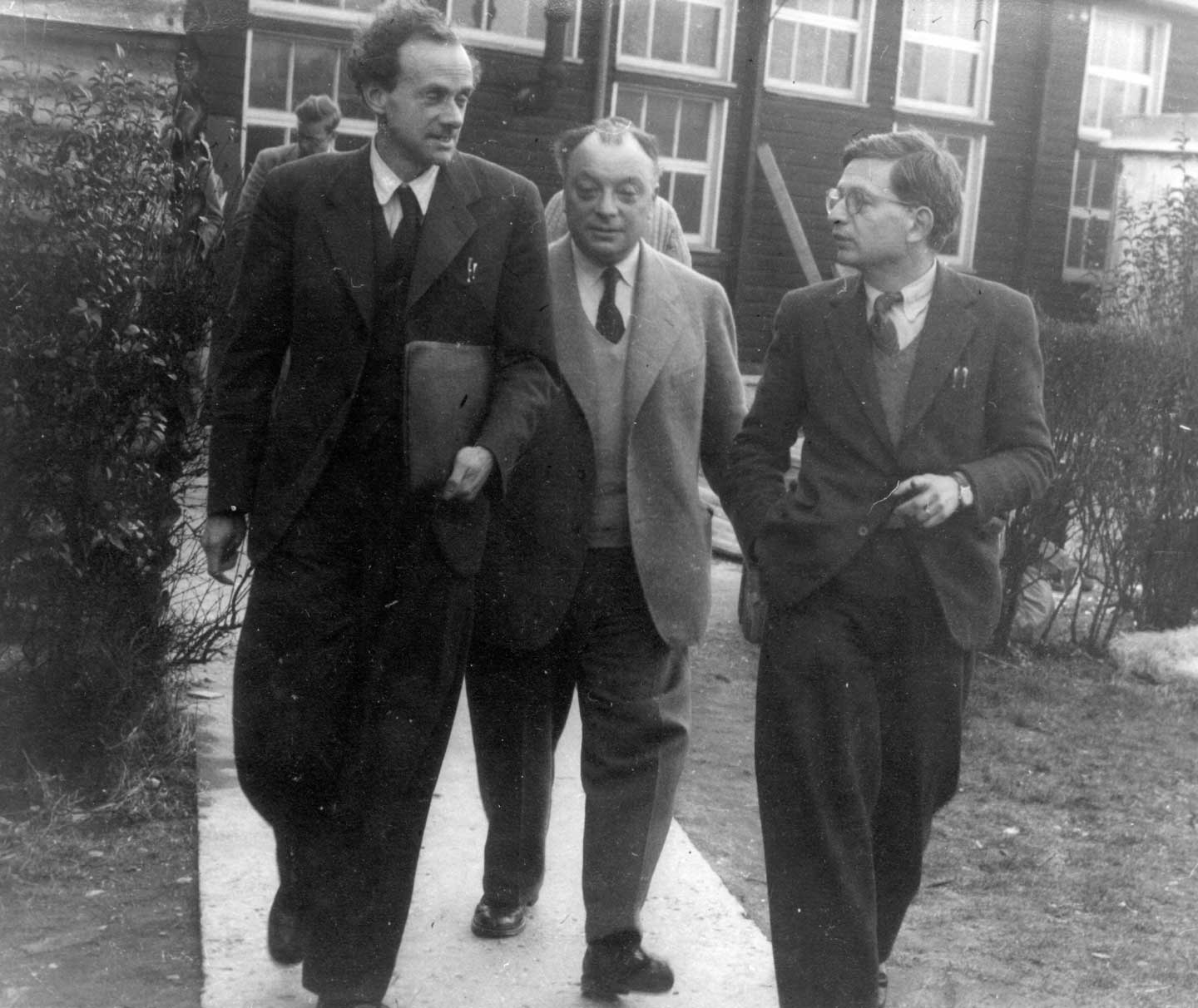}
\includegraphics[scale=0.26]{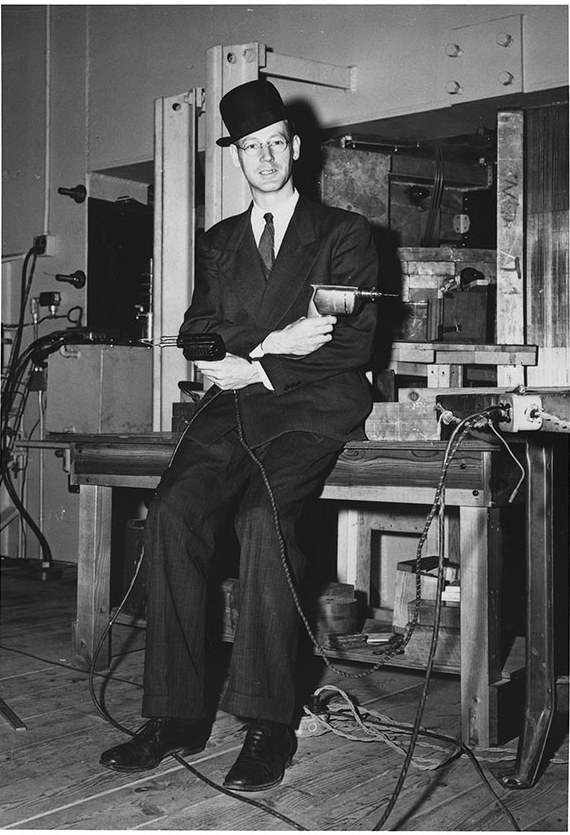}
\caption{Left panel:   Rudolf E. Peierls, at right,  with Paul A.M. Dirac, at left, and Wolfgang Pauli, at the 1948 International Conference on Nuclear Physics, in Birmingham, from \url{https://cds.cern.ch/record/42760}. See also CERN Courier vol 42 no 7: September 2002 (P.A.M. Dirac's obituary).
Right panel: Donald Kerst around 1950 wielding a soldering gun and drill,  from \url{https://www.aip.org/file/donald-w-kerst}.}
\label{fig:1948Peierls-kerst}
\end{figure}
 Touschek participated to the conference, where he  met Donald Kerst, shown in the r.h. panel in Fig.~\ref{fig:1948Peierls-kerst}. 
 
  In Birmingham, he 
  might have discussed his ongoing work  with  Rudolph  Peierls, external examiner of his 1949 PhD dissertation,  and  met Maurice Pryce, a mathematician and theoretical physicist from Oxford University, and, at the time,
Max Born son-in-law \citep{Elliott:2005}.\footnote{Letter to parents on October 10th, 1948 from Birmingham,  September 28th, 1948 letter to Kerst,  September 27th, 1948  to Pryce  {and (undated)  reply from  Pryce} (BTA, Box 1, Folder 1).} In those days, Touschek was working with Sneddon on meson production with electrons and after the conference sent to Donald Kerst a preliminary version of the paper, as possible example of  meson (particle) experiments  one could do  with  300 MeV.  At Kerst's Institute at University of Illinois,  they worked on the same problem (electron excitation), but Touschek thought they were about one year behind.\footnote{Kerst was constructing such a betatron in the United States, which became operational in 1950.
See also Kerst memoir in  \url{http://www.nasonline.org/publications/biographical-memoirs/memoir-pdfs/sessler-andrew.pdf}.} The paper  was later submitted and published. He also had news of his friend Fritz Houtermans coming to England, with whom he had quarrelled for no reason Touschek could remember. It was a silly thing, and Bruno sent him a postcard with a view of the place he had been in Caithness county  and a short amicable message. This restored them to the past  friendship, 
and Houtermans, who was at the time in England,  proposed to visit Bruno in Glasgow. Bruno was elated at the idea and telegraphed back 
``By all means come!".\footnote{Houtermans' letter  to Touschek is kept in Bruno Touschek papers in Rome, dated  October 25th,1948. On the top of this letter Touschek scribbled ``By all means come!  Wire date!'' Then adds his address and the word ``sent''  pointing to  Houtermans' address. The letter and Touschek's added words suggest   the text of a telegram he sent to Houtermans.} But  the encounter fell through and they do not seem to have met in this occasion.  

In October  as the summer fun was over and  the days shortened in the approach to winter, Touschek was once again  feeling restless and  unhappy. 
He was holding    a Nuffield fellowship,  his  previous contract having expired, and he inquired from Dee what could come next.  Dee's frank answer  includes  the possibility of a professorship in two years after the doctorate, but with some {\it caveat}: Bruno needed to accept life in the UK, and feel more at ease, as he was into discussing and quarrelling too often for his comfort. The real point is that he was anxious to progress, to be closer to the places where theoretical physics was taking giant steps forward, such as indeed was happening in particular in the United States.  He   feared  remaining isolated in Glasgow, without being able to keep the needed intellectual connection to other theorists. He felt that perhaps he should be moving out again,   and that he may have been wasting  his time. It should be added that he may very well have been going through some  exhaustion and its consequent depression state. He was in fact still continuously travelling, like having to be in   Malvern on  Monday and back to Glasgow on the Tuesday.\footnote{Letter to  parents, October 5th, 1948, also about meeting Kerst in Birmingham.}

In December, painful  memories  were coming back, as a new movie from Germany was released and shown in Glasgow's Cosmo theatre to a packed audience. {\it Die M\"order sind unter uns},\footnote{The title of the movie, {\it The murderers are among us} in English,  recalls the  title of the 1921 Fritz Lang's movie {\it M}, originally {\it M\"order unter uns}.} was shown in various countries around the world, and was seen by Touschek together with the whole physics department. The action of the film,  by the German director Wolfgang Staudte, took place in allied-occupied Berlin, where Touschek had lived during the war. It was one of the first post WWII German films, the first to use as setting for the story the  consequences of the bombings, with piles of rubble and destroyed buildings.
It was   produced by a company, DEFA, established in the Soviet occupied zone. Its aim    was to urge the public to see and judge  those responsible for the atrocities committed during the war.\footnote{The movie is available through YouTube. Another, almost contemporary, movie on the same subject is Roberto  Rossellini's 1948 movie {\it Germania Anno Zero}.  }  

None of this could be soothing Bruno's anxiety and possibly incoming depression. In addition, as the year 1948 drove to its end, Touschek went through one more change of lodgings. The occasion amounts to  an almost comic story, with a landlady quarrelling with her landlord husband, the husband hitting the wife, Touschek trying to defend the wife  and being hit by the husband, who finally called  the police. The story is related by Bruno in a letter to his parents, but  in later years he  narrated it to his friends in Rome, who picked it up to become an  often relished anecdote about Touschek's  life in the UK In Carlo Bernardini's version, Touschek described Mrs. Boyle's house in Ashton Road  as {\it una casa piena di generali}, a house full of generals.\footnote{Personal communication by Carlo Bernardini (1931-2018), Professor of Physics in Rome Sapienza University, close friend and collaborator of Bruno in the AdA adventure.} In his letter to Amaldi in 1981, Dee describes the episode  as follows:
\begin{quote}
[\dots] on a Sunday morning [\dots] during my lunch, I answered the door to find Bruno on the doorsteps, very dishevelled and agitated and exhibiting a severely bruised eye. It transpired that during lunch his host had spoken very rudely to his wife and Bruno's attempts to teach him marital civility had ended in a violent physical encounter.
\end{quote} 
It is at this point that Professor Dee and his wife, always very kind and  affectionate as Curran says in Dee's biography \citep{Curran:1984aa}, offered Touschek to move in their house. And this is why  the end of the year 1948  finds Touschek  settled  into the top floor of Dee's home, 11 University Square.\footnote{Letter to parents from Glasgow, 6th December, 1948.}

 Dee and his family lived in one of the 13 townhouses 
  built for the University Professors by the famed architect George Gilbert Scott. Scott built a large number of institutional and domestic  buildings in the {\it gothic revival} style, such as The Midland Hotel near St. Pancras Station in London. Number 11   was especially reserved for the Professors of Natural Philosophy.  Its first occupant was William Thompson, Lord Kelvin, who lived in the house from 1870 until 1899, when he retired. It was entirely lit by electricity, probably the first in the world to have such futuristic installation. It still houses a clock, especially designed by Kelvin,  which spans two floors.\footnote{See  \url{https://universitystory.gla.ac.uk/building/?id=85}.} Philip Dee was the fifth resident in the house, from 1943 to 1972, when he retired. 
In Fig.~\ref{figs:11University} we show a front  
view of Dee's  house and, at right, the plaque commemorating Lord Kelvin's residency.  
 \begin{figure}[htb]
 \includegraphics[scale=0.197]{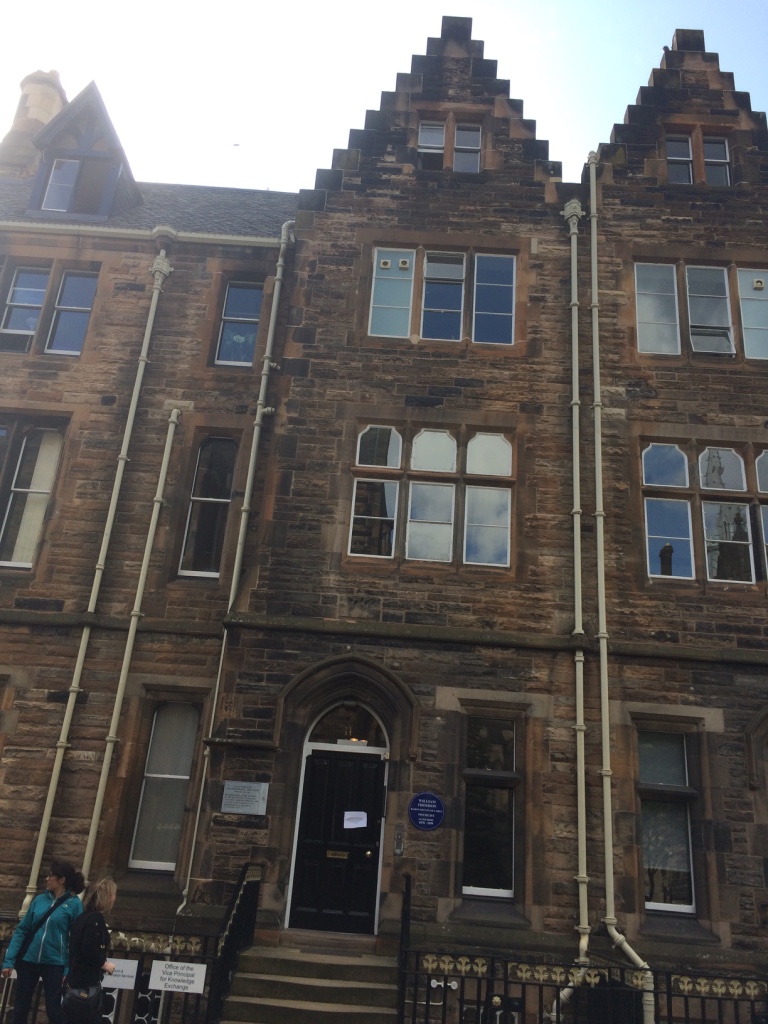}
 \includegraphics[scale=0.278]{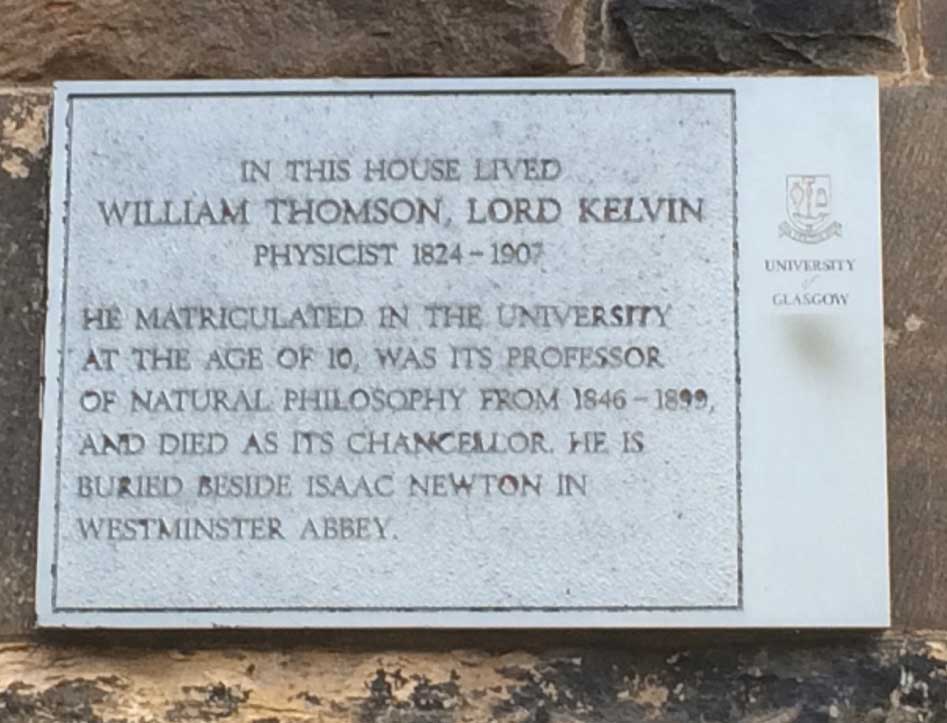}
\caption{The Square, 11 University: left  photo shows 
a  front view of  the house, where \BT\ lived in the top floor with Prof. and Mrs. Dee, at right  the plaque commemorating Lord Kelvin.}
\label{figs:11University}
\end{figure}

Touschek lived with the Dees for almost two years  in the top floor of  their  house.\footnote{Letter to parents, December 6th, 1948: "At the Dees I live now in top-flat [of their house]", letter to Max Born from Oakfield Avenue on October 26th and November 1950 to parents with description of moving and room plan. }  Professor  and Mrs. Dee were remembered by Bruno as well as by many other colleagues as  exceptionally kind. As soon as Bruno moved in their house, Mrs. Dee took care of buying proper furniture for his room, treating him almost like a son. In his 1979 letter to Amaldi, Dee remembers:
\begin{quote} 
Our house in the University was an old one on five floors, with rather steep communicating stairways. During the year or two which followed, I never met  Bruno on these stairs. His transit time from top to bottom and in reverse were  always so short that there was negligible probability for an encounter.
\end{quote}

 Dee remembers warmly Touschek's period and Touschek found a family atmosphere in Mrs. Dee's kind attentions to his needs. In the left panel of Fig.~\ref{fig:Dee-solvay1948}, we show a contemporary photograph of Philip Dee together with Wolfgang Pauli, taken at the 1948 Solvay Conference with the official photo of the Conference also shown  in the right panel.\footnote{The 1948 Solvay Conference took place after a hyatus of 15 years, the longest interval since its beginning in 1911. Through WWI, the conference series had also seen  a long interruption, from  the second Solvay conference in 1913 until the third in 1921. Thus the eighth Solvay conference, held in Bruxelles on October 28th,   saw gathered together all the protagonists of modern physics before the war and some new entries as well. The 1948  conference followed two important meetings held in the United States
 a few months before in the same year, the Annual APS meeting in New York  in January, and the Pocono Manor meeting, April 30- May 2nd, in Pennsylvania.  At these two meetings, there are some of the  first public appearances of what we now call QED, Quantum Electrodynamics, with  Julian Schwinger, at both  the APS  and the Pocono Manor meeting,  giving lectures on the new method to solve and calculate problems in particle scattering \citep{Kaiser:2005}.} 
 \begin{figure}
 \includegraphics[scale=0.0795]{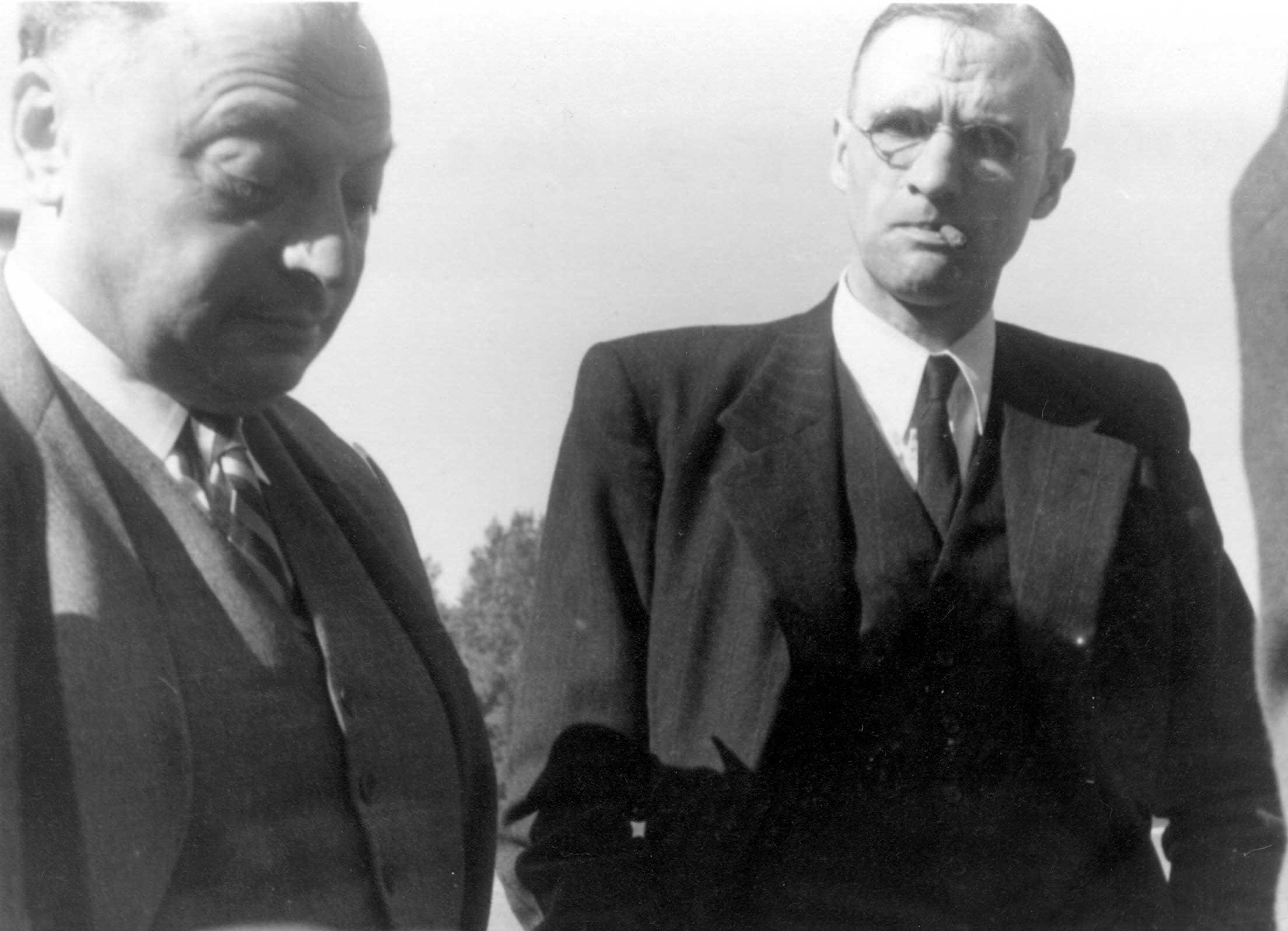}
 \includegraphics[scale=0.36]{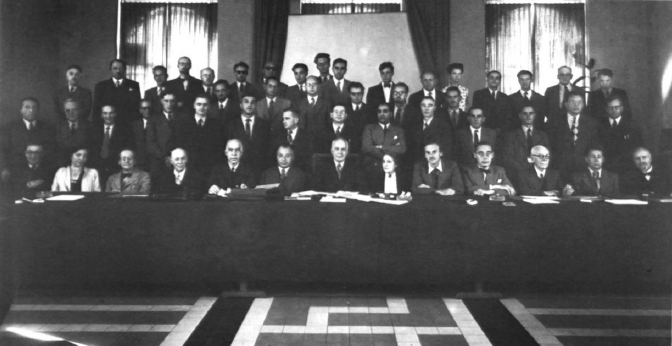}
 \caption{Left panel: Wolfgang Pauli, at left,  and Philip Dee at the 1948 Solvay Congress in Brussels, from  \url{https://cds.cern.ch/record/42755}. Right panel:
 The 1948 Solvay Conference  photograph showing participants. Philip Dee, Rudolph Peierls and Homi Bhabha are shown in second row, fifth, ninth and tenth from left respectively,  Bruno Ferretti in third  row, behind  Peierls and Bhabha.  Photograph from \url{https://upload.wikimedia.org/wikipedia/commons/8/82/Solvay_conference_1948_g.jpg}.}
 \label{fig:Dee-solvay1948}
 \end{figure}

  The wounds of the past could start to  mend.  The  more stable situation brought by  comfortable lodgings, more akin to  the well-to-do  Vienna homes, with his parents or  his grandmother's,   or in  Rome   with aunt Ada in the '30s, would now help him to recover from the traumas of his past.

 \subsection{ 1949: Getting the Doctorate}
 {1949 was a very important year in Touschek's progression towards becoming a {\it true blood}  theoretical physicist. As we shall see, he  started to work   with Born  on the new {(fifth)} edition of his {\it Atomic Physics} book,  travelling often to Edinburgh and   writing the appendix  on $\beta$-decay, a subject Touschek  knew well from his \Gott \ days and also including  some  nuclear physics.\footnote{For this see the Appendix on  meson physics,  as it called at the time, 
 pag. 415 of  the 1962 Edition \citep{Born:1962}.} Then, at the end of the year, he obtained his Doctorate and became Nuffield Lecturer. }
 
 As the new year started, Touschek was still  involved in collaborations  with experimentalists, 
 { but was more and more turning to theoretical physics, where his output is quite intense, mostly in collaboration with Ian Sneddon.} In addition  to 
 the papers published with Sneddon more directly focused on nuclear physics proper  \citep{TouschekSneddon:1948aaa,TouschekSneddon:1948ab} as well as on the ``excitation of nuclei by electrons'' \citep{TouschekSneddon:1948aa,Touschek:1950aa}, they wrote a preliminary short note on the interaction between electrons and mesons submitted in October 1947 and published in April 1949  \citep{TouschekSneddon:1949ab}, and soon after, on January 20, submitted a more complete paper on the results of their investigations on the ``probability of producing mesons by electron bombardment'', a very important question in view of the recent developments in the design of synchrotrons expected to produce  high-energy electrons. 
 These results were presented by Bruno in  February 1949 at the Annual meeting of the Physical Society, held that year at the University of Edinburg, where he gave a paper ``on electrons as nuclear projectiles'', while Curran presented work ``on the use of proportional counters to investigate $\beta$-disintegration'' \citep{Edinburg:1949aa}.
 
All this was at the core of Touschek's PhD dissertation, which he submitted in May, and whose title was  {\it Collisions between electrons and nuclei}, and represented ``a review of the work on electron excitation carried out in collaboration with Sneddon during the period 1947-1949".\footnote{A copy of the dissertation was kindly provided by Prof. D. H. Saxon to one of us (L.B.) in 2010.}
 
  In March he was still living with the Dees, not having yet found   suitable accommodations elsewhere, partly because of the cost of a reasonable serviced lodging, but very likely because Dee's  hospitality  was providing  him with a very comfortable home and had a calming effect on him, which he appreciated.  He continued travelling to Edinburgh, where he was an active participant to Born's Seminar. Max Born was fond of Bruno and appreciated his presentations, even when not all of his ideas would turn out to be correct.   In one occasion, Bruno gave there a lecture, and two days later, to his great embarrassment,  discovered    that part of the arguments he had presented was wrong.
 But Born  knew how to encourage students whose capacities he valued (he had been one of Heisenberg's teachers, the other had been Sommerfeld) and kindly told Bruno that he had really liked the other part (which was correct). 

 As he started feeling more at ease with  his research,  and confident of  being able to  keep ahead with the requirements for the PhD, he decided to have a real summer vacation.
  The Dees were going to spend their  holidays  on the island of Skye and the department would accordingly enter into some lethargic state.  Bruno's summer plans were to be with his parents. As a Nuffield research fellow, he could take a full month vacation, quite enough for   relaxing  in a place  nice and warm before returning to  Glasgow 
   and the Scottish winter. The plan was to be near some mountain lake or pool in the Austrian Tyrol, where he could find a place  to swim, as he loved, and take excursions, walking  through the woods and mountains of famous Alpine resorts, such as Kitzb\"uhel,  usual vacation site for the Thirring family,  or Alpbach.  He started proposing the idea to his parents  in March, since restrictions still applied for Visas to enter Austria from abroad and travelling documents would take time, not to mention planning for the money to finance the trip.\footnote{Letter to parents, March 9th, 1949.}

 A  month later  the question of the summer stay  was becoming a full-blown problem.  One month vacation  in Tyrol would not come  cheap of course. His parents in fact, did not think such vacation could be afforded.  But Bruno felt confident it could be handled, even recurring to a loan from his bank if needed. He also considered the possibility to go further South, and visit 
 {his maternal  aunt Ada and her husband, the  Italian industrialist Gaetano Vannini. Aunt Ada and her husband were  the only ones in the family who were in good financial conditions.They had no children and were very attached to Bruno, who had often spent his school vacation with them, in Rome, before the war. More recently, he had written to aunt Ada  from an airplane, probably while going to Vienna a  year before at Christmas time, and she had been duly  impressed by the fact that he was an air traveller. In a letter to his parents, Bruno, jokingly, muses   that    his  work with pump equipments (for the planned synchrotrons) may have contributed to impress his aunt, who ran  a pumping business with her husband. Aunt Ada and her husband 
  had also started  building a house near Lake Albano. If the house were finished, visiting them would be an attractive, inexpensive prospects and he  wrote her a letter.}\footnote{Letter to parents, April 23rd,1949.}

Summer plans remained   undefined until late May, but after debating whether to go to St. Ulrich or to St. Johann or some similar place,  and whether  to rent a house or stay in a hotel, he finally convinced his parents and decided on Flecken, a small village with  the attraction to be near  swimming possibilities, such as  the  Pillersee could offer,   in addition to  being close to the Thirring family, also holidaying near by, in Kitzb\"uhel some  20-30  kilometers away.
 
During  this first part of the year, he was also working hard,  travelling, occasionally   racing all over England. In April he had to go to Oxford and then to Harwell where he met Richard Becker, his professor from \Gott. Becker was  quite optimistic about the situation in Germany  and inquired if Bruno would have liked to go back. The idea was appealing, but Bruno did not come to a definite conclusion, perhaps for the poorer financial prospects which Germany still offered, and the occasion to return to Germany slipped away. Later, he regretted not to have come to the opposite decision, but subsequent events in his life may indicate otherwise: Italy, where he settled in January 1953, offered him the way  to combine  theoretical physics  with his  knowledge of  particle accelerators,  and thus   build  the world's first ever electron-positron collider, AdA,  in 1960. 

After  the intense travelling to all the places in England where synchrotrons were being built, and the hard focusing on his theoretical work for the approaching dissertation, the long sought summer vacation in Tyrol finally came. August went by and Bruno enjoyed  his parents company. In Fig.~\ref{fig:Pryce-twowalking-hat-plume} we show one of Touschek's companions during the Tyrol vacation, the Oxford physicist and mathematician Maurice Pryce,  and a  drawing  with  the silhouette of   two figures, an older one with a plume in the hat (a characteristic Tyrolean headgear),  and a younger  slim one,  which may be  a sketch  of Bruno and his father, during the Tyrol vacation.
 \begin{figure}[htb]
\centering
 \includegraphics[scale=0.45]{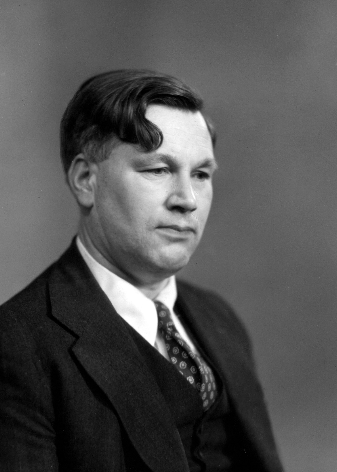}
 \fbox{\hspace{+3cm}
\includegraphics[scale=0.099]{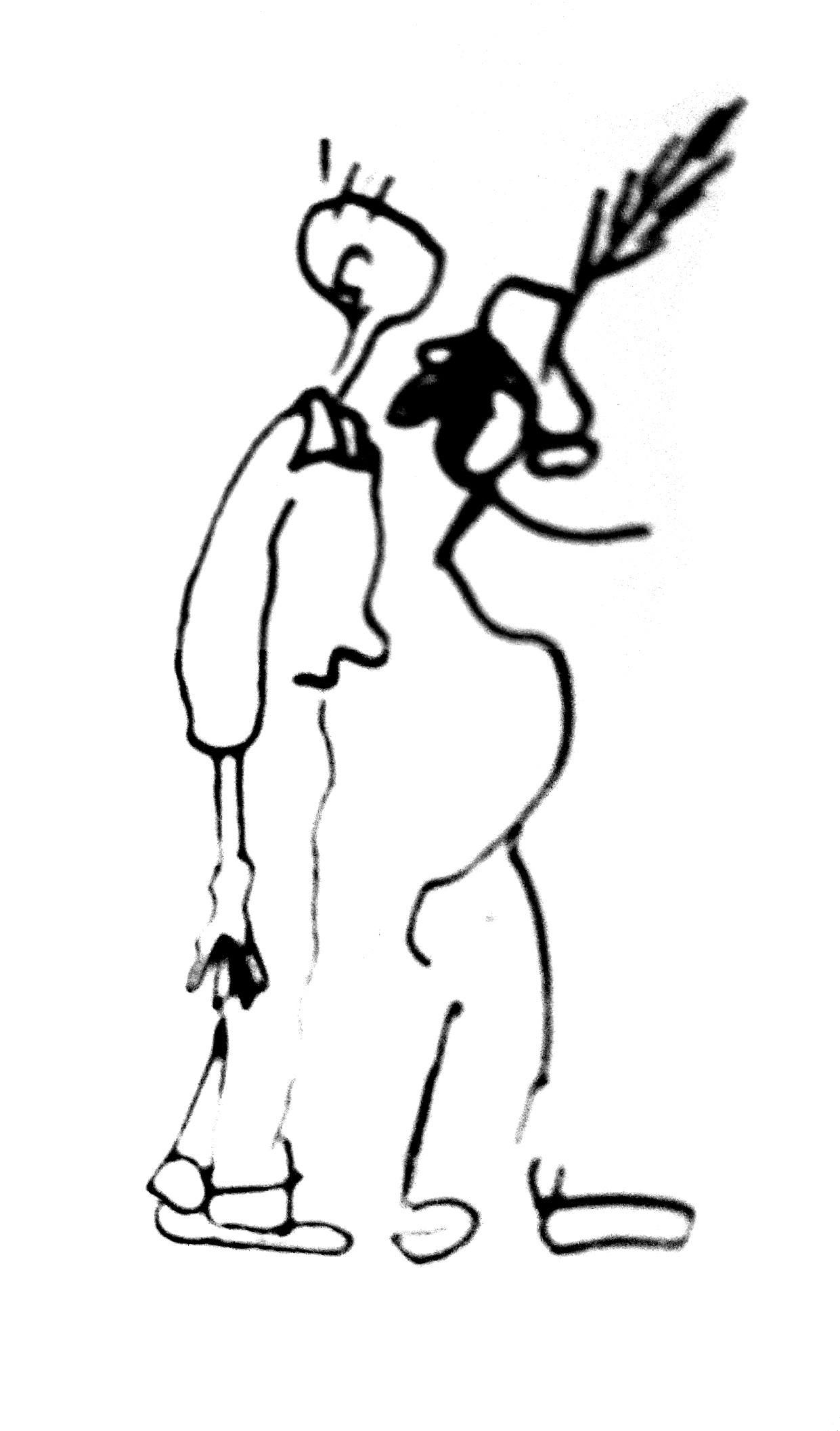}}
\caption{Left panel: Maurice Pryce, with whom Touschek shared a mountain excursion in Tyrol during summer 1949, from \citep{Elliott:2005}. Right panel: One of Touschek's drawings placed together with   fall 1948 letters, but possibly referring to 1949 Tyrol vacation with his father (from  \RSUA,  BTA, Box 1, Folder 1).   Graphics by A.Ianiro.}
\label{fig:Pryce-twowalking-hat-plume}
\end{figure}
After his parents  left, he remained in Tyrol for an extra 5 days in the company of other physicists, among them Pryce,   with whom he had a very pleasant  mountain tour, and  L\'eon Rosenfeld,\footnote{L\'eon Rosenfeld was a Belgian physicist  born into a secular Jewish family. He was a polyglot who knew eight or nine languages and was fluent in at least five of them. Rosenfeld obtained a PhD at the University of Li\`ege in 1926, and  was a close collaborator of  Niels Bohr.} 
 with whom he shared his return home through Switzerland.\footnote{Letter to parents on September 20th, 1949.}
 
 Back to the UK, there  came both the return to physics and the  shocking news of  
 devaluation of the pound! On September 19th, the pound sterling was devalued  from 4,03 dollars to 2,80 dollars. This was announced by the Chancellor of the Exchequer, Sir Stafford Cripps, in a broadcast to the nation. The devaluation was enormous, almost 30\%, resulting  in increasing costs for everything imported from abroad.
 For Bruno this meant that whatever money he could send to his parents would be decreased by almost a third through the new exchange rate. He was very upset and felt 
 {he had been  betrayed by the country he was now living in}. He felt insecure about future prospects, while the devaluation was also threatening his commitment to his parents financial comfort. Thoughts of going abroad, in the United States or Canada, for 1 or 2 years crossed his mind. Or, since preparations for such move could take time, other possibilities in England or, more likely, in  Ireland could be explored. He had a year-long invitation by Janossy in Dublin, which had graduated from the same high school  in Vienna as Bruno, the Schottengymnasium, but four years before him. Visiting   Dublin  was an attractive possibility,  also because  Walter Thirring, Hans Thirring's son, whom Touschek knew well, and considered an outstanding young theorist, was also going to be there in 1950.   

All these plans of course would have to wait until Bruno received his PhD, but in fact this was going to happen soon. In mid September  he was informally told that the dissertation he had submitted in May had come back  with a very good rating from Rudolph Peierls (the external examiner in Birmingham).\footnote{Letters to parents on September 20th, 1949.} The graduation  ceremony was going to  be held in early November, and he expected to receive a new contract with  an increased salary, which might offset the pound devaluation. 

 While pleased with looking forward to receive finally his official entry pass into being a  theoretical physicist, Bruno could not avoid feeling that Glasgow
 was as dirty as ever, looking  even blander and dirtier after a summer break.    The prospect of intensified austerity did not  make things any more pleasant, but he hesitated in running away, yet. He had  done this all his life, and thus   escaped stagnation and probably death. But now, he had  gone far, beyond his early dreams, and the next move  needed to be carefully thought out. It is quite possible that similar considerations were in his mind when, earlier in the year in Harwell with Becker, he did not follow up with the \Gott\ offer.
  
  In October,   he received an official letter informing him that his thesis has been approved and no corrections were requested, intimating that he would receive his Doctorate in the coming session,  the letter also  including the address of the robe-makers, in case he planned to be present at the official Graduation Cerimony (which he did). He also received the assurance that he could continue staying in Glasgow as ``Official Lecturer in Natural Philosophy"
  \citep[12]{Amaldi:1981}. With the salary coming with his prospect of improved status as a Lecturer, he could feel financially more secure. Emboldened by  the official confirmation that his application for  doctorate had been approved,  he  sought to buy   a car, even though he still  had  neither a  driving license, nor  the money. Indeed his impatience with the slow process of learning to drive, led him to potential troubles, as he related his first driving attempt, with the car to be purchased,  in a humorous letter to his parents. Fig.~\ref{fig:robeletter-car} reproduces a small drawing included in the letter, showing the effect of his driving the B.S.A. - sport - (170 pounds),   as he was rushing out of the shop and  trying it.\footnote{Letter to parents, October 24th, 1949.}
   We show Touschek's description of this adventure in the left panel of  Fig.~\ref{fig:robeletter-car}, and, at right,  the  Senate letter he had received just a few days before. 
  \begin{figure}[hbt]
  \centering
 \includegraphics[scale=0.7]{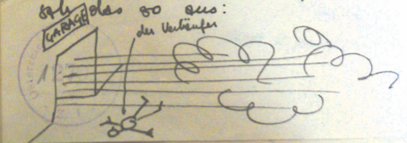}
  \includegraphics[scale=0.11]{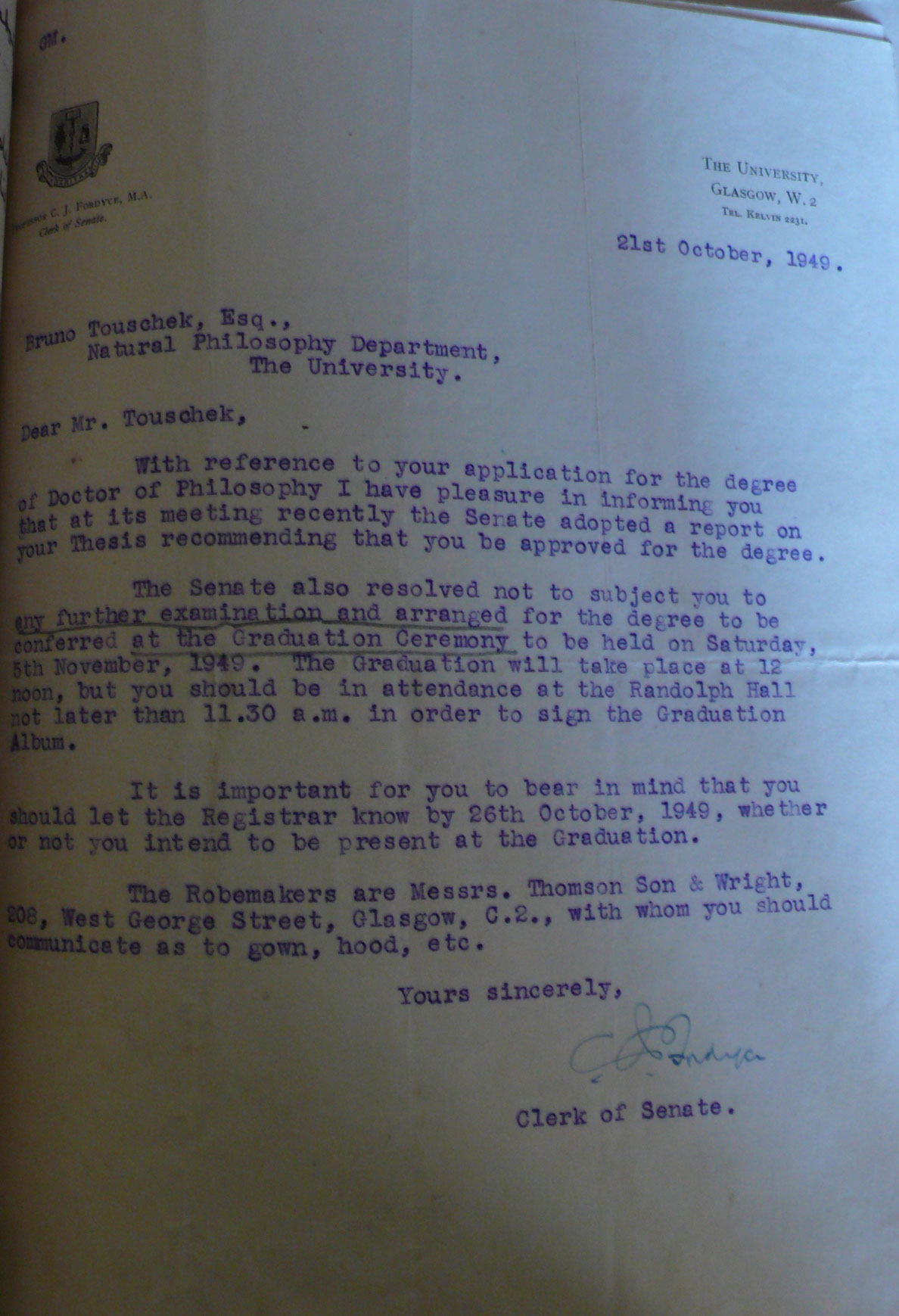}
   \caption{At left:  Touschek's   drawing about   leaving the garage where he had just wanted to buy a car, and almost crushed  the attendant seller. At right, the Senate letter informing Touschek that his application for PhD had been accepted,  October  21st, 1949. Courtesy of  Mrs. \EYT.}
     \label{fig:robeletter-car}
 \end{figure}

As we have seen, Touschek's doctoral thesis was on the interaction of electrons with mesons, a topic 
 on which he wrote several papers  with  Sneddon. John C. Gunn, who held the newly established Chair of theoretical physics,  was his internal examiner.\footnote{John C. Gunn was a Professor of Natural Philosophy at University of Glasgow. He was born in Glasgow, and studied at St. John's College in Cambridge. After the war he was lecturer in Applied Mathematics at the University of Manchester and then started research in nuclear and particle physics  at University College in London, as from \url{https://www.universitystory.gla.ac.uk/biography/?id=WH1433&type=P}.   In London,  Gunn  published  a paper on ``Interaction of Mesons with a potential field" \citep{GunnMassey:1948}. The paper, published by the Royal Society,  was presented  by Harrie Massey,  one of the active supporters  of CERN from the UK side.  Gunn  was appointed to the Chair of Theoretical Physics in University of Glasgow in 1949. See also \url{https://www.universitystory.gla.ac.uk/biography/?id=WH1433&type=P}.} 
The external examiner  was Rudolf Peierls, from University of Birmingham, with whom  Touschek shared his life-long  interest in the problem of {\it Radiation Damping}.\footnote{We know that, according to what \RW \ wrote to Amaldi, Touschek had worked on radiation damping during his prison days, even writing a paper, `in invisible ink' on the copy of Heitler's book {\it Theory of Radiation}.}
{A 1944 photograph of Touschek's s internal examiner appears  in Fig.~\ref{fig:Gunn-PhDrecord}. The official record of awarding of Touschek's  PhD is shown in the right panel.  }
\begin{figure}
\includegraphics[angle=90,scale=0.187]{Gunnwedding44.pdf}
\includegraphics[scale=0.083]{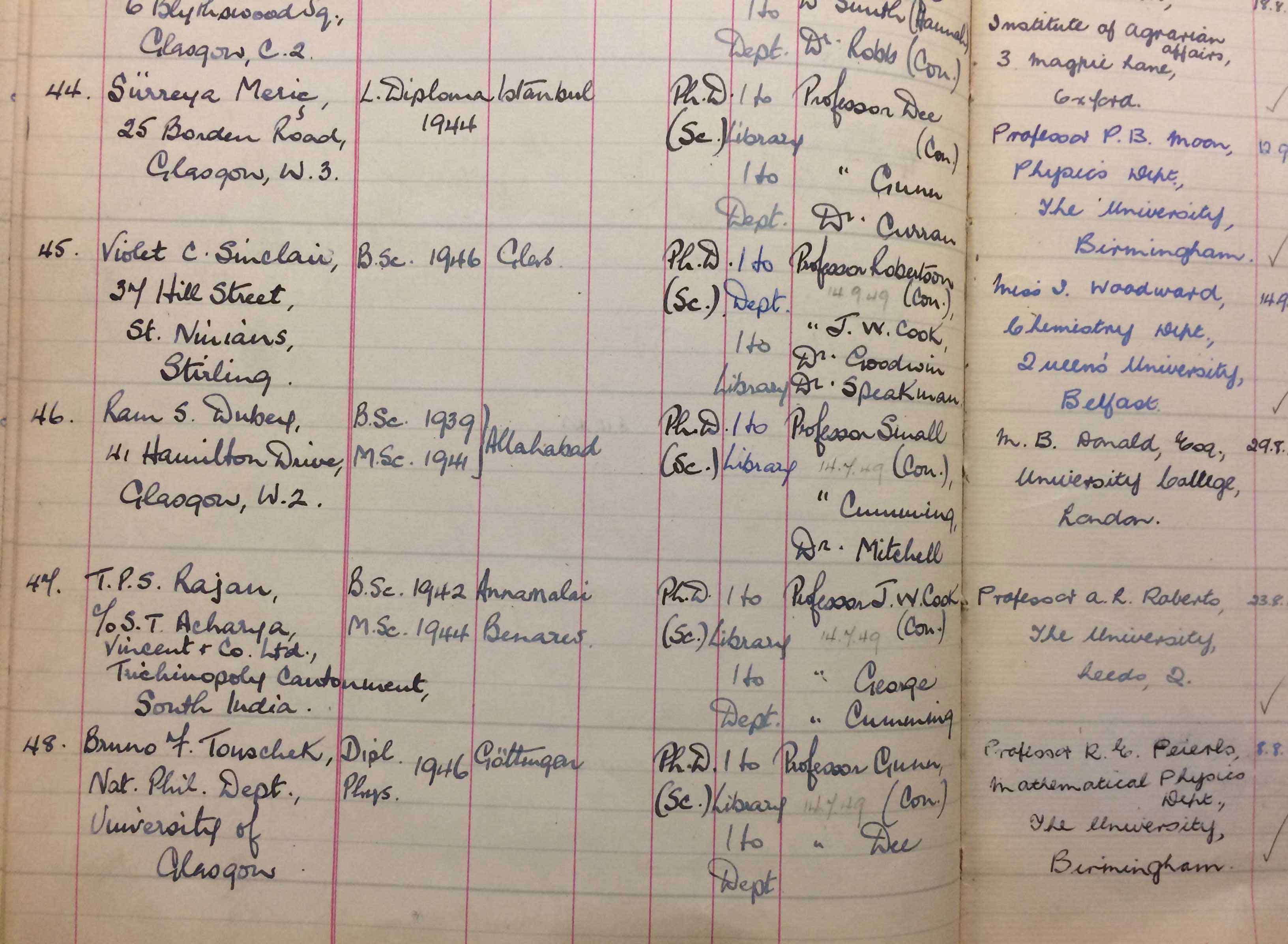}
\caption{ At left, we show a  photograph of John C. Gunn  and his new bride on their  wedding day, 1944, courtesy of Prof. J.M.F. Gunn, from Birmingham University. At right,  pages from   Touschek's official PhD record    from Glasgow University Record of Higher Degrees, courtesy of \UGA. }
\label{fig:Gunn-PhDrecord} \end{figure}

Touschek received his degree on November 5th, 1949. We show in Fig.~\ref{fig:Dphil} the certificate by the Academic Senate and Touschek's  official photograph on the occasion of the doctorate.   
\begin{figure}
\begin{centering}
\includegraphics[scale=0.108,angle=90]{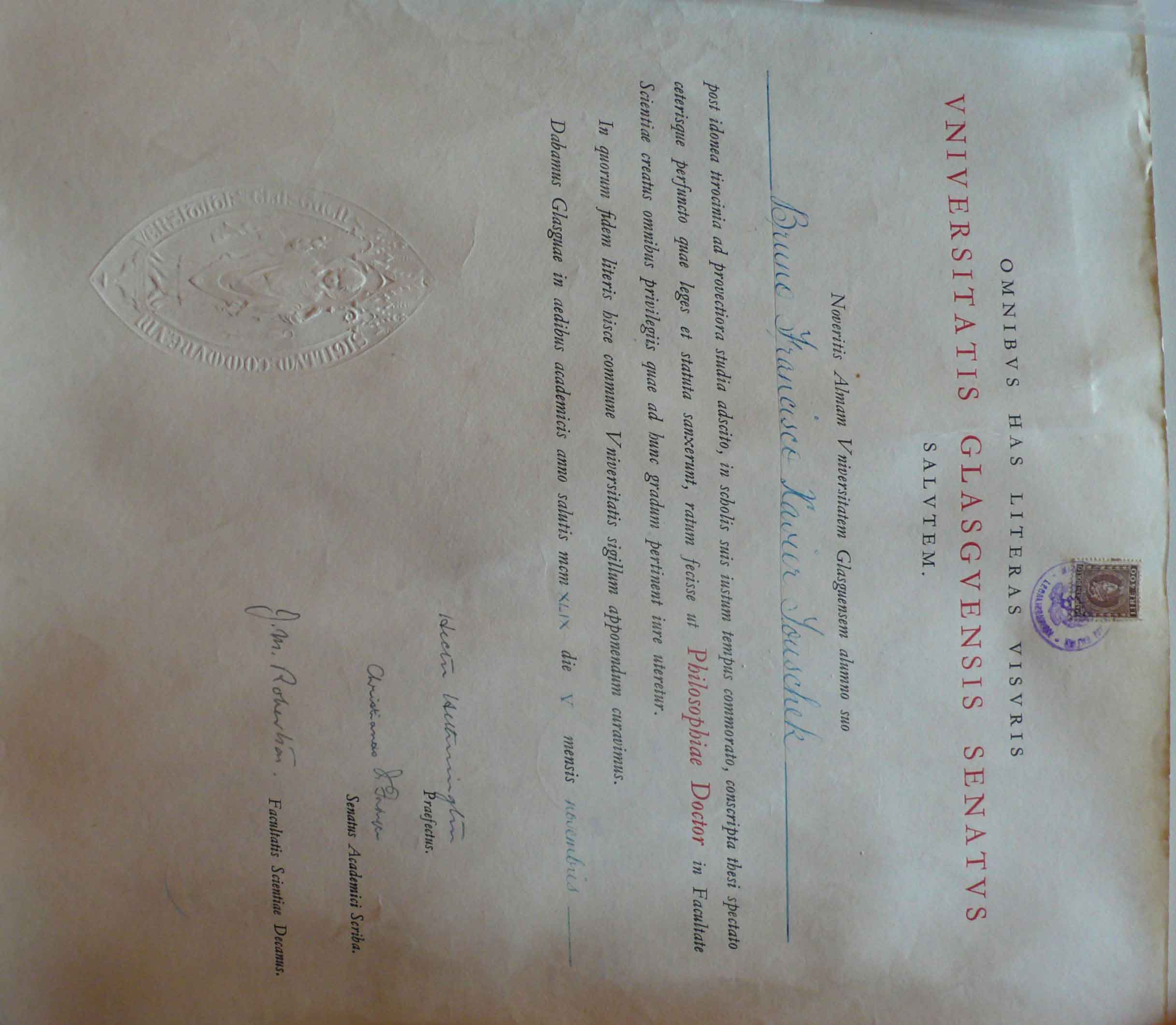}
\includegraphics[scale=0.1]{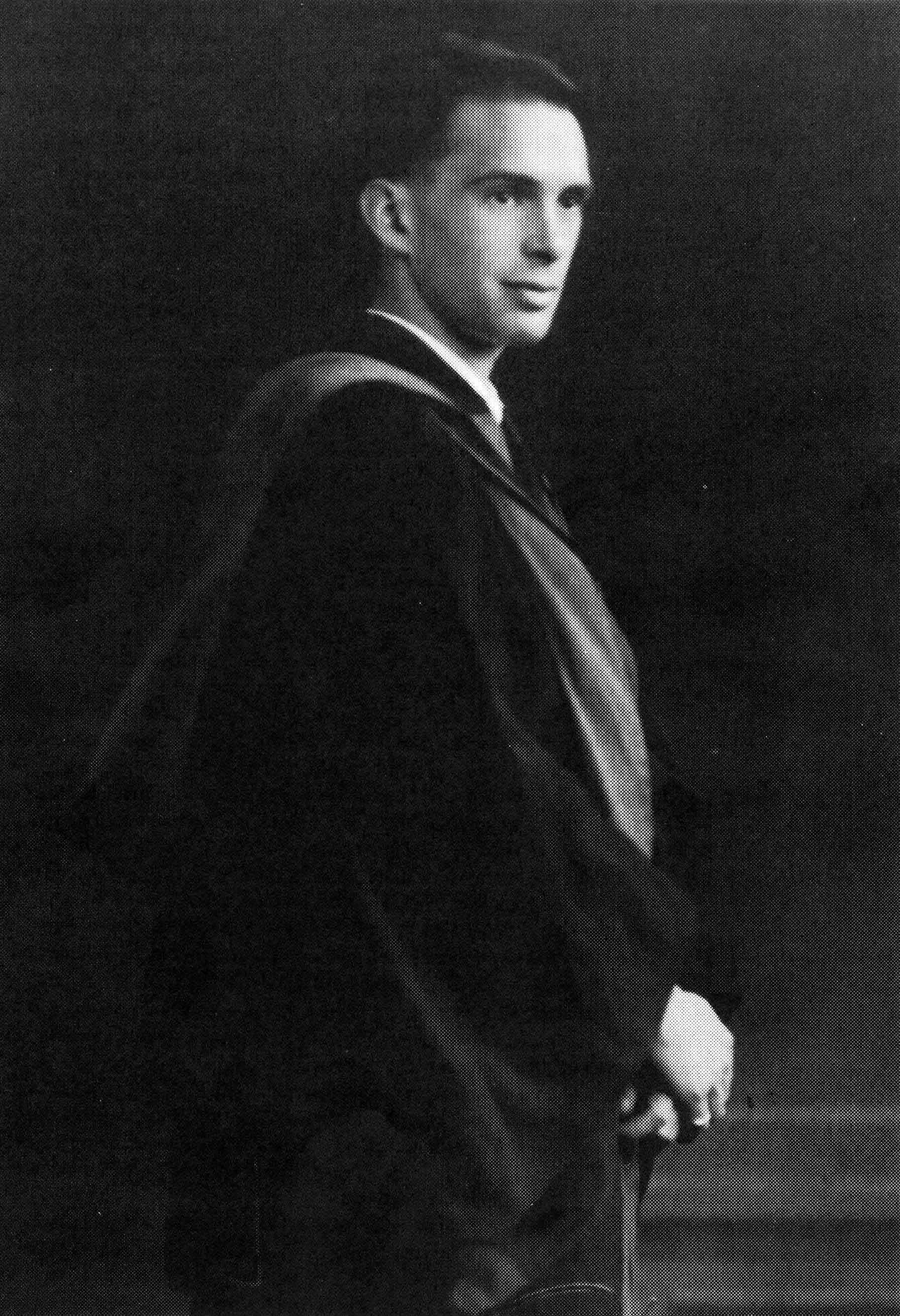}
\caption{Touschek's certificate of award of Doctor of Philosophy  from University of Glasgow and his official Graduation photograph  \citep{Amaldi:1981}. Courtesy of Mrs. \EYT.}
\label{fig:Dphil}
\end{centering}
\end{figure}
Amaldi writes that, immediately after his doctorate,   Touschek was appointed `Official Lecturer in Natural Philosophy' at University of Glasgow \citep[12]{Amaldi:1981}, a position he held for three years.

Shortly after Touschek's doctorate cerimony,    a conference was organized by Max Born at the University of Edinburgh on November 14-17 \citep{Edinburgh:1949}. Through the year, Bruno had often visited Max Born in Edinburgh and, at some time,  started working on an Appendix to the 
new  edition of Born's famous {\it Atomic Physics} book with a contribution on the process of   $\beta$-decay.
The subject
had interested Touschek 
since his \Gott\ days.  It was now understood that the underlying  mechanism could be used to distinguish  the  so-called  $\mu$-meson, discovered in 1936 in cosmic rays, from the proposed carrier of nuclear forces (successively named the $\pi$ meson). The subject of particle decays  was   at the center of debates, as one can see from  the  Conference preliminary program   shown in Fig.~\ref{fig:Edinburgh-conference}, with  cosmic rays results  still the winner of the day in terms of experimental high energy particle physics. 
 One notices   that among the speakers 
 there was  Bruno Pontecorvo, who presented a talk  on {\it The decay products of the $\mu$-Meson}.
  He had spent the war years mostly in Canada, 
  and was then based at AERE.
\begin{figure}
\centering
\includegraphics[scale=0.4]{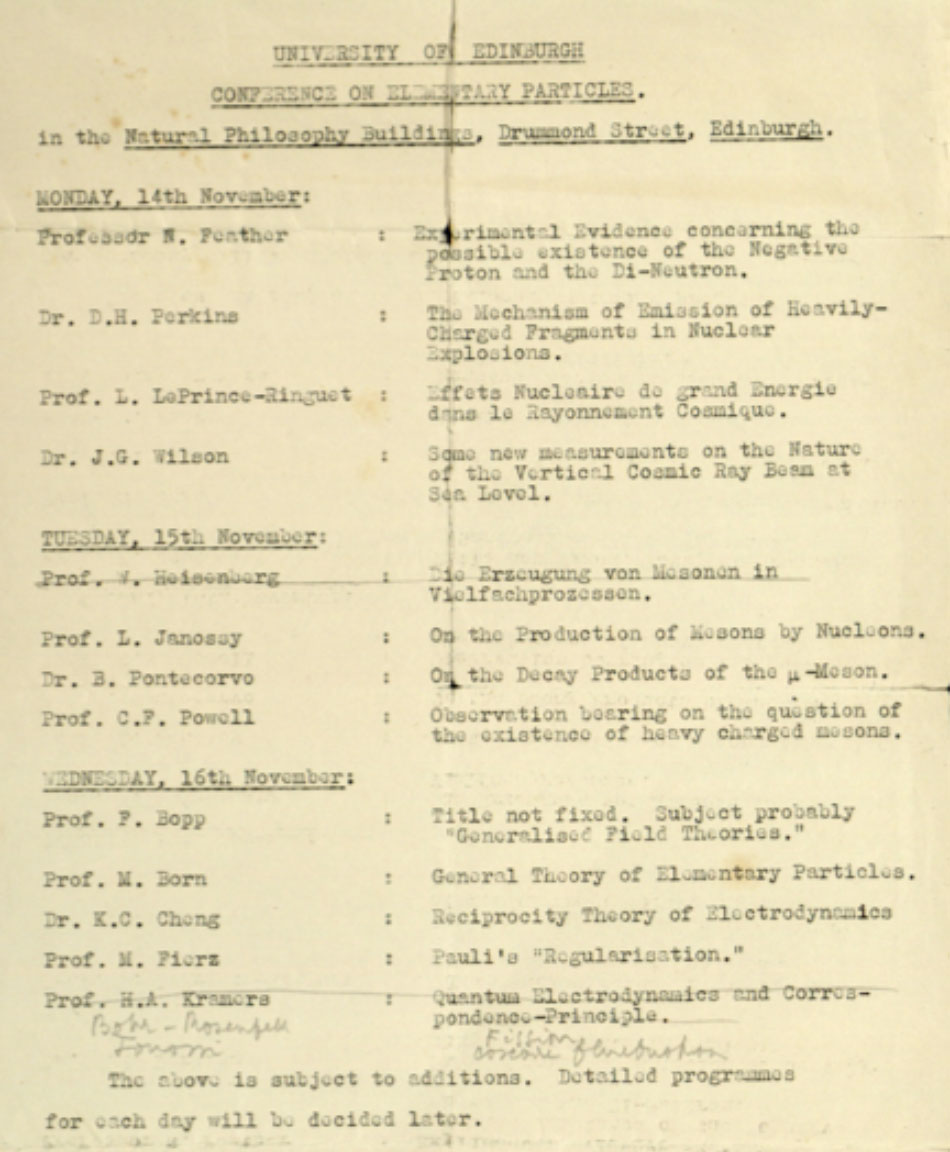}
\caption{Preliminary program of the 1949  Conference on Elementary Particles,  held in Edinburgh 14-16 November 1949,  from {Bruno Pontecorvo papers at Churchill Archives, Cambridge University.}}
 \label{fig:Edinburgh-conference}
\end{figure}

The conference had been organized by  University of Edinburgh to take advantage
of the presence in Scotland of Niels Bohr, who gave that year's \href{https://www.giffordlectures.org/lecturers}{Gifford Lectures}, held
bi-annually in Scottish Universities (Anonymous 1950).\footnote{ Niels Bohr gave the 1949 Gifford Lectures at the University
of Edinburgh under the title \href{https://www.giffordlectures.org/lectures/causality-and-complementarity}{\it Causality and Complementarity}. The lectures remain unpublished,
but the audio recordings of 9 of the 10 lectures [lecture 2 is unfortunately missing] are maintained at the Niels Bohr Archive (http://nbarchive.dk). A manuscript entitled ``Summary of Gifford Lectures is reproduced in \href{https://www.nbarchive.dk/publications/bcw/}{\it Niels Bohr Collected Works}, vol. 10, {\it Complementarity beyond Physics (1928-1962)}; edited by David Favrholdt.}
  The Glasgow particle physics group, including Dee, Touschek and students, drove everyday to Edinburgh, with Bruno holding the wheel of Dee's car, and the students huddled in the back of the car, frozen  with fright at every turn of the road, since Bruno was still a  rather inexperienced driver. After the Conference,  the big event of the year in the Glasgow Physics Department took place  on Thursday and Friday, 17 and 18th November:    Niels Born came visiting  on his Gifford Lecture tour. The highlight of Bohr's visit was an inaudible lecture at the physical institute during which Bohr moved from the  lectern to the blackboard, to clarify a point which nobody could follow anyway.\footnote{Bohr's lectures were famously long   and often not understandable. One such reaction is reported from the   Pocono Manor meeting, March 30--April 2nd, 1948, in Pennsylvania.  At this conference, Feynman gave the first public introduction to his method, since then  referred to as  {\it  Feynman diagrams} and universally used to calculate particle interactions.  An interesting article related to \href{https://merakienterprisesllc.com/2015/03/06/the-unveiling-of-feynman-diagrams-at-the-pocono-manor-conference/}{The unveiling of Feynman diagrams at the Pocono Manor Conference} tells how Feynman's talk, presented at the end of the day,  was poorly received. In particular, `Bohr leapt to the mistaken conclusion that they represented a violation of Pauli's exclusion principle.' and after more questions were asked which Feynman appeared unprepared to answer, `Bohr rose and approached the blackboard where he delivered a long speech on the Pauli exclusion principle'. At the end of the session, it seems that almost nobody had understood what Feynman's method could do. See also a 2018  article by Ashutosh Jogalekar on \href{https://www.3quarksdaily.com/3quarksdaily/2018/05/the-birth-of-a-new-theory-richard-feynman-and-his-adversaries.html}{The Birth of a New Theory:  Richard Feynman and His Adversaries}, in {\it 3 Quarks Daily on-line magazine}.}
The  talk was given to a packed audience of students and professors alike, who had all come to listen and see the {{\it der Physik papst}, the  pope of physics}, as he was called.  A visit from the King of England  would not have given   half as much trouble. Bohr was accompanied by his assistant  Stefan Rozenthal and his  long-time collaborator  L\'eon  Rosenfeld, whom Touschek  knew from Birmingham and Alpbach. Touschek drily reflected that Bohr's esthetically supreme choice in his adepts' name allowed no choice for someone with a non descript name such as his own. But while Bohr could neither  catch nor pay attention to Touschek's name, {he must have been sufficiently impressed by the young man's intelligence that} he invited him to come to Copenhagen on his next trip. This was an invitation which  Touschek was glad to accept for  the next fall.
Heisenberg, Bohr, Heitler and Rosenfeld are shown in Fig.~\ref{fig:Heis-Bohr-Heitler-Rosenfeld}, from around 1934.
\begin{figure}[h]
\centering
\includegraphics[scale=0.97]{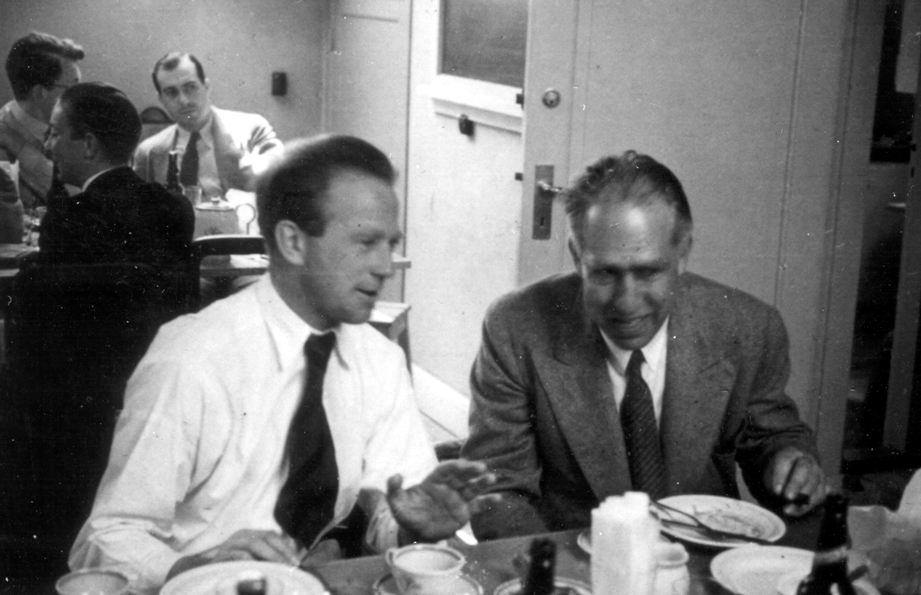}
\includegraphics[scale=0.1]{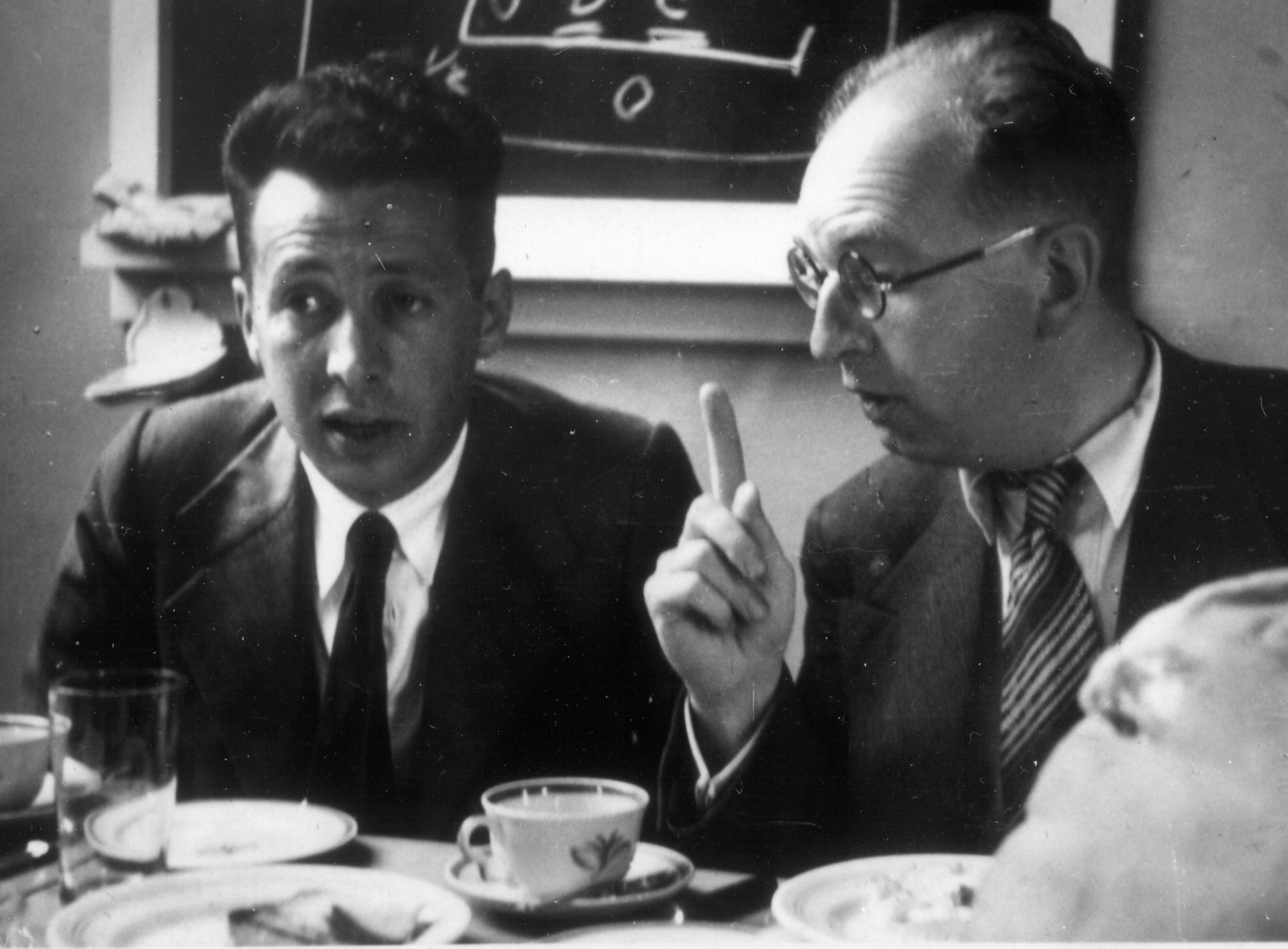}
\caption{From left: Werner Heisenberg and Niels Bohr, Walter Heitler and L\'eon Rosenfeld, in photos of  the mid 1930s (Wolfgang Pauli's Archive at CERN,  \url{https://cds.cern.ch/record/42960} and  \url{https://cds.cern.ch/record/42896}). For other photos of interest see 
\url{https://cds.cern.ch/search?cc=Pauli+Archive+Photos&ln=en}. Photograph at right is also to be found in \url{https://calisphere.org/item/ark:/28722/bk0016t4h8s/} where it is  indicated as  around 1934, Copenhagen conference.}
\label{fig:Heis-Bohr-Heitler-Rosenfeld}
\end{figure}

Bohr and  his wife arrived 
with   something like 71  pieces of baggage which Dee personally brought down from the third floor into his car, when they left. At the train station, Dee was having great difficulty keeping track of all the  porters   running in different  directions with that much luggage, not to mention the enormous tip to give them after the Bohrs left. All through this,  Bohr very quietly and steadily was  clarifying the subtleties of the uncertainty principle.\footnote{Letter to parents,  November 21st, 1949}

For Xmas, Touschek decided he had had enough of Glasgow and took a break to the North of Scotland, planning to go to Glencoe and Fort William in Inverness-shire, climb Ben Nevis and visit the island of Skye, enjoying the North in the heart of winter.  In Fig.~\ref{fig:inverness}, we show the sites he visited during this Christmas vacation.\footnote{Letter to parents, 
December 25th, 1949.}

\begin{figure}
\centering\includegraphics[scale=0.4]{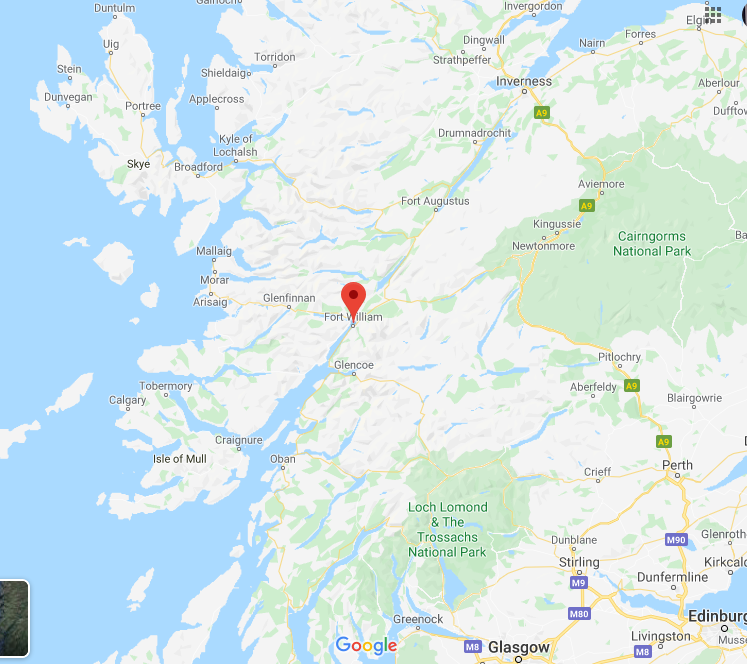}
\caption{The places \BT \ toured during his Xmas vacation in 1949.}
\label{fig:inverness}
\end{figure}



As the year was coming to an end,  from the wilderness of the North,  he was thinking of his future,  facing  possible new research directions or places to go and  his thought wandered to Rome and aunt Ada. The idea of   a summer vacation with  a trip to Italy and to the South started appealing to him once more. He thought of writing to her, but did not have her address and postponed.


\section{Lecturer in Glasgow: 1950-1952}
\label{sec:lecturer} 
While Bruno in Glasgow had been  working on his doctorate and, afterwards, continuing as ``Official Lecturer in Natural Philosophy'', in continental Europe the massive process of reconstructing European science was  undertaken. In France, in Italy, in Germany, in all the universities which the war and fascism had decimated of its scientists and laboratory equipment, younger scientists joined forces with the few who had remained and {had started to} rebuild  Europe. Such massive effort, which culminated with the foundation of CERN, but also with the launch of national laboratories and research institutions such as the \LAL \  in Orsay or the National Institute for Nuclear Physics (INFN) in Italy, led to the construction of new powerful  particle accelerators, which, unexpectedly, could compete with the American ones.
  {Indeed} at the end of the 1950s, a proton synchrotron was working at CERN, a linear electron accelerator in Orsay, an electron synchrotron in Frascati. {While} all  this was in the future,  the year  1950  saw Europe on the stepping stone of  
{an epochal change}, which was only partly reflected in the UK. Britain had won the war,  and throughout it had maintained -- indeed fostered and increased --  its intellectual and technological power,  but  the drive to start anew  in a joint effort, as  the continental nations were doing,  was  not as strong. One recalls that Britain itself 
{criticized the project of a European laboratory, even considering it a crazy idea \citep[114]{HermannEtAl:1987aa}, and hesitated in joining CERN at the beginning. Only in 1954 did Britain become one of CERN member states, after its scientists fully endorsed the idea and lobbied the government  \citep{Amaldi:1979aa}. }


Only  part of this was resonating  in Glasgow, as the world entered into the 1950s.  At this time, 
 Glasgow appeared stagnant to the young Touschek and he began actively  looking for a place to go and continue his path in theoretical physics. 
 And then, as 1949 ended and 1950 took over, there  unfolded  events which would affect nuclear physics in the UK, and, with it, Touschek's life. These events were the consequence of actions which had started 8 or 10 years before, and had been known only to a handful of people, from  both sides of the Atlantic. In February 1950 they became public, after having been brought to the attention of the US and UK highest political and scientific authorities  in January.  The  knowledge of what had happened   shook the world and changed the course of many people's lives, as well as the direction of nuclear physics research in UK. We refer here to so called {\it Fuchs affair}. Rudolf Peierls, Touschek's external PhD examiner,  was most directly affected, but Bruno's life as well  was indirectly influenced by it.

  \subsection{Fallout  from the Fuchs affair}
\label{ssec:fuchs} 
  
In early 1950,  the British scientific community was shaken by the discovery that the theoretical physicist in charge of the Nuclear Physics program at Harwell, Klaus Fuchs, was, and had been, a spy, who had passed crucial information about the making of the atomic bomb to the Soviet Union \citep{Close:2019}. Fuchs was German born, and had left Germany for the UK before the war. After getting his PhD at the University of Bristol in 1937, he  went to   Edinburgh, where he worked with Max Born and was  awarded  a Doctor of Science degree. He was  interned as an enemy alien  at the start of World War II, but was soon  released as his known enmity to  the nazi regime cleared his political allegiance.  His importance as a theoretical physicists   was underlined by  Max Born, who  wrote that Fuchs  ``was the soul of my research group [\dots] He  is in the small group of theoretical physicists in this country." \citep[45]{Close:2019}. He was  also highly considered by Peierls and in May 1941 he  became Rudolph Peierl's assistant in Birmingham, working with him on ``Tube Alloys'', the British atomic bomb project, beginning to pass information to the Soviets. In 1942 he became a British citizen,  and could thus move with Peierls to work on the Manhattan Project, later joining the Los Alamos top secret laboratory in New Mexico.\footnote{From \url{https://www.britannica.com/biography/Klaus-Fuchs}: ``Klaus Fuchs, in full Emil Klaus Julius Fuchs, (born December 29, 1911, R\"usselsheim, Germany, died January 28, 1988, East Germany), German-born physicist and spy who was arrested and convicted (1950) for giving vital American and British atomic-research secrets to the Soviet Union."} 
After the war, Fuchs  returned to England, and went  to the
nuclear research centre at Harwell.

When Fuch's betrayal become public, suddenly, in the UK,  all foreign born  scientists came under suspicion.  As \href{https://news.harvard.edu/gazette/story/2018/06/professor-richard-wilson-dies-at-92/}{Richard Wilson} remembers in his memoirs: ``The problem began in December 1949 when the nuclear physicist Alan Nunn May was arrested as a spy at McGill University in Canada just after a lecture. He had been working on the atomic bomb project and was accused of giving information to the USSR some five years before. 
The USA panicked. All foreign [born] nuclear physicists were suspect.'' \citep[111]{Wilson:2011}.\footnote{Richard Wilson (1926-2017) was an English born experimental physicist, who was Professor of  Physics at Harvard University, USA, and
 designed, constructed, and used the Cambridge Electron Accelerator 6 GeV synchrotron, which, from 1962 on, further probed nucleonic structure.}
{ Fuchs was one of them. His espionage activities were detected, and he was arrested on February 2nd 1950, upon which he admitted passing information to the Soviet Union since 1943}.\footnote{ He was sentenced to 14 years in prison. After his release in 1959 for good behaviour, he went to East Germany, where he was granted citizenship and was appointed deputy director of the Central Institute for Nuclear Reactions.} {We show a photograph of Karl Fuchs in the left panel of Fig.~\ref{fig:Fuchs-PonteFermi}.}

 Rudolf Peierls, Touschek's external examiner, was particularly shaken by the uncover of Fuchs as a Soviet spy. Peierls, also German born, had  supported Fuchs' eligibility for the Manhattan project. The two scientists were very close friends, having shared the experience of leaving Germany as Hitler came to power, and gone to the UK to continue their work. Peierls may have felt not only betrayed, but also himself in danger.\footnote{In his memoirs {\it Birds of Passage} \citep[223]{Peierls:1985aa} Peierls recalls: ``Our most dramatic experience was the Fuchs case. [\dots]  On the day I heard of his imprisonment under the spy charges  I went on to Brixton prison to see Fuchs. We had a long talk. Yes, he had given secret information to Soviet contacts.''} Peierls'  wife was Russian, and he had frequently visited the USSR before the war, also had collaborated and would do more so later, with prominent Russian scientists such as Lev Landau. In the witch hunt atmosphere which would soon engulf the United States  with the rise of McCarthysm, German scientists in the UK would also feel insecure. Peierls  himself was the target of suspicions and criticism \citep[392-401]{Close:2019}.\footnote{As Close writes: ``With his phone and mail continually monitored, Rudolph Peierls became part of the communist witch-hunt until 1954''  when the British security closed their file on him. But  the United States did not relent and in 1957 asked  the British Department of Atomic Energy that Peierls be given no access to American secret documents. At which point, Peierls decided to resign from his consultancy at Harwell.}

Everywhere in Europe and the US, the fear of a communist threat to Western society led to major changes to some scientists life. One can remember that on April 26th, 1950, \FJ \ \citep{Pinault:2000aa} was made to resign from the Chairmanship  of the French Atomic Energy Commission   for his sympathies for the Communist Party and activities in favour of an international ban on nuclear weapons \citep[157]{Close:2015}. And, in early September 1950, the Italian physicist Bruno Pontecorvo, shown in the right panel of Fig.~\ref{fig:Fuchs-PonteFermi}, one of Fermi's collaborators before the war,  suddenly disappeared from  Italy  with his wife and children, joining the Soviet Union, as it became known only five years later \citep{Turchetti:2012,Close:2015}.\footnote{Pontecorvo was  born in Pisa in 1913 from a prominent Jewish family. He  moved to the USSR in 1950,  
returning  to Italy  for the first time only   28 years later, in 1978,  on the occasion of Edoardo Amaldi's seventieth  birthday celebrations. 
He died in Dubna in 1993. See also \citep{Mafai:1992,Turchetti:2007aa}.}

\begin{figure}
\hspace{-1cm}\includegraphics[scale=0.433]{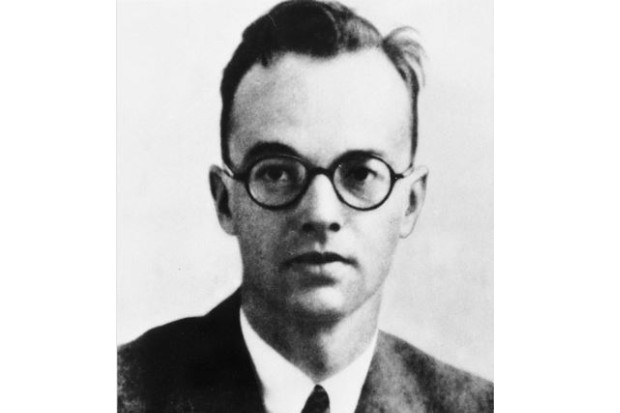}
\includegraphics[scale=0.445]{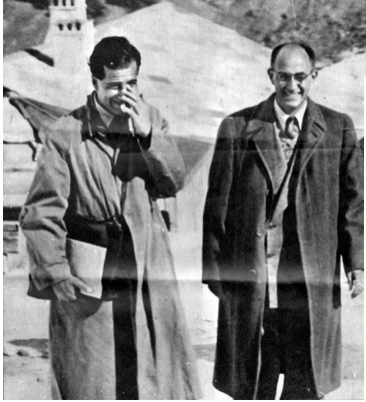}
\caption{Left panel: Klaus Fuchs, at the time of his arrest for espionage in favour of the Soviet Union, \copyright ImperialwarMuseum, from \url{https://history.blog.gov.uk/2020/03/02/whats-the-context-sentencing-of-atomic-spy-klaus-fuchs-1-march-1950/}. Right panel: Bruno Pontecorvo with   Enrico Fermi, shown at right, in a photograh taken on  occasion of the International Conference on Nuclear Physics, Quantum Electrodynamics and Cosmic Rays held in Basel, CH,  and Como, Italy,  in September 1949, from a contemporary newspaper.}
\label{fig:Fuchs-PonteFermi}
\end{figure}
The cold war had started. This atmosphere weighted heavily on the morale of UK scientists, and had strong impact on  career prospects  for non-British citizens working in the UK 

The impression from the Fuchs affaire  is  present in Touschek's letters in early 1950. Fuch's trial started on February 12th, very soon after he had been arrested, only one week after he had confessed \citep[334]{Close:2019}.  Touschek mused on the prosecutor's description of Klaus Fuchs as Dr. Jekyll and Mr. Hyde's personality, and drily observes that a double personality has nothing to do with nuclear secrets. He did not sympathize with the Soviet Union, but felt that the prosecutor's hype was excessive.\footnote{Letter to parents, February 14th, 1950.}

Unfortunately, the Fuchs affair was not a passing episode.

\subsection{Bruno Touschek and Max Born -- 1950-52}

During  his stay in Scotland, Bruno  exchanged letters with Max Born in Edinburgh and frequently visited him. From this correspondence, one can see that Touschek's  formation into  a theoretical physicist owes much to the relationship with  the great scientist. The letters often include questions of theoretical physics, articles to be discussed, even some fatherly advice, such as we glimpse from  a letter where  Touschek acknowledges some errors he had made and accepts Born's suggestion  that 'a little Puritan classical electrostatics would do me no harm'.\footnote{Touschek to Born, September 25th, 1950, 
Churchill Archives Center.}

Most of Born's correspondence with Touschek is kept in the Churchill Archives Centre, in Churchill College, Cambridge, UK.\footnote{We were made aware of this correspondence by Ms. Antonella Cotugno, from Rome University Library,  and the quoted text comes from 23 letters she copied with the kind permission
of  Churchill Archives Centre, Churchill College, Cambridge University, UK, where  the original letters are kept and are    
available in the    \href{https://janus.lib.cam.ac.uk/db/node.xsp?id=EAD\%2FGBR\%2F0014\%2FBORN}{Cambridge University, Churchill Archives Centre, The Papers of Professor Max Born}. These copies are kept in Bruno Touschek papers at Sapienza University in Rome.} 
The sequence of letters kept in Churchill Archives Center starts with May 18th, 1950, with Born informing Touschek 
{that he had just come back from Cairo finding the galley proofs of his book {\it Atomic Physics}. 
 He could not yet 
 work on them, because of a commitment to write an article on the physics of the last 50 years for the \textit{Scientific American}, 
 and asked Bruno to help him with the proofs, a copy of which he should have received as well, and invited him to come over for a day to Edinburgh. Bruno replied ten days later,} sending the corrected proofs. 
After mentioning an interesting effect observed by his experimentalist colleagues and for  
which he was trying to find a theoretical explanation, he inquired about  a position in Cairo:  ``On Monday you mentioned a possible vacancy in Cairo and I could not help thinking about it''.  As we know, since  passing his  doctorate  in November, Touschek had been looking for   a way out of Glasgow,  and started some fantasy about  such  position. Uncertain about writing directly to the Physics Department in Cairo, he mused whether this would   be a form  of 'escapism', born from  growing  doubts about  his future in the UK. He had  loved Scotland, but not anymore now that he had gotten used to it. This is an interesting observation, since it reflects an internal pervasive restlessness, which may have had a remote origin, probably when his mother died and, as a young boy before the war,  he would leave Vienna for periods of time to  stay with aunt Ada, in Rome. 

In any case, he reassured Born that  he won't want to leave for another year, because of work he had started and wanted to finish.  He closed the letter, asking Born if he would be angry at him if he applied for the position in Cairo. Shortly after, on May 31st,  Born thanked for the proofs, hoping to see him before Bruno  leave for Germany, and promising  to inquire about the position in Egypt.   

 After the summer, and visits to \Gott, Copenhagen and Austria (vacationing in Tyrol),  Touschek picked  up again the correspondence on September 29th, apologizing for some mistake in a paper he had been working on, and informed Born of a pending visit by Walter Thirring in October-November. During the summer Touschek had attended a small private conference organized  in Kitzb\"uhel, at  Hans Thirring's  house. This may have prompted plans  for  a visit to Glasgow by Thirring's son, Walter,  in the coming fall. In Thirring's autobiography, the tour which he took in fall 1950, first to Dublin invited by Schr\"odinger  and then to Glasgow, is seen as an apprenticeship and travel tour \citep{Thirring:2008}, before Thirring would  go back to  \Gott \ for the rest of the academic year. In this letter, Touschek proposes to take Thirring with him to Edinburgh, upon his arrival in Glasgow. 
 
 When Touschek wrote next, on October 26th, he apologized for the silence,  partly due to  have been very busy  leaving Dee's house to move to his own place in Oakfield Avenue, and mentioned work he had started with Thirring.   At the time of this visit, Walter Thirring  was interested  in going through electrodynamics with the covariant formalism of Schwinger and Feynman, in particular doing so through Heitler's book, Touschek's {\it bible} during the war years.  From this letter, the why and when of Touschek and Thirring's paper  \citep{Thirring:1951cz} on the Bloch and Nordsieck (BN)  problem \citep{Bloch:1937pw}, first appears.  
  Born welcomed the idea of Thirring  coming  to Edinburgh and on November 10th  invited both  Touschek and Thirring to come on the following Thursday for discussions, 
  'high tea', or possibly for lunch. During November, Touschek and Thirring wrote the paper on the covariant formalism of the BN problem and then, after having submitted it to the {\it Philosophical Magazine} Touschek sent a copy of the manuscript to  Born on December 18th. By that time, in early December,  Thirring had gone  back to \Gott. There is no mention of this paper in Thirring's memoirs, whereas for  Touschek it would later become a  milestone in 
   his formulation of infra-red radiative corrections to electron positron experiments \citep{Etim:1967}. 

The next letter in the Churchill Centre Archives is  dated April 3rd, 1951. Touschek apologizes for the silence, and, implicitly, for  lack of visits to Edinburgh 
during these 4 months, informs Born of his recent theoretical physics output and asks him for a referral for a position in Oxford as Senior Lecturer. He had written to 
{Maurice Pryce, Born's former student in the 1930s and his son in law, having married one of Born's daughters, but apparently Pryce was in Princeton, according to Sneddon,  and there was no reaction.} Born replied immediately, after a couple of days, that he was happy to give a referral and  thought that Pryce, who should come back from Princeton next July, should be delighted to have Touschek in Oxford.  He was happy to have heard from Touschek, inviting him to come again over to Edinburgh. 

As of  May 28th, Touschek had not received  any reply from Oxford, neither from Pryce nor from  the  University Registrar, and  became worried that  there would not be enough time to give notice to Glasgow University, in case of a  positive answer from there. One reason for this could be that  
Pryce,  after  a one-year sabbatical leave at Princeton, on his return would become head of the theoretical physics division at the Atomic Energy Research Establishment at Harwell, \citep{Elliott:2005}, replacing Klaus Fuchs who had been arrested on February 2nd 1950 and convicted, on March 1,  on a charge of spying \citep{Close:2019}.
In the same letter,  Touschek mentions his latest physics paper, a work which he calls `no more than a patent application', meaning that he had an idea and wanted to have it down in print, to establish  his priority. We know from correspondence with his family that he felt he often worked out  results which were  developed  around the same time by other physicists, but which did not get attention because he had not published them, or, more likely, did not propagate outside his own restricted circle. In a later CV, Touschek mentions one such case concerning  the Appendix  about  meson theory he wrote for  Born's book.  In the appendix,  he anticipated the universality of weak interactions, but  had not published his intuition about the subject, having discovered that it had been developed elsewhere, 
in particular by G. Puppi  \citep{Puppi:1948}.\footnote{Puppi  is credited for having been the first to discuss the muon-electron universality, which actually had been already mentioned by Bruno Pontecorvo \citep{Pontecorvo:1947}, but seldom recognized. Oskar Klein, too, had realized that all weak processes investigated thus far seemed to be due to the same  universal Fermi interaction  \citep{Klein:1948aa}, as did Tiomno and Wheeler \citep{RevModPhys.21.144}.} 

The preoccupation  to establish a priority   explains  the term  `a patent application'  for the paper he submitted on May 3rd, 1951. This paper \citep{Touschek:1951}  
is of note because it includes citation of  
 article by the Italian theorist Bruno Ferretti, submitted  from Rome and published in February 1951  \citep{Ferretti:1951}, which may have attracted Bruno's attention to the Physics Institute in  University of Rome. 

In the May 28th, 1951 letter,  Touschek also informed  Born of his  summer plans, 
which included driving the motor cycle he had just bought, and travelling with a companion, Dr. Rae, through the continent. But Born had also already left for the Continent, and would only be returning at the beginning of August. The department, in Edinburgh, offered to forward Touschek's letter to Born in \Gott. 

No more letters are recorded until the next year. What prompted such long silence?  Letters between September 1951 and January 1952,  when  Touschek was back in Glasgow  after the summer leave,   may have been lost. But it is just as likely   that Touschek, after  unsuccessful attempts to find an alternative position in Oxford, or Germany, turned his attention to Italy, starting his contacts during summer 1951, as we shall see later. This preoccupation would engulf him.  Back to Glasgow, he was also very busy with  teaching, working with his  research students, and  worrying about  a (never published) book with Ian Sneddon.

The next group of letters starts in January 1952, but they are   less frequent. From January through March, there are various exchanges  about slides and photographs of notable scientists in Born's possession which Touschek wished to use for 
`a chatty talk' on the historical development of quantum mechanics he had been asked to give at the Glasgow Physics Institute. In Fig.~\ref{fig:QM_Lecture_BT} we show the scheme  he sent to Born, together with a Glasgow Herald cutting about the talk. A long silence follows after this, as no other letters appears to have been exchanged. 
\begin{figure}
\centering
\includegraphics[scale=0.18]{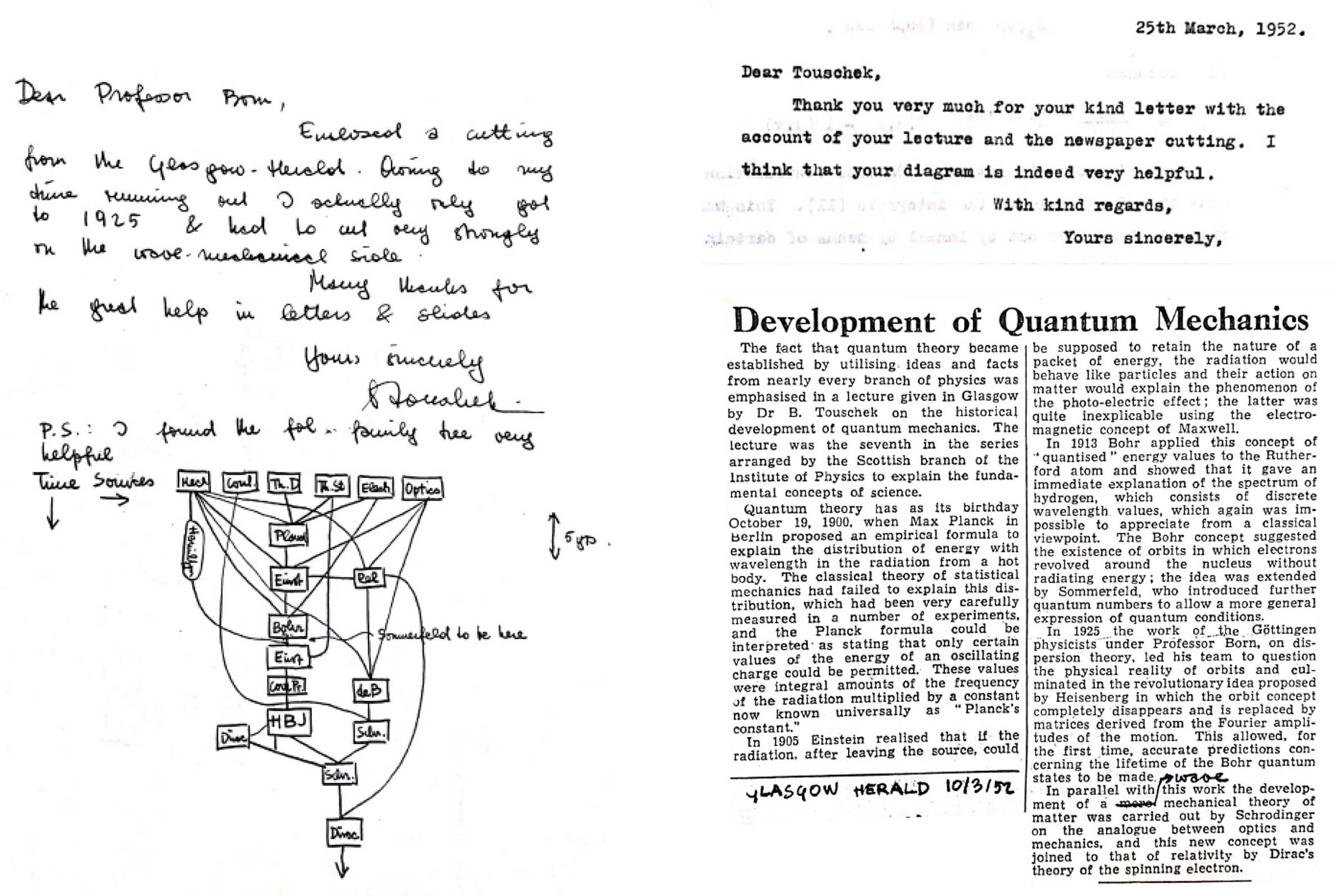}
\caption{ Touschek's handwritten letter to Max Born, and a  Glasgow Herald  article about  the lecture he held  at the Glasgow Physics Institute. Courtesy of \CHA. }
\label{fig:QM_Lecture_BT}
\end{figure}


On November 6th, Touschek writes that he is moving to Rome shortly, and would like to visit Born and his wife, once more. To which Born  replies on November 10th, in a somewhat  reproachful way, that he had heard of Touschek's move,  and congratulates him for the new position. Adding  his hope that Touschek  will be happy there, he also comments that 'to live in Rome, alone, is a great privilege'.  He will  try  to find a time to see him, among all his engagements, such as the Gifford lectures taking place in Edinburgh. Born's letter ends by saying that they (himself and Mrs. Born) are `quite well, apart from getting old and always being tired'. Then, in a last letter written just on the following day, concerned that his previous message 
 had  not been sufficiently welcoming, he urges Touschek to come, adding, as an inducement, that he would like Touschek to explain him a recently published  paper by Heisenberg \citep{Heisenberg:1952zz}. Born was going  to give a lecture on Heisenberg's work in the coming month of December, in London,  and since Heisenberg referred to Born's old non-linear electrodynamics, he wished to mention something about it. 
  Born confesses that despite his efforts he has not been able to understand Heisenberg's work, but hopes Touschek knows something about the matter.
It is amusing  to compare  Born's comments  with the interest these papers have generated through the years:   Heisenberg's  paper has received continuous attention since the time of his publication,  and  Born's  non linear electrodynamics work  \citep{Born:1934gh,Born:1935ap}, has recently been subject of  growing interest, as one can see from the 2020 citation results for Heisenberg and Born's papers  in  Fig.~\ref{fig:Heisenberg-Born-citations}.
\begin{figure}
\includegraphics[scale=0.68]{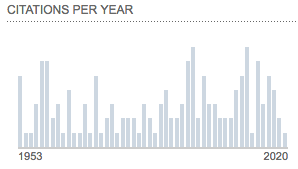}
\includegraphics[scale=0.33]{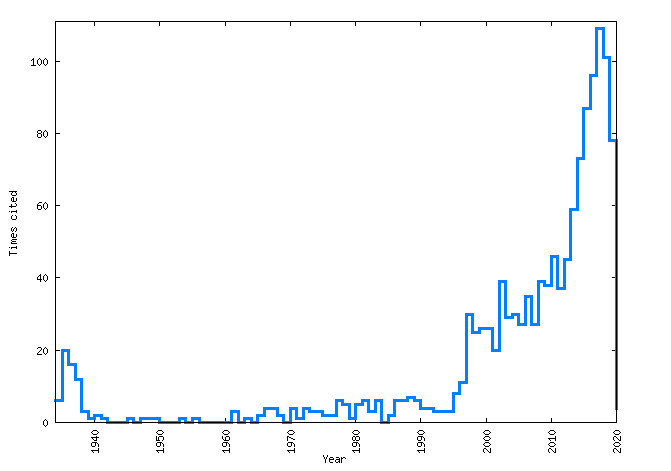}
\caption{The figure shows, at left, the impact of Heisenberg's 1952 paper \citep{Heisenberg:1952zz} mentioned by Max Born in the 11th November, 1952 letter to Bruno Touschek, and, at right,  the impact of Born and Infeld's article \citep{Born:1935ap},  cited by Heisenberg. The plots are  a 2020 citation analysis from the repository \url{http:inspirehep.net}.}
\label{fig:Heisenberg-Born-citations}
\end{figure}

In conclusion, as  we shall see in detail in the following pages,  the long intervals in the Born-Touschek correspondence both in the summer of 1951 and more so in summer  1952, 
may have been  related  to Touschek reluctance to tell his friend and mentor that he was going away,  abandoning Scotland, the frequent visits to Edinburgh, the warm friendship between them, the  train rides through the countryside  from West to East and back,  afternoon `high tea' at the Borns' home, the intellectual stimulus of Born's genius and his Seminar. 
He  was  focusing to find  a position in Italy and leave Glasgow. As he wrote to his parents, he was planning `betrayal' and could not face to tell his mentor. The Oxford option had  not worked out, and possibilities  to go to Germany had either dried out or were not sufficiently appealing. For reasons  which included both his physics interests and personal story, it was to Italy that he chose to go. In the sections to come, we shall try to unravel the motivations and the possibilities  which  made him decide to go South.

 \subsection{Planning betrayal}
 \label{ssec:planningbetrayal}
In this section we shall return to 1950, and  to   the three full years Touschek spent in Glasgow after his doctorate and how his decision to leave the UK matured.
 
 Indeed, only a few months into his position as Lecturer,  Bruno was finding the atmosphere in Glasgow rather stifling and was trying to see how to escape from it.  His  physics interests were also leading  him more and more into theoretical particle physics,  and Glasgow was not offering sufficient stimulus.   He had slowly lost interest in the Glasgow synchrotron program, which was stagnating.  The building was growing and growing, but Touschek had no more interest in just having the synchrotron built, probably for lack of  enthusiasm in the physics program. He could envisage the synchrotron to be finished in one or, rather,  two more years, and then what?  What to look for? Which new programs could be seen as natural follow-up to the 300 MeV synchrotron? In Glasgow, Touschek mused, only Dee was still interested in the project, everybody else was reading {Nature's jobs openings   and  looking for positions  elsewhere. }

Touschek was right in feeling that the UK was not offering  him much of  a way forward. 
In two years, by 1952, when the Glasgow synchrotron could be close to start operating,   European physicists in the continent would be ready for   the challenge  of  building  a large international laboratory, such that it could possibly house much more powerful accelerators. In the department in Glasgow, the atmosphere was changing as well. Touschek's friend and collaborator Ian Sneddon was leaving, first for the United States, then for a new position in Staffordshire. We show a photograph of Ian Sneddon in 1951 in Fig.~\ref{fig:sneddon}.
\begin{figure}\centering
\includegraphics[scale=0.5]{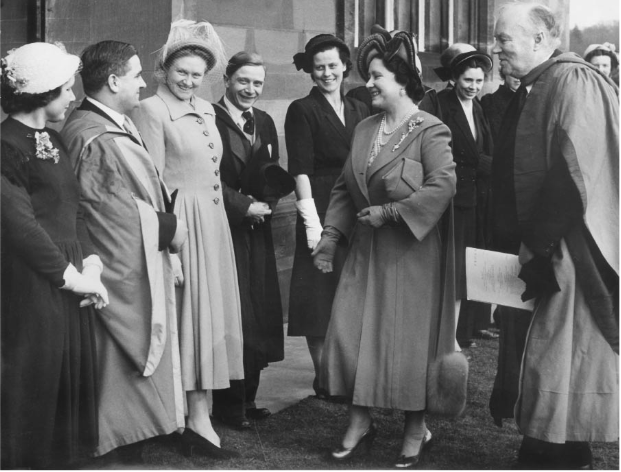}
\caption{Ian Sneddon being presented to H.M. Queen Elizabeth by the Principal, Lord Lindsay, at the formal opening of the University College of North Staffordshire, later to become the University of Keele,  on 17 April 1951 from \citep{Chadwick:2002aa}. }
\label{fig:sneddon}
\end{figure}
 Mostly because of the Fuchs affair,  this road  was {hardly open to Bruno}. He regretted to have refused a position in \Gott \ the previous year, but he now  decided he did want  to leave  and would use the summer to look for a job in the Continent.   The winter climate did not help, of course, nor did the town, which  did not offer much entertainment, the only exception being the occasional evening at the cinema, and this  as well was often depressing if not  making him dizzy with boredom.\footnote{Letter to parents  March 21st, 1950. He also mentions having been rather annoyed by the movie {\it Giovent\`u perduta}, by Pietro Germi (1947).}

To get over his low spirits, Bruno started a detailed planning for the summer, something which always helps to overcome  winter blues. In 1947, he had stayed  around Glasgow for his first summer (occasionally swimming in Loch Lomond and nearby places), in 1948 he had gone harvesting to the  Northernmost  part of Scotland, while   in  1949, as travel restrictions to Austria were eased,  he spent a full month in Tyrol with his parents. Now it was the time  for his first vacation as a Lecturer, and not as a research student.  The plan was to  be again  in Tyrol,  even organize a stay for  Professor and Mrs. Gunn, but not only. He was really planning betrayal.\footnote{Letter to parents on March 21st, 1950.} Soon after obtaining his doctorate he had already started thinking about looking for a position in the continent, now he was going to work on it.  In June the summer program was completed: 20.6--9.7 in \Gott; 10.7--18.8: Tyrol and surroundings; 18.8--25.8: Hamburg  and surroundings [Umbegeung]; 26.9--18.9: Copenhagen.\footnote{Handwritten letter to parents June 23rd, 1950.} 

The 1950 summer plans included, for the first time, a very long absence from Glasgow,  a three months leave, of which   almost one month to be spent in \Gott, and then visiting Hamburg and  Denmark, following Bohr's invitation in November 1949.  In Tyrol he was with his family, but was also part  to a small private conference in Hans
Thirring's house in Kitzb\"uhel, as mentioned in one of Touschek's letters to Born.\footnote{Letter to Born, September 29th, 1950.
Courtesy of the Churchill Archives Center, Cambridge, UK.}

The Gunns' planned  vacation in Tyrol  took  place during this extended visit as well.\footnote{In November 1950 letter to his parents, Touschek mentions that the Gunns had very much enjoyed the visit.}  M. J. Gunn, Gunn's  son born in 1954, remembers his father  often talking  about Touschek: ``-- it was clear he
liked Bruno and he had been a lively figure in the department in Glasgow. I heard the story about his escape from the Gestapo (or SS?) during that
period.  I [also] remember hearing about the Tyrol trip. It was my mother's first trip abroad. My father was more adventurous than sensible -- so one of their walks was on their hands and knees across a glacier ".\footnote{Private e-mail communication to G. P. by Prof. M.J. Gunn, February 17th, 2020. }

 Bruno  arrived in Copenhagen on August 31.\footnote{See letter to parents from Copenhagen, with handwritten date hard to read,  1950,  September or October 6th, likely to be September 6th, since he writes to Born from Glasgow at the end of September,.}
But  he was unwell, having caught a very annoying ear infection, and could hardly enjoy his visit. Notwithstanding  the rain and himself  being sick, he liked very much the city and was anxious for a bit of good weather to tour around. Unfortunately the language was a hard barrier  to overcome, as neither his native German nor his quite good  English could help.\footnote{Letter to parents from Copenhagen, September 6th, 1950, as discussed in previous note.}
{During all this time, Touschek was also keeping a close correspondence with Arnold Sommerfeld.   He had planned to visit him in Munich, on his way back  from the Austrian vacation, but it did not happen, and Touschek apologized to him in letter on October 5th.}\footnote{\BT\ to Arnold Sommerfeld, October 5th, 1950 Deutsches Museum Archiv, NL 089, 013}

While planning betrayal, Touschek had also decided it was time to move out of  Dee's house and find his own home. Since he was leaving for the entire summer, this was particularly easy. Thus, back from his extended leave, while he is still writing to Max Born from   11, The University, on September 29th,   we find him writing from 51 Oakfield Avenue on October 26, announcing Thirring's presence and his house move. A detailed drawing of his room is   shown in Fig.~\ref{fig:piantina50}, together with a contemporary view of 61 Oakfield Avenue, the residence listed in the University records for the 1950-51 academic year.

\begin{figure}
\centering
\includegraphics[scale=0.633]{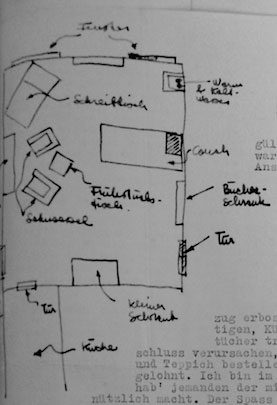}
\includegraphics[scale=0.25]{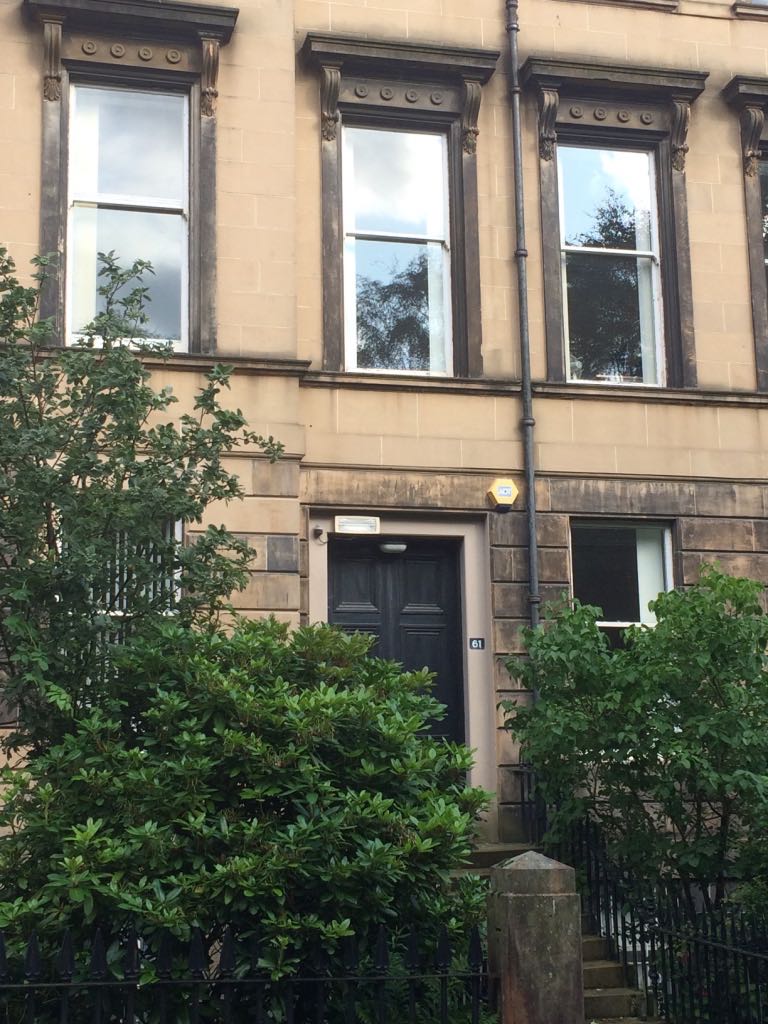}
\caption{At left we show a drawing  by \BT \ which describes the layout of  his new lodging at  61 Oakfield Avenue, in Glasgow, extracted from a November 1950 letter to his parents. Courtesy of  Mrs. \EYT. The right panel shows a contemporary view of 61 Oakfield Avenue, photograph by G.P.  August 8th, 2019. }
\label{fig:piantina50}
\end{figure}

\subsection{A visit from Walter Thirring}
In Fall 1950, an important development in Touschek's theoretical works, took place. Such development was stimulated by  a visit by the  young theorist, Walter Thirring. 

Since his arrival, Touschek's scientific interests had shifted from accelerator physics towards particle physics, where a new world was opening through the experiments  made possible by post-war  accelerators, such as synchrotrons. 
The new experiments  allowed to observe  particle interactions in  laboratory settings. Unlike the case of cosmic rays physics, where particles come from the sky and their origin is outside the experimenter's control, accelerators  allowed a choice of projectile and target, electrons or  protons,  and their energies could be  controlled. As physicists say, the initial state is  known, and the observation of the  scatterred particles (the final state),   can give precise information on  what happened during the transition.  Thus,  the interactions  among elementary particles  in the newly opened  high energy regime could  be studied with greater accuracies. New  theoretical formulations and  techniques were developed, and  took central stage, as
 infinities plaguing the calculation of scattering processes were being cured by
including virtual scattering processes in the calculation, in a combination of relativistic field theory
and perturbation theory.\footnote{See also David Kaiser's {\it Drawing Theories Apart: The Dispersion of Feynman Diagrams in Postwar Physics} \citep[53]{Kaiser:2005}.}  
 The covariant formalism, in which relativity is formulated,
made calculations more transparent and became the new language of particle physics.

 It is in this scenario, that Touschek started looking again at an old problem, the so called ``infrared catastrophe", discussed in a seminal paper by Felix Bloch and Arnold  Nordsieck, in 1937. This problem needed to be  reformulated in modern language, namely through the new covariant formulation, and the occasion to do so was the visit of  Walther Thirring  from Vienna, with a Nuffield Fellowship. Walther Thirring was the son of Hans Thirring,  whose lectures  Touschek had attended during the war, in occasion of his visits to Vienna.\footnote{For Walther Thirring's comments abot Touschek, see Thirring's autobiography \citep{Thirring:2008}. } During the summer, Bruno had learnt of the young man's travels from Vienna, first   to Dublin with Schr\"odinger, and then to Glasgow. He   held him in great consideration,  as the most brilliant  young theoretician in all of Europe, and had been looking forward to  welcome him in Glasgow.\footnote{Letter to parents,  November 1950.}

Walter Thirring arrived 
 in October 1950.\footnote{See Born-Touschek correspondence vs September or October, according to Amaldi.} 
It was an extremely busy time for Bruno.  
He had never before taken full responsibility for his lodgings, the furnishing, the cleaning, procurement of food. The many  house caring activities  engulfed him during October and November. All this had to be done in addition to looking after his research student, going to Edinburgh with Thirring, and, most of all, working with Thirring on the problem of the Bloch and Nordsieck method, writing the paper. They  submitted it  on November 29th  to the {\it Philosophical Magazine}, where it was published in March 1951 \citep{Thirring:1951cz}. This paper is very important in Touschek's future thinking about electron accelerators: after AdA's   proof of feasibility in 1961,  Touschek  planned the construction of a much more  powerful electron positron accelerator, ADONE. Because of the much higher energy range of the new machine,   the extraction of   physics results was  connected  to  the Bloch and Nordsieck theorem, and, in 1964,  Touschek started on the problem of summing the infinite number of   soft photons emitted before and after  the collisions.\footnote{{This problem  led Touschek to  train a group of young theoretical physicists recently graduated from  University of Rome, as described (in Italian) in \href{http://www.analysis-online.net/wp-content/uploads/2013/03/greco_pancheri.pdf}{Frascati e la fisica teorica}}.} 

Thirring left in early December 1950, barely  after only 2  months, and Touschek started again feeling that Glasgow could not offer the type of environment he sought.  The joys and pains of teaching {were also hitting}
 him. Students sometimes complained about difficulties in his course, or he felt that his research students were too slow and lazy to carry through some new research lead he gave them. In February, he observed with some amusement  the  frolics on the occasion of the 500th anniversary of the University of Glasgow foundation, when the students behaved `pleasantly ghastly'  and threw eggs and rolls of toilet paper at the Senate.\footnote{The letter  to parents is  dated  February 12th, without the year, but   mentions the 500 year anniversary of the university, which was founded in 1451.}



\subsection{The Southern way}
After Thirring left, Touschek started looking for a position elsewhere,  in Switzerland, with a new institution created by UNO -- CERN,  perhaps --  or in Germany,  in Munich,  but  only if Heisenberg were to move there. He expressed all these hopes and plans in his letters home. The stays in Europe during the summer made winters in Glasgow harder to bear. Max Born was going to retire in one or two years and planned to go back to Germany, so did many others.  As we have seen from the correspondence with Born, in the spring, Touschek aimed for a Senior Lectureship position in Oxford, but nothing came of it.  The solution were to come from a different direction,   through the Italian theorist Bruno Ferretti, {who, at the turn of the 1950s, was already deeply involved together with Edoardo Amaldi in discussions about the foundation of CERN \citep[67]{HermannEtAl:1987aa}.}

  \begin{figure}
\centering
\includegraphics[scale=0.6]{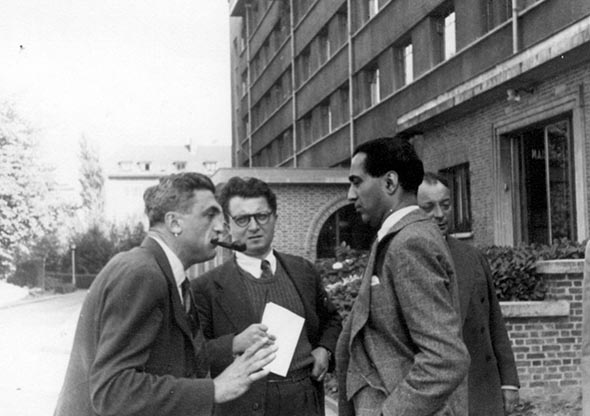}
\caption{From left, Felix Bloch, Bruno Ferretti, Homi J. Bhabha, Wolfgang Pauli, at the 1948 Solvay Congress. CERN Pauli-Archive-Photos PHO-063-1 \url{https://cds.cern.ch/record/42752/files/063.jpg?subformat=icon-1440}.}
\label{fig:Ferrettii-1948}
\end{figure}

The association between Touschek and Ferretti follows   rather  standard research channels. It is possible that the two of them met in 1947 in Birmingham, where Ferretti was giving a course and working with Peierls.

Bruno  spoke Italian, an easy way to start talking to Ferretti.  The occasion could  have been the fact that Ferretti and Peierls were working on radiation damping, the phenomenon by which an accelerated electron looses energy by radiation \citep{FerrettiPeierls:1947aa},  and a problem which had interested Touschek since the war days. Bruno  may have  read and studied Ferretti and Peierls' paper and approached them  about the article.  However in summer 1947, when  the article had appeared,  Touschek may not have been thinking much about Italy, and he was still a first year research student, hardly in confidence  with a  well known senior scientist such as Rudolf Peierls, later his PhD external examiner, or his guest Bruno Ferretti. Ferretti  appears  in Fig.~\ref{fig:Ferrettii-1948}, a photograph taken at the 1948 Solvay Congress, attended by all  the best theoretical physicists of the time.

The connection between Touschek and Ferretti was certainly established or enforced   in spring 1951,  when Touschek  saw one of Ferretti's papers in {\it The Nuovo Cimento} \citep{Ferretti:1951}, which particularly interested him. Soon after,  he wrote, and submitted for publication, his own  article  citing Ferretti's \citep{Touschek:1951}.  From this to start writing to Ferretti about physics is just an obvious step, and
either before or after Touschek's article,  the possibility of meeting Ferretti in Rome during the upcoming summer vacation for a scientific discussion, arose. A visit to University of Rome was likely included in his summer plans.  At the same time, independently  of physics interests, the  visit to Rome would allow   to resume contact with his aunt Ada,  as he had been thinking of Italy and the holidays  spent  in Rome before the war, at her beautiful apartment in the Parioli neighbourhood.

The visit  to  Rome took place in   mid  July 1951,\footnote{Letter to parents, June 25th, 1951.} and  it is conceivable that a meeting between Touschek and Ferretti in the Physics Institute in Rome had been   foreseen   to discuss physics questions.
 A description of their first meeting in Rome is given  in \citep[13]{Amaldi:1981}: ``[In September  1952] 
 Touschek went to visit Ferretti at University of Rome. A few hours after their first meeting, spent discussing mutual scientific questions, they established such marked professional respect and personal attachment for  each other that Touschek decided to remain permanently in Rome.'' These lines give a very vivid  picture of Touschek and Ferretti's encounter, except that Amaldi places it in September 1952. However, from Touschek's accurate descriptions of the 1951 and 1952 summer travels in letters to his parents,  a visit to Rome in 1952 is unlikely,
 whereas he was certainly planning to be in Rome  in July 1951.
Our conclusion is that   Touschek and Ferretti met in Rome in July 1951, discussed physics together and the idea of applying for a position in Rome  came up. As recalled by Amaldi, ``It was mainly Ferretti's personality that, in 1952, attracted and permanently fixed in Rome an occasional visitor, Bruno Touschek [\dots] \citep[441]{Amaldi:1979ab}. Following  the extended European tour,  a correspondence took place between the two of them about such  possibility,
but the thing needed time to be  perfected.\footnote{Letter to parents,  November 8th, 1951} There was in fact  the question of giving notice to University of Glasgow, and properly tie the end of Touschek's  commitment to Glasgow with the new position he was hoping to have in Rome. The details were probably worked out through such correspondence,  mentioned in Touschek's letters home, but  with no other records. 


 Christmas and New Year came, and brought very cold days. In    January 1952
 it was so cold that the water in the bathroom would freeze.

His unhappiness about the weather was not made better by his getting a cold around Christmas time. It forced him to stay home, and not be able to attend  the Dees' party.   Christmas and New Year Day were thus  terribly boring, and many small problems aggravated his spirit. Taxes to prepare, rations to obtain, writing a chapter of a book  with Gunn he could not really enjoy doing, and again the cold, made him unhappy and discontent. In between, always a keen sportsman, he would not neglect some winter outing, such as going to ski with Mrs. Gunn. Fig.~\ref{fig:sciatore} shows his visual commentary of the snowy winter.

\begin{figure}
\centering
    \includegraphics[width=0.48\textwidth]{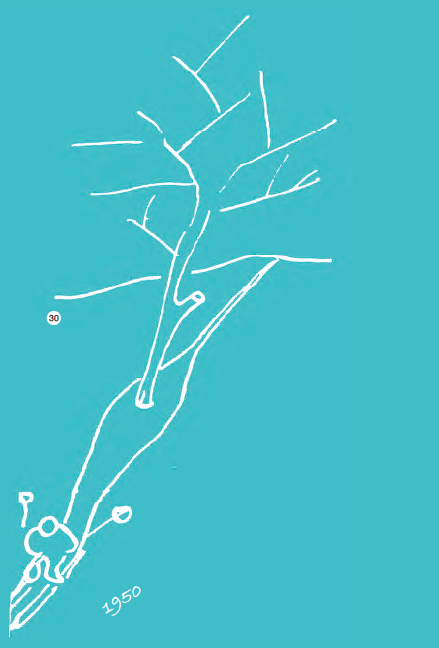}
    \caption{A drawing from a January 27th, 1952 letter.
    Graphics by A. Ianiro. Courtesy of Mrs. Elspeth Yonge Touschek.}
 \label{fig:sciatore}
\end{figure}

He went on with his teaching, and caring for his students, but as the spring came he  knew he would be going away, either to Italy or  somewhere else. 
In late August, possibly after the usual vacation in Tyrol, Touschek went down  to Milan from Austria,  through the Brenner pass.   The program,  included a visit to University of Milan. \ No explanation   is given  as to the reason for such visit, however, some physicists of that University had established a laboratory devoted to the development of applied nuclear physics, CISE (Centro Informazioni, Studi ed Esperienze),  in which Ferretti  and other physicists from University of Rome, where he was a Professor,  were also involved \citep[58]{Amaldi:1979aa}.
Thus, Touschek's visit appears to have been planned to  take advantage of  Ferretti's presence in Milan at this time, and  to discuss  the  details of the planned engagement with  the University of Rome.   Touschek had not yet decided whether to get one year unpaid leave from Glasgow, or resign from his Lecturer position  and he needed to clarify salary questions related to either  option.
After a couple of days in Milan, he crossed  into Switzerland through the Gotthard Pass, and reached Bern, where he received a letter from Ferretti who wrote he would  agree to the  postponement demanded by Glasgow until the first of January.\footnote{Letter to parents, September 10th, 1952.} In the meanwhile, Touschek was also seeking some other possibility, like a fellowship with UNESCO, which Heisenberg could try to get for him in case of need,  or a Lectureship in Bonn, which Wolfgang Paul could support. 
 On September 15th,  the situation cleared  up when Amaldi, then head of the Rome branch of the  National Institute   for Nuclear Physics (INFN), wrote officially to Bruno   offering a one year (renewable) position, adding that  in case of need,  one month salary could be advanced to him.\footnote{The newly founded institute was bypassing some of the burocracy inherent at the time in University administration. Such possibilities, one month advanced salary,  would in fact be unheard of in the regular university administration. For the history of INFN see \citep{Battimelli:2001aa}.} It is of interest to see  Amaldi's  list of   Touschek's expected duties, namely: a) research work in theoretical physics, b) discussions and advice to experimental physicists working on cosmic rays and possibly accelerators; c) not more than two hours per week of  teaching to advanced students in theoretical physics. 
 
Touschek  in the meanwhile had   decided to resign from Glasgow, thus solving the problem of missing salary and, on September 23rd, 1952, Touschek replied to Amaldi accepting the offer.
The final exchange of letters   put Bruno  on his way to Rome, sealing his future and leading him to conceive AdA less than 10 years later. 
Bruno's  future  was now traced. He would be immersed in a unique moment of hopes and dreams, when all seemed possible, and when much was indeed accomplished. In  the 1950's Italy, which had been unified from North to South only in 1870, which had given access to vote to every citizen,  women and men,   only in 1946, entered   into  a period of great cultural, economic and political development, which led to what was later   known as the {\it Miracolo italiano} of the 1960's.\footnote{
 Italy had  been unified into a political entry from North to South in 1870, Rome becoming the capital city of the  new Italian state on February 3rd, 1871. The Italian Miracle, as it was called, signalled the full transformation of  Italy into  a modern state, economically competing with other European nations.}

As one  leaves this part of Touschek's life, which represents one of the many roads ultimately leading to particle colliders, the UK road, 
we show a 1952 photo of the Physics Faculty from the Department of Natural Philosophy of University of Glasgow, with \BT\ at right in the front row,  in Fig.~\ref{fig:glasgow}. A few months later, he left.
 \begin{figure}
 \centering
 \includegraphics[scale=0.35]{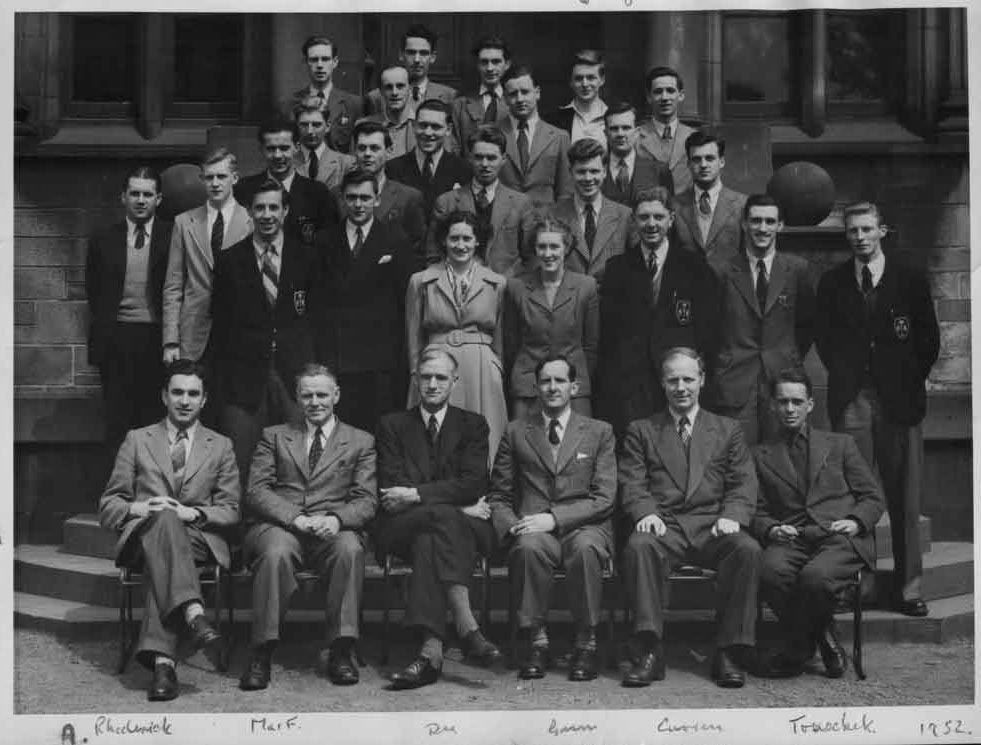}
  \caption{The physics faculty at University of Glasgow in 1952, with Philip Dee at the center, and \BT \  seated at right, from  
  \url{https://universitystory.gla.ac.uk/images/UGSP01004.jpg}. }
  \label{fig:glasgow}
 \end{figure} 
 

 When Bruno Touschek arrived in Rome,  it was the end of the year. He had left  Glasgow where winter brought early darkness  and freezing cold, and after a  brief stay in Bern,
 crossed the Alps through Switzerland, and was charmed by warmer weather and  the Roman  food. As it often happens to travellers from Northern lands, he found he did not need to walk fast to keep warm, rather he needed to slow down his nordic walking, and must have welcomed  the  lighter days of the Rome  winter.  On his arrival he dutifully visited his aunt Ada, and was housed  in  a  nice family pension near the University, booked  for him through  the Physics Institute. He found the Institute really  excellent, with no feeling of estrangement, as  he met there Patrick Blackett whom he knew from Glasgow and Wolfgang Pauli whom he knew from \Gott.  He also found the colleagues from the Institute very  interesting and welcoming.\footnote{Letter to parents, December 30th, 1952.}
 
 Bruno's life had been broken, but it was now being pieced together again. Physics in Italy was fully flourishing and the future there looked really exciting.   
 He felt ready for a great adventure.

\section{Sources and acknowledgements}
 Since 2002, a main primary source has been the Bruno Touschek papers kept at the ``Edoardo Amaldi Archives'' of the Physics Department at Sapienza University in Rome as well as contacts with Bruno Touschek's widow, Mrs. Elspeth Yonge Touschek, who gave us access to  important personal  material  in her possession about Touschek's life. In particular, until 2013, when she unfortunately passed away,    Elspeth  shared with us  many of the  letters  Touschek  wrote to his parents during most of his life. L.B. had been visiting her and discussing a project for a docu-film \href{https://www.youtube.com/watch?v=R2YOjnUGaNY}{Bruno Touschek and the Art of Physics} about Bruno's life since 2003, one of us, G.P., had known her since 1966.  

New information came from recent searches, and we acknowledge the courtesy of reproduction or consultation of the Archives of the  Physics Department of Sapienza University in Rome (SUA), in particular Edoardo Amaldi (EAA) and Bruno Touschek (BTA) papers,  and copies of Max Born correspondence with Bruno Touschek from   \href{https://janus.lib.cam.ac.uk/db/node.xsp?id=EAD\%2FGBR\%2F0014\%2FBORN}{Churchill Archives Centre, The Papers of Professor Max Born}  
of   Churchill College,  Cambridge University (CHA), the Archives of the Deutsches Museum, Munich, Arnold Sommerfeld papers, 
the  University of Glasgow Archives \& Special Collections, University collection, GB 248 DC 157/18/56 (UGA) 
and Special Collections, and the Archives of Keele University Library.
  
G.P. is grateful for the hospitality from  the Center for Theoretical Physics of the  Massachusetts Institute of Technology during the time this note was prepared. L.B.   gratefully acknowledges support from the Max Planck Institute for the History of Science during the preparation of this work.

We acknowledge collaboration and advice from many sources. We thank  Michael C. Gunn from Birmingham University for contributions about his parents, including memories of the Tyrol vacation and photograph of his parents on their wedding day,   David Saxon of Glasgow University for a copy of Touschek's PhD thesis. We thank for consultion Gianni Battimelli  and Maurizio Lusignoli of Sapienza University in Rome,  and Fred Jegerlehner  from Humboldt University in Berlin. We  thank Yogendra N. Srivastava from University of Perugia  for  theoretical physics advice, Robert MacLaughlan of Newcastle University  and Maria Daniella Dick of Glasgow University for  advice about Glasgow places and  a tour searching for Touschek's different house locations in Glasgow.  We are indebted to  Cinzio  and Antonio Ianiro for help in  graphical elaborations,  
Amrit P.  Srivastava for advice and collaboration  in translating Touschek's letters. Thanks are due to  Neelam F. Srivastava from Newcastle University for continuous advice and  checking of the text. L.B. is especially grateful to Juan-Andres Leon from Max Planck Institute for the History of Science for engaging and insightful discussions.

We are also grateful to  the staff of the Library and Archive of the Physics Department of Sapienza University in Rome, and in particular Antonella Cotugno,  for collaboration on the Touschek-Born correspondence and   to the \LNF \ staff, in particular Orlando Ciaffoni  for technical support, the Library and Educational Services  staff members  Debora Bifaretti, Tonino Cupellini, Claudio Federici and Lia Sabatini   for continuous assistance and collaboration.


\bibliographystyle{plainnat}
\bibliography{Touschek_book_May_8}
\end{document}